\documentclass[superscriptaddress,preprint,nofootinbib,nobibnotes,eprint,amsmath,amssymb,aps,prb,citeautoscript,titlepage,nobalancelastpage,floats,final,eqsecnum]{revtex4-2}

\usepackage[utf8]{inputenc}

\usepackage{amsmath}
\usepackage{amssymb}
\usepackage{amsfonts}

\usepackage{latexsym}
\usepackage{slashed}
\usepackage[svgnames]{xcolor}
\usepackage{soul}
\usepackage{braket}
\usepackage{MnSymbol}
\usepackage{wasysym}
\usepackage{mathrsfs}  
\usepackage{physics}
\usepackage{enumitem}         

\usepackage{graphics}
\graphicspath{ {graphs/} }

\usepackage{microtype}
\usepackage{multirow}

\usepackage{hyperref}
\usepackage{natbib}

\hypersetup{
	pdfencoding=auto,
	psdextra,
	colorlinks=true,
	linkcolor=Blue,
	citecolor=DarkRed,
	urlcolor=Blue
}

\usepackage{mathtools}
\DeclareMathOperator*{\osum}{\mathrlap{\hspace{0.00ex}\bigcirc}{\sum}}
\usepackage{easyReview}

\begin{document}
	
	\title{\Large{{\sc Fractal Subsystem Symmetries, Anomalies, Boundaries, and Effective Field Theory }}}

	\author{Heitor Casasola}
	\email{heitor.casasola@uel.br}
	\affiliation{Departamento de Física, Universidade Estadual de Londrina, Londrina, PR, Brasil}
	
	\author{Guilherme Delfino}
	\email{delfino@bu.edu}
	\affiliation{Physics Department, Boston University, Boston, MA, 02215, USA}
	
	\author{Yizhi You}
	\email{y.you@northeastern.edu}
	\affiliation{Department of Physics, Northeastern University, MA, 02115, USA}

	\author{Paula F. Bienzobaz}
	\email{paulabienzobaz@uel.br}
	\affiliation{Departamento de Física, Universidade Estadual de Londrina, Londrina, PR, Brasil}
	
	\author{Pedro R. S. Gomes}
	\email{pedrogomes@uel.br}
	\affiliation{Departamento de Física, Universidade Estadual de Londrina, Londrina, PR, Brasil}

\date{\today}
		
\begin{abstract}

This work reports an extensive study of three-dimensional topological ordered phases that, in one of the directions behave like usual topological order concerning mobility of excitations, but in the perpendicular plane manifest type-II fracton physics dictated by a fractal subsystem symmetry. We obtain an expression for the ground state degeneracy, which depends intricately on the sizes of the plane, signaling a strong manifestation of ultraviolet/infrared (UV/IR) mixing. The ground state degeneracy can be interpreted in terms of spontaneous/explicit breaking of fractal subsystem symmetries. We also study the boundary physics, which in turn is useful to understand the connection with certain two-dimensional phases. Finally, we derive a low-energy but not long-distance effective field theory, by Higgsing a fractal $U(1)$ symmetry and taking the deep IR limit. This description embodies in a natural way several aspects of the phases, such as the content of generalized global symmetries, the role of fractal symmetries on the mobility of excitations, the anomalies, and the boundary physics.

\end{abstract}

\maketitle

\tableofcontents

\section{Introduction}

Understanding the nature of entanglement has led to significant advancements in the comprehension of quantum matter \cite{Wen_2017}. Long-range entangled phases arise in strongly interacting many-body systems, and exhibit a plethora of emergent rich properties. These states typically arise in the low-energy physics of microscopic Hamiltonians with underlying generalized symmetries \cite{Nussinov_2009,Gaiotto:2014kfa} (recent reviews can be found in \cite{McGreevy_2023,Gomes:2023ahz,SchaferNameki2023,Brennan:2023mmt,Bhardwaj:2023kri,Luo:2023ive,Shao2023}). The ground state wavefunctions may exhibit spontaneous symmetry breaking of such generalized symmetries \cite{Wen_2019,Qi:2020jrf,Rayhaun:2021ocs}, and the ground state degeneracy on closed manifolds can be interpreted as a consequence of 't Hooft anomalies \cite{Gaiotto:2014kfa}.

Over the past decades, the study of quantum stabilizer codes, which possess highly entangled ground states, has greatly contributed to our understanding of topological ordering through microscopic Hamiltonians. Recently, a class of stabilizer codes with fracton topological order in three spatial dimensions and higher has garnered attention from both high-energy and condensed matter communities \cite{Chamon_2005,Bravyi_2011,Castelnovo_2012, Yoshida:2013sqa, Vijay:2015mka,Haah_2011}.  
The infrared (IR) sector of these stabilizer codes is captured in terms of generalized gauge theories possessing subsystem symmetries \cite{Vijay:2016phm}, which are a type of higher-form symmetries but with charge conservation along rigid subspaces.

Fractonic phases are broadly classified as type-I and type-II, according to the mobility of the underlying excitations. In all these phases, a single excitation, called a fracton, is completely devoid of mobility. Type-I refers to phases in which bound states are mobile, while type-II refers to phases in which not even bound states are able to move. (see Refs. \cite{Nandkishore:2018sel} and \cite{Pretko:2020cko} for reviews on fracton physics). Relatedly, while type-I fracton physics is associated with conservation laws along subspaces of integer dimensions (e.g. lines and planes), type-II fractons are often associated with conservation laws in subsystems of fractal dimension. Due to their intricate arrangements, these fractal symmetries strongly constrain the mobility of excitations, leading to the type-II fracton-like physics and to exotic entanglement behavior \cite{Haah_2014,San_Miguel_2021,Casasola:2023tot}.

It is quite remarkable that fractal structures take place in many modern physical applications. In particular, these self-similar patterns have attracted significant attention in areas such as quantum glassiness \cite{Castelnovo_2012,Nandkishore2017}, quantum dynamics in fractal geometries \cite{Kempkes_2018,Pai_2019a,St_lhammar_2023}, quantum computing \cite{Devakul_2018,Tantivasadakarn_2022}, and error-correction codes \cite{Haah_2011,Bravyi_2013}. Interestingly, these systems provide condensed matter platforms in which properties of fractals are reflected in physical observables.

In this paper, we provide an extensive characterization of a class of fractal topological ordered phases introduced in Refs. \cite{Castelnovo_2012,Yoshida:2013sqa}. These are three dimensional phases with very peculiar features: the physics in one of the directions is like in a usual topological order in the sense that the excitations are mobile along it, whereas the physics in the perpendicular plane is like in a  type-II fracton system, due to the existence of fractal subsystem symmetries in this subspace. The pattern of symmetries provides the system with very rich physical properties, which turn out to be extremely sensitive to the lattice size, implying a strong form of UV/IR mixing. These features challenge the characterization of such phases. Nevertheless, pursing our studies initiated in \cite{Casasola:2023tot}, we present here a reasonably complete description of various aspects involved in these intriguing fractal topological ordered phases. 

To be more specific, in Ref \cite{Casasola:2023tot} we have studied the following aspects:  we computed the ground state degeneracy for sizes $L_x\times L_y\times L_z$, with $L_x=L_y$, and the $xy$-plane corresponding to the subspace where the fractal subsystem symmetries reside. Then, we studied the symmetry operators and the corresponding 't Hooft anomalies between them, dictating the ground state degeneracy, and which can be understood in terms of spontaneous/explicit breaking of fractal subsystem symmetries. We also analyzed the mutual statistics between excitations. 

As mentioned above, the fractal symmetries are very sensitive to the sizes $L_x$ and $L_y$, so that the study of sizes $L_x\neq L_y$ corresponds to a highly nontrivial extension of that results, which we carry out in this manuscript. We address several additional new aspects. We present a detailed study of the boundary theory and the connection with two-dimensional systems with fractal symmetries. We also carry out a substantial study of these phases from the perspective of effective field theory, which makes the role of the generalized global symmetries very transparent and nicely incorporates the edge physics.

The work is organized as follows. In Sec.~\ref{sec:model}, we simply present the model and proceed to Sec.~\ref{sec:degeneracy} to compute the ground state degeneracy. In Sec.~\ref{sec:sym}, we discuss the line and fractal symmetries and examine the mixed t’ Hooft anomalies. The mutual statistics of the excitations is discussed in Sec.~\ref{sec:mutualst}. In Sec.~\ref{sec:SSB}, we discuss the ground state degeneracy in terms of spontaneous breaking of the subsystem symmetries. Sec.~\ref{sec:boundary} is dedicated to the boundary physics and in Sec.~\ref{sec:comparison} we analyze the connection with two-dimensional models. In Sec.~\ref{sec:eft}, we derive an effective description for the model and discuss several related aspects. We conclude in Sec.~\ref{sec:conclusion} with the final remarks. Subsidiary computations are presented in Appendices \ref{sec:MS1} and \ref{sec:MS2}.


\section{The Model \label{sec:model}}

\begin{figure}
	\includegraphics[width=7cm]{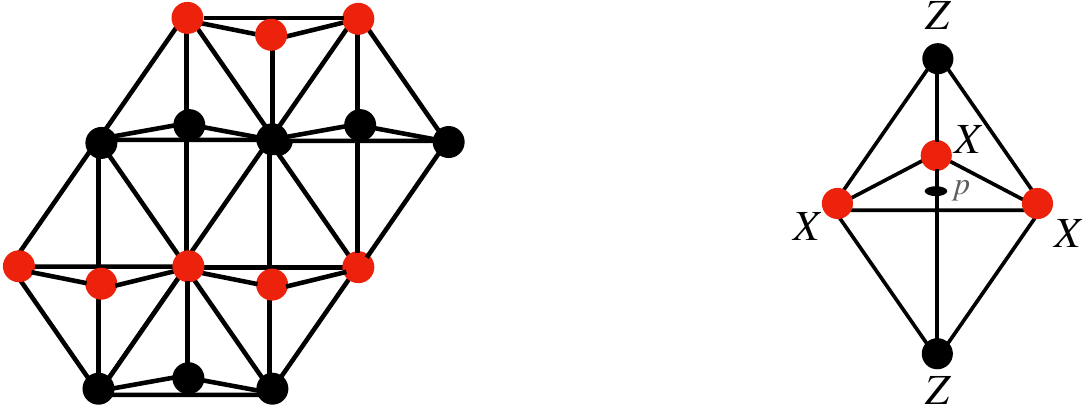}
	\caption{In the left  a representation of hexagonal close-packed lattice and in the right a TD operator.} \label{fig:Lattice}
\end{figure}

The model, originally proposed in \cite{Castelnovo_2012}, is defined in a hexagonal close-packed (HCP) lattice,  that is a structure composed by two triangular sublattices, $\Lambda = \Lambda_{1} \oplus \Lambda_{2}$, which are not perfectly aligned. Instead, these sublattices exhibit a relative shift, as depicted in Fig.~\ref{fig:Lattice}. On this lattice, we define the operators 
	\begin{equation} \label{eq:TDoperator}
	\mathcal{O}_{p,a} \equiv Z_{p + \frac{1}{2} \hat{z}} \,
	X_{p + \hat{e}_{1}^{a}} \,
	X_{p + \hat{e}_{2}^{a}} \,
	X_{p - \hat{e}_{1}^{a} - \hat{e}_{2}^{a}} \,
	Z_{p - \frac{1}{2} \hat{z}},
\end{equation}
where $X$ and $Z$ are Pauli matrices, $p$ is the center of the operator, and
\begin{equation}
	\hat{e}_{1}^{a} \equiv \frac{1}{2} \hat{x} + \left( -1 \right)^{a} \frac{ \sqrt{3} }{ 2} \hat{y} \quad \text{and} \quad
	\hat{e}_{2}^{a} \equiv -\frac{1}{2} \hat{x} + \left(-1\right)^{a} \frac{ \sqrt{3} }{ 2 } \hat{y}.
	\label{basisdef}
\end{equation}
The index $a=1,2$ specifies the two sublattices $\Lambda_{1}$ and $\Lambda_{2}$, respectively. In situations where the distinction is not essential, the index $a$  may be omitted. With this definition of $\hat{e}_{i}^{a}$, the operators on the $xy$-plane of different sublattices point in opposite directions. Figure~\ref{fig:Lattice} provides a visual representation of the operator defined in Eq.~\eqref{eq:TDoperator}, featuring a geometric structure known as a ``Triangular Dipyramid'' (TD). Thus, we shall refer to such operators as TD operators.

\begin{figure}
	\centering
	\includegraphics[width=0.6\linewidth]{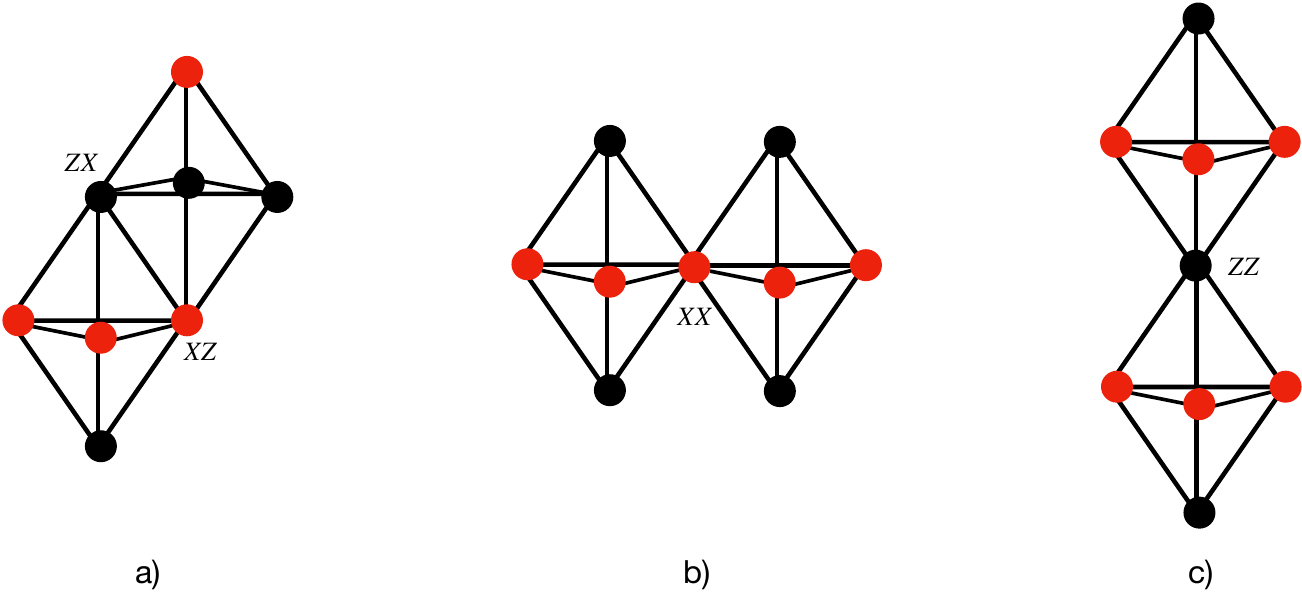}
	\caption{In a) the TD operators has two sites in common, whereas in b) and c) they share a single site.} \label{fig:neighbors}
\end{figure}

The Hamiltonian is defined as a sum of commuting projectors,
\begin{equation} \label{eq:hamiltonian}
	H =  - J \sum_{a = 1,2} \; \sum_{p \, \in \, \Lambda_{a}} \mathcal{O}_{p},  \qquad (J > 0),
\end{equation}
which is exactly soluble. It is simple to check that $\left[ \mathcal{O}_{p}, \mathcal{O}_{q} \right] = 0$, due to the fact that these operators can share either one or two sites, as illustrated in Fig.~\ref{fig:neighbors}. Given that $\left( \mathcal{O}_{p} \right)^{2} = \openone$, it follows that the eigenvalues of the TD operators are restricted to $\pm 1$. Consequently, the ground state of the system corresponds to
\begin{equation} \label{eq:gscondition}
	\mathcal{O}_{p} \ket{GS} = + 1 \ket{GS}, ~~~\text{for all}~~ p.
\end{equation}

\section{\label{sec:degeneracy} Ground State Degeneracy}

Considering periodic boundary conditions (PBC), there is a one-to-one correspondence among spins and TD operators. As a result, the condition in Eq.~\eqref{eq:gscondition} labels $N$ degrees of freedom, except for constraints among the TD operators. Hence, the ground state degeneracy is associated precisely with the number of constraints, which we denote as $2N_{c}$, 
\begin{equation} \label{eq:gsd1}
	GSD = \frac{2^{N}}{2^{N - 2N_{c}}} = 2^{2N_{c}}. 
\end{equation}
The factor of 2 in $2N_c$ is convenient because in this way $N_c$ will correspond to the number of constraints for each sublattice. 
The number of constraints depends dramatically on the linear size of the lattice in the $xy$-plane, reflecting a severe form of UV/IR mixing. The computation of the ground state degeneracy for linear sizes $L_x=L_y=L$ was presented in \cite{Casasola:2023tot}. We shall firstly review this intricate computation and then discuss how it can be fully generalized to arbitrary sizes with $L_x\neq L_y$.


\subsection{Sizes with $L_x = L_y=L$}

\begin{figure}
	\centering
	\includegraphics[width=4cm]{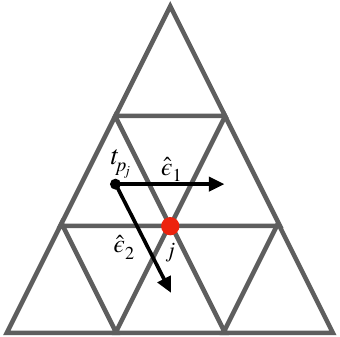}
	\caption{The $xy$-plane and the basis vectors $\hat{\epsilon}_{1}$ and $\hat{\epsilon}_{2}$.} \label{fig:new_basis}
\end{figure}

The constraints can be expressed in terms of a set $\{ t \}$ of variables defined in $\mathbb{Z}_{2}$ as
\begin{equation}
	\prod_{p} \mathcal{O}_{p}^{t_{p}} = \openone.
\end{equation}
The number of linearly independent nontrivial solutions of $t_{p}$ is precisely $2N_{c}$. This product can be rewrite in the spins lattice $\tilde{\Lambda}$, instead of the original lattice of the operators $\Lambda$,
\begin{equation} \label{spin_lattice}
	\prod_{p\, \in\, \Lambda} \mathcal{O}_{p}^{t_{p}}
	= \prod_{j \, \in \,\tilde{\Lambda}} Z_j^{t_{p_j - {\hat{z}} /{2}}} \, X_j^{t_{p_j}} \,
	X_j^{t_{p_j + \hat{\epsilon}_1}} \,
	X_j^{t_{p_j +  \hat{\epsilon}_2}} \,
	Z_j^{t_{p_j + {\hat{z}} / {2}}}
	= \openone,
\end{equation}
where $p_{j}$ is a site in $\tilde{\Lambda}$ that corresponds to $p$ in the original lattice\footnote{We will typically use the $i$, $j$, and $k$ for sites in $\tilde{\Lambda}$ and $p$, and $q$ for sites in $\Lambda$.}, as depicted in Fig.~\ref{fig:new_basis}. The above expression yields to two conditions that must be satisfied. From the $Z$-operators, we obtain
\begin{equation} \label{eq:equality1}
	t_{p_j + \frac{\hat{z}}{2}} + t_{p_j - \frac{\hat{z}}{2}}=0 \pmod{2} 
\end{equation}
and from the $X$-operators we have
\begin{equation} \label{eq:equality2}
	t_{p_j} + t_{p_j + \hat{\epsilon}_1 }+ t_{p_j + \hat{\epsilon}_2}=0 \pmod{2}.
\end{equation}
The condition \eqref{eq:equality1} implies that all $t_{p_j}$ in a straight line along $z$-direction must have the same value. Therefore, it is sufficient to consider only two $xy$-planes, one from each sublattice, to completely define the whole set $t$.

\begin{figure}
	\centering
	\includegraphics[height=8cm,angle=90]{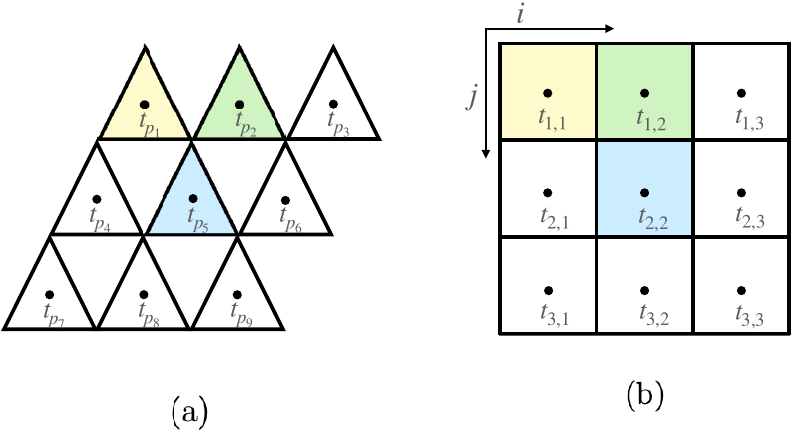}
	\caption{In Figure (a) the $t_{p_{j}}$ are represented in the original triangular lattice, and in Figure (b) the positions new notation of $t_{j, i}$.} \label{fig:new_position}
\end{figure}

Eq.~\eqref{eq:equality2} leads to constraints in each one of the $xy$-planes, which are closely tied to the fractal features of the model  \eqref{eq:hamiltonian}. To further understand this, it is convenient to adjust the notation and rewrite Eq.~\eqref{eq:equality2} as
\begin{equation} \label{eq:coefficients}
	t_{j + 1, i + 1} = t_{j, i}+t_{j, i + 1} \pmod{2},
\end{equation}
where $i$ and $j$ denote orthogonal directions in the plane, as illustrated in Fig.~\ref{fig:new_position}. As discussed in \cite{Devakul_2019}, in the context cellular automaton, this equation defines a Sierpinski rule. It works as a ``translationally-invariant local linear update rule'', in the sense that a state can be determined by the configuration of the previous time step. In the present case, this constraint has a similar role. Given a line configuration\footnote{In this context we will use the term ``configuration'' to represent the arrangement of the variables $t_{j, i}$.} $\left\{t_{j, i}\right\}$,  Eq.~\eqref{eq:coefficients} allows us to fully determine the subsequent line $\left\{t_{j + 1, i}\right\}$ and, consequently, the entire plane.

To find the constraints in the $xy$-plane, we use a polynomial representation, as introduced in \cite{Martin1984}, so that a line configuration can be represented by a Laurent series as
\begin{equation}
	S_{j}(x) = \sum_{i = 1}^{L} t_{j, i} \, x^{i},
\end{equation}
where the coefficients are the variables $t_{j, i}$ defined mod 2.  By incorporating into this definition the update rule \eqref{eq:coefficients}, we obtain
\begin{equation} \label{eq:evolution}
	S_{j + 1}(x) = \sum_{i = 1}^{L} \left( t_{j, i}+t_{j, i - 1} \right) x^{i} = \left( 1 + x \right) S_{j}(x),
\end{equation}
with $\left(1+x\right)$ being the polynomial representation of the Sierpinski evolution. Given an initial line configuration $S_{1}(x)$, any subsequent line $S_{j+1}(x)$ can be recursively determined by applying Eq.~\eqref{eq:evolution}, 
\begin{equation} \label{eq:evolution2}
	S_{j + 1}(x) = \left( 1 + x \right)^{j} S_{1}(x).
\end{equation}

When the Sierpinski evolution is compatible with PBC, the product of TD operators arranged in a fractal structure results in an identity, leading to certain constraints among $\mathcal{O}_{p}$'s in the $xy$-plane. In  terms of the polynomial representation, PBC translate into $S_{j + L} (x) = S_{j} (x)$ and $x^{i + L} = x^{i}$. Therefore, our goal is to identify configurations $S_{1} (x)$ that satisfy the condition
\begin{equation} \label{eq:pcondition}
	S_{1 + L} (x) = \left( 1 + x \right)^{L} S_{1} (x) = S_{1} (x).
\end{equation}

With this purpose, we note that $S_{1} (x)$ must contain an even number of non-zero $t_{1, i}$'s to be compatible with PBC. This requirement becomes evident by considering a simple configuration  $S_{1} (x) = x^{i}$. Then, the evolution governed by \eqref{eq:evolution2} implies that all subsequent lines $S_{j} (x)$ have an even number of $t_{j, i} \neq 0$. Consequently, it becomes impossible to find $S_{L + 1} = x^{i}$, to satisfy the PBC.

For a consistent first line configuration consider
\begin{equation} \label{eq:firstline}
	S_{1}(x) = \left( 1 + x \right)^{2^{m}}\sum_{i = 1}^{L} t_{1, i} \,  x^{i},
\end{equation}
with $m=0,1,2, \ldots$. The term included in this equation follows the ``Freshman's dream'',
\begin{equation}
	\left( 1 + x \right)^{2^{m}} = 1 + x^{2^{m}} \pmod{2},
	\label{eq:fd}
\end{equation}
for any integer $m$, and ensures an even number of non-zero coefficients. This result follows from the binomial expansion,
\begin{equation}
	\left( 1 + x \right)^{2^{m}} = \sum_{l = 0}^{2^{m}}\binom{2^{m}}{l}\, x^{l},
\end{equation}
where the coefficient is even for any $l$, except when $l$ equals to $0$ or $2^{m}$. The configuration defined in Eq.~\eqref{eq:firstline}, under periodic boundary conditions, implies
\begin{equation} \label{eq:sizes}
	S_{L + 1} (x) = \left( 1 + x \right)^{L + 2^{m}} \sum_{i = 1}^{L} t_{1, i} \, x^{i} = \left( 1 + x \right)^{2^{n}}  \sum_{i = 1}^{L} t_{1, i} \, x^{i}. 
\end{equation}
Consequently, any linear size of the form 
\begin{equation}
	L\equiv 2^{n}-2^{m}, \qquad n > 2 \quad \text{and} \quad 0\le m < n - 1,
	\label{sizes}
\end{equation}
has constraints in the $xy$-plane.

\begin{figure}
	\centering
	\includegraphics[scale=.35]{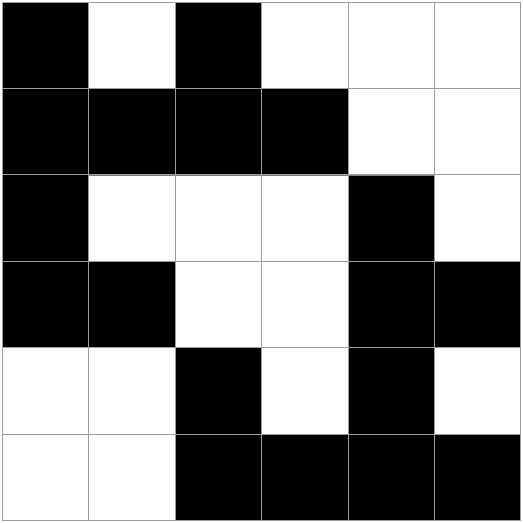}
	\caption{Example of a periodic configuration generated by $S_{1} (x)$ according to Eq.~\eqref{eq:basis} in a lattice of size $6\times 6$.}
	\label{6by6}
\end{figure}

To specify all possible configurations of the set $\left\{t\right\}$, we can define the basis elements as
\begin{equation} \label{eq:basis}
	S_{1}^{i} (x) = \left( 1 + x^{2^{m}} \right) x^{i}, \qquad \text{for} \quad i = 1, 2, \ldots , 2^{n} - 2^{m + 1}. 
\end{equation}
This basis defines the dimension of the set of first lines as $N_c=2^{n} - 2^{m + 1}$ for each plane (in the Supplementary Material \ref{sec:MS1} we have included some examples to illustrate this discussion). An example of a configuration generated by \eqref{eq:basis} is shown in Fig. \ref{6by6}.

Equation~\eqref{eq:equality1} defines two independent planes, so $N_{c}$ is the number of constraints of a single plane. As discussed in Eq.~\eqref{eq:gsd1}, the degeneracy corresponds precisely to $2N_{c}$. Therefore, a size $L = 2^{n} - 2^{m}$ has a ground state degeneracy 
\begin{equation} \label{eq:deg}
	GSD =2^{2N_c}= 2^{2^{n + 1} - 2^{m + 2}}. 
\end{equation}
For linear sizes which are not of the form \eqref{sizes}, the ground state is unique.

\subsection{Sizes with $L_x\neq L_y$}

We discuss now the generalization of the previous computation for cases with $L_x\neq L_y$. There are some compatibility conditions between the sizes $L_x$ and $L_y$ in such a way that the system can exhibit nontrivial ground state degeneracy. When at least one of the linear sizes is not of the form \eqref{sizes}, the ground state is unique. On the other hand, assuming that $L_x= 2^n-2^m $, we will show that the system supports a degenerate ground state if $L_y$ is a divisor of $L_x$, i.e., $L_x=r L_y$, for positive integers $r$	\footnote{We are considering $L_x \ge L_y$ without loss of generality.}. 

\begin{figure}
	\centering
	\begin{minipage}[c]{0.46\textwidth}
		\centering
		\begin{minipage}[c]{\textwidth}
			$S_{1}^{1} (x)$
			\includegraphics[width= \textwidth]{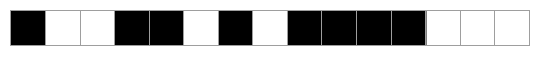}
		\end{minipage} 
		
		\vspace{4pt} \begin{minipage}[c]{\textwidth}
			$S_{6}^{1} (x)$
			\includegraphics[width= \textwidth]{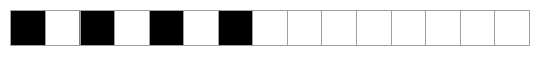}
		\end{minipage} 
		
		\vspace{4pt} \begin{minipage}[c]{\textwidth}
			$S_{11}^{1} (x)$
			\includegraphics[width=\textwidth]{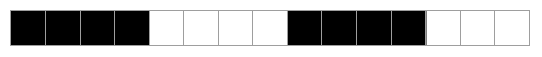}
		\end{minipage} 
	\end{minipage} 
	\hspace{0.06\textwidth}
	\begin{minipage}[c]{0.46\textwidth}
		\centering
		\includegraphics[width=\textwidth]{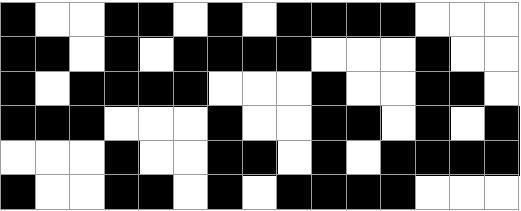}
	\end{minipage}
	\caption{For $L_{y} = 5$ and $L_{x} = 15$, defining $\bar{S}_{1}^{1} (x) = S_{1}^{1} (x) + S_{6}^{1} (x) + S_{11}^{1} (x)$, the first line respects the PBC $\bar{S}_{1}^{1} (x) = \bar{S}_{6}^{1} (x)$.} \label{fig:stilde}
\end{figure}

To understand this, we notice that the first line configuration $S_{1}^{i} (x)$, as defined in Eq. \eqref{eq:basis}, does not respect the perodicity in the $y$-direction, $S_{1 + L_{y}}^{i} (x) \neq S_{1}^{i} (x)$. However, we can use it to construct another first line configuration that has the proper periodicity, 
\begin{equation}
	\bar{S}_{1}^{i} (x) \equiv \sum_{l = 0}^{r - 1} \left( 1 + x \right)^{l L_{y}}  \left( 1 + x^{2^{m}} \right) x^{i}, 
 \label{eq:basis2}
\end{equation}
where $i = 1, 2, \ldots , 2^{n} - 2^{m + 1}$. Now we proceed to check that this configuration indeed satisfies the PBC $\bar{S}_{1 + L_{y}}^{i} (x) = \bar{S}_{1}^{i} (x)$. We can write 
\begin{equation}
	\bar{S}_{1 + L_{y}}^{i} (x) = \left( 1 + x \right)^{L_{y}} 	\bar{S}_{1}^{i} (x) 
	= \sum_{l = 1}^{r} \left( 1 + x \right)^{l L_{y}} S_{1}^{i} (x).
\end{equation}
In the term of the sum with $l = r$, we can use the property~\eqref{eq:fd} and the fact that $r L_{y} = L = 2^{n} - 2^{m}$, to express
\begin{eqnarray}
	\begin{aligned}
		\left( 1 + x \right)^{r L_{y}} S_{1}^{i} (x)  
		& = \left( 1 + x \right)^{2^{n}} x^{i} = \left( 1 + x^{2^{n}} \right) x^{i}  \\
		& = \left( 1 + x^{L + 2^{m}}\right) x^{i} = \left( 1 + x^{2^{m}} \right) x^{i}  \qquad \\
		& = \left( 1 + x \right)^{2^{m}} x^{i} = S_{1}^{i} (x).
	\end{aligned}
\end{eqnarray}
With this, we find
\begin{eqnarray}
	\begin{aligned}
		\bar{S}_{1 + L_{y}}^{i} (x) & = \sum_{l = 1}^{r} \left( 1 + x \right)^{l L_{y}} S_{1}^{i} (x) \\
		& = \sum_{k = 0}^{r - 1} \left( 1 + x \right)^{l L_{y}} S_{1}^{i} (x)  = \bar{S}_{1}^{i} (x).
	\end{aligned}
\end{eqnarray}
In Fig.~\ref{fig:stilde}, we show an example of $\bar{S}_{1}^{i} (x)$ for $L_x = 15$ and $L_y=5$.

We can determine the number of independent first line configurations directly from \eqref{eq:basis2}, which corresponds to the maximum value of $i$ before it starts to repeat under PBC. It is given by
\begin{equation}
i_{max}=L_y-2^m.
\label{maximum}
\end{equation}
However, we need some attention to properly use this result. This is so because it may happen that 
\begin{equation}
L_x = k L_x',
\label{001}
\end{equation}
with $L_x=2^n-2^m$, $L_x'=2^{n'}-2^{m'}$, and $k$ a positive integer. Then we can write
\begin{eqnarray}
L_x \times L_y &=& L_x \times \frac{L_x}{r}\nonumber\\
&=& k L_x'\times   \frac{k L_x'}{r}\nonumber\\
&& k L_x'\times   \frac{L_x'}{r'},
\end{eqnarray}
where $r'\equiv \frac{r}{k}$ must be a positive integer\footnote{As both $r$ and $k$ are divisors of $L_x$, there is at least one value of $k$ so that $r'=\frac{r}{k}$ is a positive integer, namely, $k=r$.}. Notice that $L_y$ is preserved under such redefinitions, i.e., $L_y= \frac{L_x}{r}=\frac{L_x'}{r'}$. In this case, the maximum number of independent first line configurations is dictated by the smallest value among $m$ and $m'$. For example, consider a system $14\times 7$, whose linear size in the $x$-direction can be written as $2^n-2^m=2^4-2^1$ or $k(2^{n'}-2^{m'})=2(2^3-2^0)$. The maximum number of independent first line configurations is given by \eqref{maximum} with $L_y=7$ and $m'=0$, which leads to $7-1=6$. Therefore, the building-block associated with the zise $14\times 7$ is that one of size $7 \times 7$.

Using this reasoning, we can understand that all the cases where $L_y< 2^m$, in which \eqref{maximum} would lead to a negative value, are actually  constructed from smaller building-block sizes. To see this, we parametrize the divisors of $L_x=2^n-2^m = 2^m (2^{n-m}-1)$ as
\begin{equation}
r(l,q)= 2^l q,  
\end{equation}
where $q$ is a divisor of $2^{n-m}-1$ and $l=1,\ldots, m$. Note that as $2^{n-m}-1$ is an odd number, then the maximum value of $q$ is $(2^{n-m}-1)/2$. Now, the condition  $L_y< 2^m$ can be written as $r(l,q)> 2^{n-m}-1$, which implies
\begin{equation}
2^l > 2~~~\Rightarrow ~~~l>1.
\end{equation}
Thus we can re-express
\begin{eqnarray}
L_x \times L_y &=& (2^n - 2^m) \times  \frac{(2^n - 2^m)}{2^l q }\nonumber\\
&=& 2^l  (2^{n-l} - 2^{m-l}) \times  \frac{(2^{n-l} - 2^{m-l})}{ q } \nonumber\\
&=& k (2^{n'}-2^{m'}) \times \frac{2^{n'}-2^{m'}}{r'}\nonumber\\
&=& k L_x'\times L_y,
\end{eqnarray}
with $k\equiv 2^l$, $n'\equiv n-l$, $m'\equiv m-l$, and $r'=q$. As $m> m'$, the number of linearly independent first line configurations is dictated by $L_y-2^{m'}> 0$. As an example, let us consider a system with size $24\times 6$, which gives $r=4$. We first note that 24 is of the form $2^n-2^m$, namely, $24=2^5-2^3$. A naive application of \eqref{maximum} leads to $i_{max}=6-2^3=-2$. However, the linear size along the $x$-direction can be decomposed in the form \eqref{001} in several ways
\begin{eqnarray}
24&=& 2 (2^4-2^2)\nonumber\\
&=& 3 (2^4-2^3)\nonumber\\
&=& 4(2^3-2^1)\nonumber\\
&=& 6 (2^3-2^2)\nonumber\\
&=& 8 (2^2-2^0)\nonumber\\
&=& 12 (2^2-2^1).
\end{eqnarray}
The only decompositions which keep $L_y$ invariant, i.e., that ones with $r' = \frac{r}{k}$ being a positive integer, are $2 (2^4-2^2)$, with $k=2$, $r'=2$, and $m=2$; and  $4(2^3-2^1)$, with $k=4$, $r'=1$, and $m=1$. As $m=1 < m=4$, the maximum number of independent first line configurations is dictated by the decomposition  $24=4(2^3-2^1)$, which leads to $6-2^1=4$. The results for the ground state degeneracy depending on the system are summarized in Table \ref{table:deg}. 

\begin{table}
	\centering
	\begin{tabular}{|c|c|}
		\hline
		~ System Size ~ &		~ $\log_2 (GSD)$ ~  \\ \hline
		$L_x$, $L_y\neq k (2^n-2^m)$  &		$0$\\ \hline
		$L_x=k (2^n-2^m)$, $L_y \neq \frac{L_x}{r}$  & 0  \\ \hline
	~$L_x=k (2^n-2^m)$, $L_y = \frac{L_x}{r}$ ~& ~$2\left( L_{y} - 2^{m} \right)$ ~\\ \hline
	\end{tabular}
	\caption{Ground state degeneracy depending on the system size. When there are different ways to decompose $L_x$ in the form $L_x=k (2^n-2^m)$ satisfying the requirement that $r'=\frac{r}{k}$ is a positive integer, then the ground state degeneracy is dictated by the decomposition corresponding to the smallest value of $m$.}
	\label{table:deg}
\end{table}

Finally, we note that the sizes in Table \ref{table:deg} can be used as building-blocks for sizes $k_i (L_x \times L_y)$, where $k_b$ is a positive integer with the subscript $b=x,y$ indicating in which direction the copies are disposed. The ground state degeneracy is dictated by the building-block $L_x \times L_y$. As an example, consider the size $30\times 20$. It is not of the form of the sizes present in the Table \ref{table:deg}, but it can be written as $2_y (30\times 10)$, where $2_y$ means that the two copies are disposed along the $y$-direction.


\section{Symmetry Operators and Mixed 't Hooft Anomalies \label{sec:sym}}

The ground state degeneracy is a direct consequence of the anomalies of the symmetry algebra. For degenerate system sizes, the model \eqref{eq:hamiltonian} exhibits two distinct types of symmetries:  Wilson lines and Fractal membranes. The Wilson lines are represented by closed loops in the $z$-direction, while fractal symmetries manifest as closed membranes in  the $xy$-plane. Both symmetries -- line and membrane -- are mutually anomalous. This is referred to as a mixed anomaly,  in the sense that there is a nontrivial commutation between two different types of symmetry operators. In non-degenerate system sizes, the fractal membranes fail to close properly, resulting in the explicit breaking of fractal symmetry due to the system size.

\subsection{Wilson Lines}

The Wilson line operators are products of $X$'s along the $z$-direction, 
\begin{equation}
	W^{a}_{i} = \prod_{j \, \in \, \mathcal{L}^{a}_{i}} X_{j},
\end{equation}
where $\mathcal{L}^{a}_{i}$ defines a straight line across all $xy$-planes at the point $i = (x, y)$ belonging to the sublattice $a$. Under PBC, it forms a closed loop and hence corresponds to a nontrivial symmetry of the Hamiltonian.

\begin{figure} 
	\includegraphics[angle = 90, width =  3.5cm]{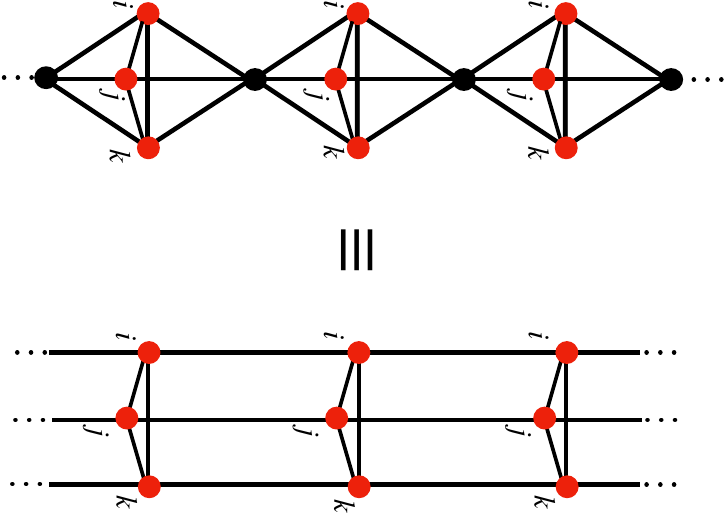}
	\caption{The product of TD operators along the $z$-direction, reducing to the product of three Wilson lines.} \label{fig:wilsonlines}
\end{figure}

There is one different $W^{a}_{i}$ operator for each point $i$ on the $xy$-plane, resulting in a total of $L_x  L_y $ operators. However, not all of them are independent. A number of constraints arises from the fact that the product of TD operators along a straight line in the $z$-direction is equivalent to the product of three Wilson lines,  
\begin{equation} \label{eq:wilsoncons}
	\prod_{p \, \in \, \mathcal{L}^{a}_{q}} \mathcal{O}_{p} = W_{i}^{a} W_{j}^{a} W_{k}^{a}.
\end{equation}
This is illustrated in Fig.~\ref{fig:wilsonlines}. As there is a one-to-one correspondence between points of the dual and original lattices, there are $L_xL_y$ constraints, but not all of them are independent because the topological constraints discussed in Sec.~\ref{sec:degeneracy}. They can be expressed as
\begin{equation} \label{eq:wilsoncons2}
	\prod_{z} \; \prod_{p \, \in \, \mathcal{M}^{a}_{I} (z)} \mathcal{O}_{p}  = \prod_{q \, \in \, \mathcal{M}^{a}_{I}} \; \prod_{p \, \in \, \mathcal{L}^{a}_{q}} \mathcal{O}_{p} = \openone,
\end{equation}
where $\mathcal{M}^{a}_{I}$, $I=1,2,\ldots,L_y-2^m$, represents a configuration of TD operators in the $xy$-plane corresponding to a constraint. Putting apart one point $p'$ from this product we find
\begin{widetext}
	\begin{equation} \label{eq:wilsoncons3}
		\left(\prod_{p \, \in \, \mathcal{L}^{a}_{p'}} \mathcal{O}_{p}\right) \prod_{\substack{ q \, \in \, \mathcal{M}^{a}_{I} \\ q \neq p'}} \; \prod_{p \, \in \, \mathcal{L}^{a}_{q}} \mathcal{O}_{p} = \openone \qquad \Rightarrow \qquad \prod_{\substack{ q \, \in \, \mathcal{M}^{a}_{I} \\  q \neq p'}} \; \prod_{p \, \in \, \mathcal{L}^{a}_{q}} \mathcal{O}_{p} = W_{i}^{a} \, W_{j}^{a} \, W_{k}^{a}.
	\end{equation}
\end{widetext}
Equation~\eqref{eq:wilsoncons3} reveals that one of the constraints defined in Eq.~\eqref{eq:wilsoncons} can be expressed in terms of the others. This process can be repeated for all the $N_c=L_y-2^m$ constraints, so that the total number of independent $W^{a}_{I}$ is precisely $L_x L_y - \left(L_x L_y - N_{c}\right) = N_{c}$. For this reason, we will label the independent Wilson lines with the index $I$,
\begin{equation}\label{eq:wilson_lines}
	W_{I}^{a} \equiv \prod_{j \, \in \, \mathcal{L}^{a}_{I}} X_{j} \, , \qquad I = 1, 2, \, \ldots \, , N_{c} \, . 
\end{equation}

For system sizes with a unique ground state, there are no independent Wilson lines. In these cases, all $W_{i}^{a}$ can be reduced to a product of TD operators and, consequently, they represent trivial symmetries.


\subsection{\label{sec:membranes} Fractal Membranes}

For a system size with a degenerate ground state, the fractal membrane operators are closely related to the topological constraints. Rewriting the constraint equation
\begin{equation}
	\prod_{z} \; \prod_{p \, \in \, \mathcal{M}^{a}_{I} (z)} \mathcal{O}_{p}  = \openone,
	\label{cmo}
\end{equation}
we note that a subset of this product, ranging from $z_{1}+a_{0}/2$ to $z_{2}-a_{0}/2 $, results in one membrane at the bottom and other membrane at the top of this product in $z$,
\begin{equation} \label{eq:subsetprod}
	\prod_{z = z_{2}  - \frac{a_{0}}{2}}^{z_{1} + \frac{a_{0}}{2}} \; \prod_{p \, \in \, \mathcal{M}_{I}^{a} (z)} \mathcal{O}_{p} = M_{I}^{b} (z_{1}) \; M_{I}^{b} (z_{2}), \qquad b \neq a, 
\end{equation}
where we have defined the fractal membrane operator as
\begin{equation} \label{eq:membrane}
M_{I}^{b} (z)\equiv\prod_{j \, \in \, \mathcal{M}_{I}^{b} (z)} Z_{j}.
\end{equation}
Note that the product of TD operators on the sublattice $a$ results in $Z$-operators acting non-trivially on the opposite sublattice. 
 
\begin{figure}
	\includegraphics[scale=0.5,angle=0]{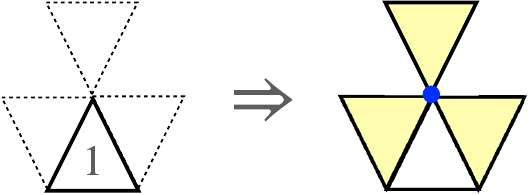}
	\caption{Map between topological constraints and membrane operators.}
	\label{rule}
\end{figure}

\begin{figure}
	\includegraphics[scale=0.45]{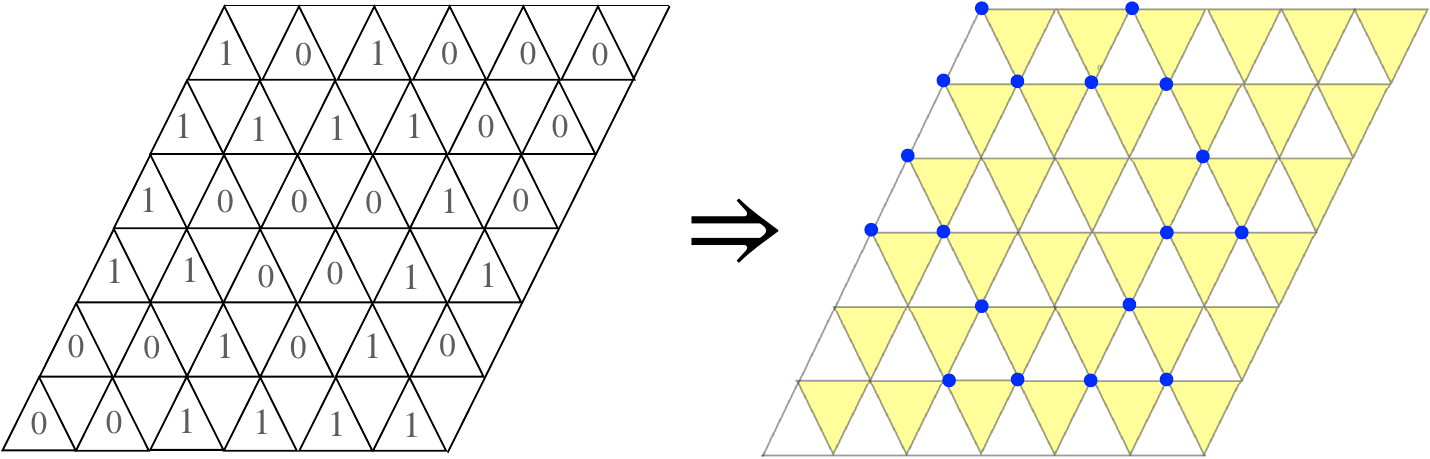}
	\caption{Membrane operator constructed from a topological constraint.}
	\label{constraint_membrane}
\end{figure}

The left-hand side of Eq.~\eqref{eq:subsetprod} trivially commutes with the Hamiltonian, since it is simply a product of TD operators. Thus the product of membrane operators in the right-hand side should also do it. 
However, as the two membranes are widely separated and given the locality of the Hamiltonian, each one should commute individually with $\mathcal{O}_{p}$. Therefore, $M_{I}^{b}$ can be treated as another symmetry operator that determines the holonomies of the system, and the degenerate ground states can be labeled as eigenstates of $M_{I}^{b}$. In effect, this discussion leads to a map between topological constraints and membrane operators, as shown in Fig. \ref{rule}. An example for a $6\times6$ lattice is shown in Fig.  \ref{constraint_membrane}.

Following the definition in Eq.~\eqref{eq:membrane},  there are $2N_{c}$ independent fractal membrane operators, one for each constraint. Notice that given a membrane $M_{I}^{b}(z_{1})$, we can move it to $z_{2}$ with a product of TD operators,
\begin{equation} \label{eq:membmob}
	M_{I}^{b} (z_{1})  \left( \prod_{z = z_{2}  - \frac{a_{0}}{2}}^{z_{1} + \frac{a_{0}}{2}} \; \prod_{p \, \in \, \mathcal{M}_{I}^{a}} \mathcal{O}_{p} \right) = M_{I}^{b} (z_{2}).
\end{equation}
If there is no defect between $z_{1}$ and $z_{2}$, both membranes are constrained to have the same eigenvalue. Due to this mobility, we do not need to specify their position in $z$.

For the cases where the ground state is unique, there are no topological constraints and, accordingly, there are no fractal membrane operators. This occurs because the fractal structures do not close properly under PBC.


\subsection{\label{sec:tHanomaly} Mixed 't Hooft Anomaly}

As discussed earlier, for degenerate system sizes, there is a one-to-one correspondence between the number of Wilson lines and fractal membrane operators. The pairs of $W_{I}^{a}$ and $M_{I}^{a}$ can be arranged to satisfy the commutation relation
\begin{equation}\label{eq:commutation}
	W_{I}^{a}M_{J}^{b} = \left( -1 \right)^{\delta_{IJ} \, \delta_{ab}} M_{J}^{b} W_{I}^{a}.
\end{equation}
All we have to do is to identify a point belonging to a membrane operator which is not shared with the remaining membranes. Then, if we choose a Wilson line intersecting that point, only this pair will have a nontrivial commuting relation. 

\begin{figure}
	\includegraphics[width=0.13\linewidth]{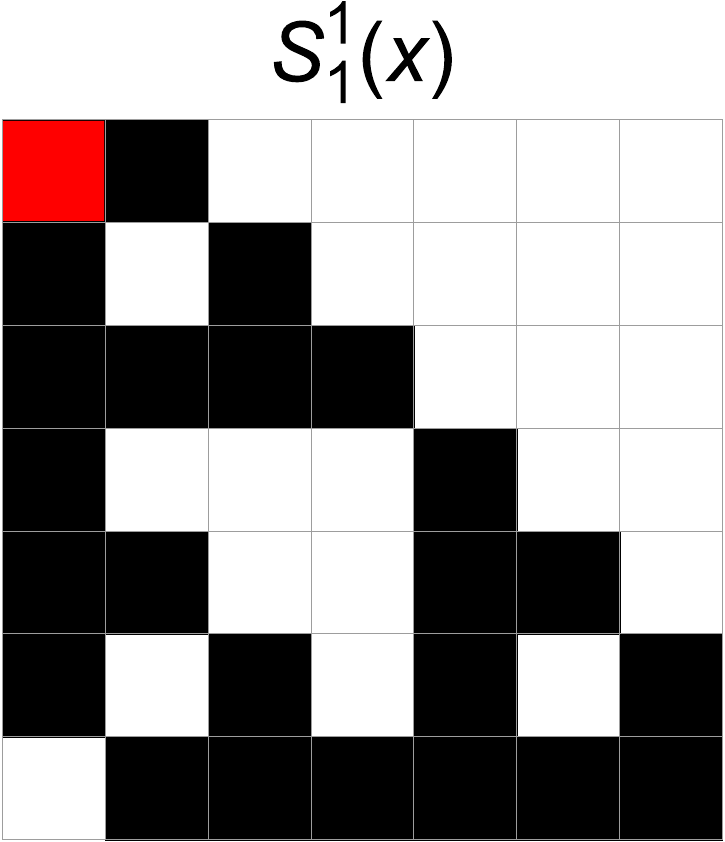} \hspace{0.01\linewidth} \includegraphics[width=0.13\linewidth]{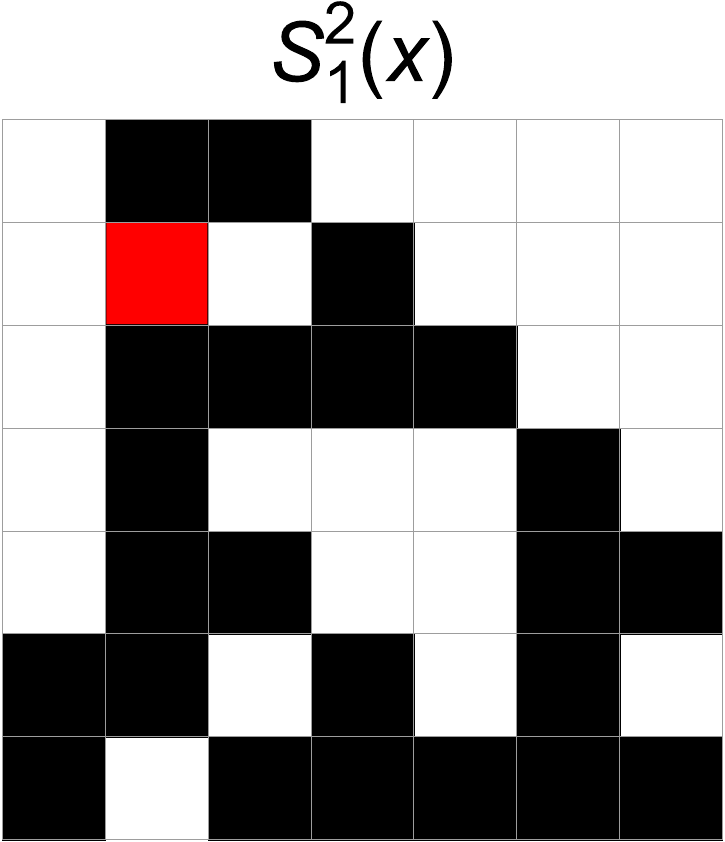} \hspace{0.01\linewidth} \includegraphics[width=0.13\linewidth]{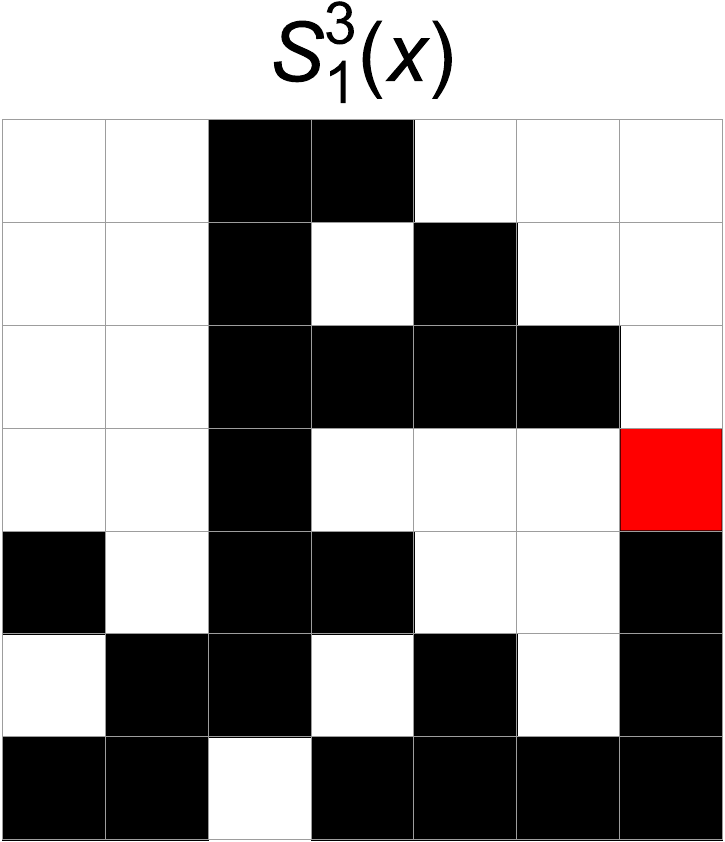} \hspace{0.01\linewidth} \includegraphics[width=0.13\linewidth]{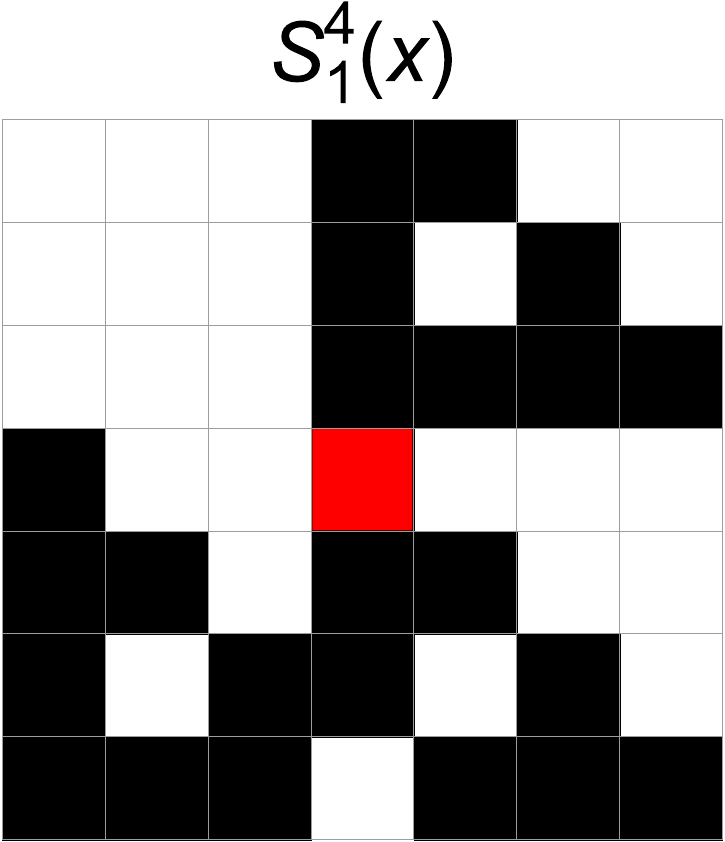} \hspace{0.01\linewidth} \includegraphics[width=0.13\linewidth]{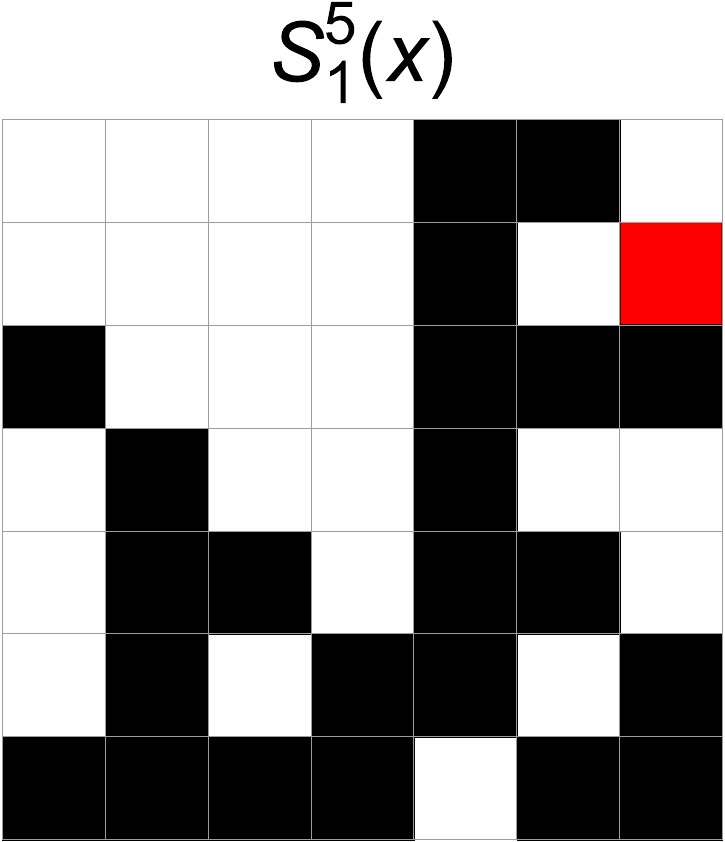} \hspace{0.01\linewidth} \includegraphics[width=0.13\linewidth]{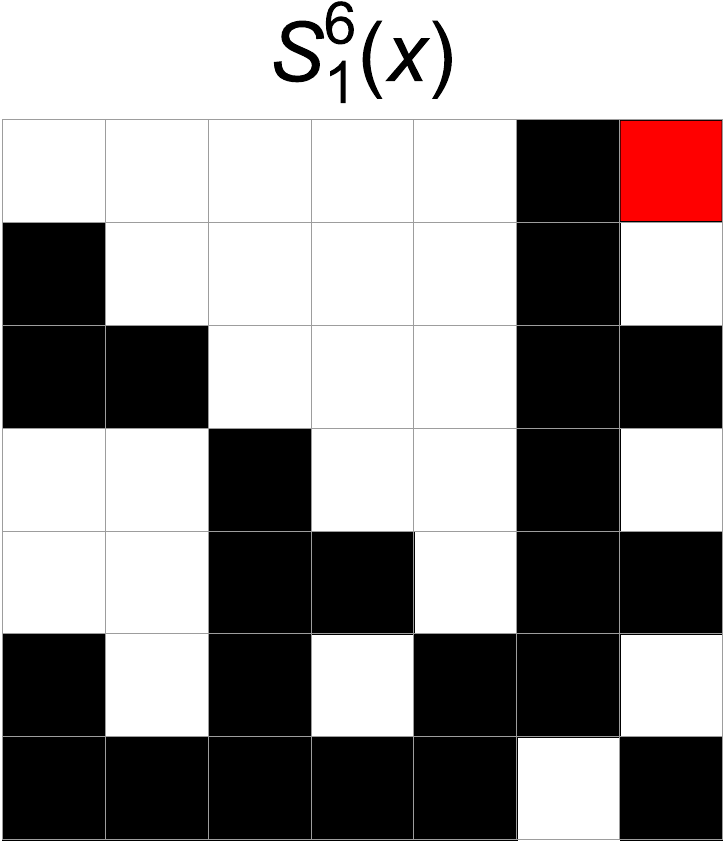} \\ \vspace{15pt}
	\includegraphics[width=0.13\linewidth]{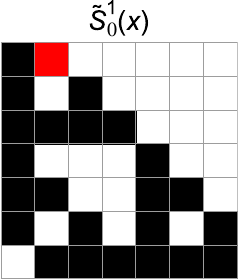} \hspace{0.01\linewidth} \includegraphics[width=0.13\linewidth]{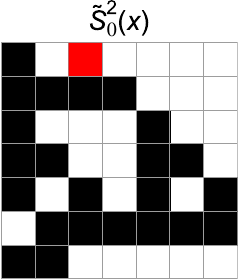} \hspace{0.01\linewidth} \includegraphics[width=0.13\linewidth]{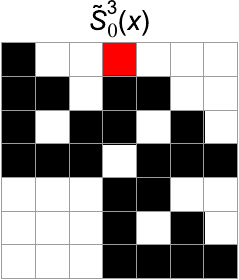} \hspace{0.01\linewidth} \includegraphics[width=0.13\linewidth]{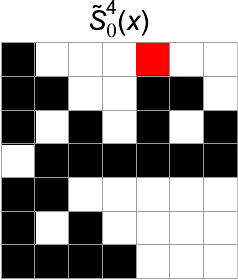} \hspace{0.01\linewidth} \includegraphics[width=0.13\linewidth]{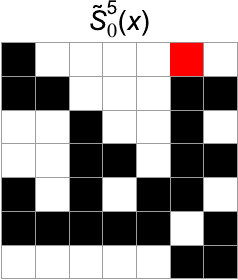} \hspace{0.01\linewidth} \includegraphics[width=0.13\linewidth]{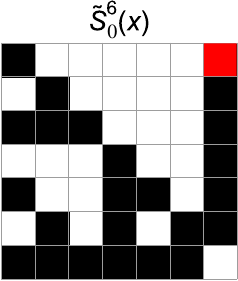}
	\caption{The membranes generated by $S_{1}^{i}(x)$ and $\tilde{S}_{1}^{i}(x)$, where  $i=1,2, \, \dots \, ,6$, in system of size $L=7$. Each membrane highlights the unique $t_{ji} = 1$ that does not repeat in other membranes.} \label{fig:tildemembranes}
\end{figure}

For the cases where $L_x=L_y=2^n-2^m$, this procedure can be done in a very explicit way. Using $S_{1}^{i}(x)$ in \eqref{eq:basis} makes it nontrivial to identify these unique points. A more convenient way to proceed is by introducing the new basis
\begin{equation} \label{eq:nfirstline}
	\tilde{S}_{p}^{l} \equiv \sum_{i = 1}^{l} S_{1}^{2^{m}i - p},
\end{equation}
where $p = 0, 1, 2, \, \ldots \, , 2^{m}-1$ and $l = 1, 2, \, \ldots \, , (2^{n} - 2^{m + 1})/2^{m}$. This redefinition ensures that all $t_{ji} = 1$ points, which do not repeat for the other membranes, are located in the first line, as depicted in Fig.~\ref{fig:tildemembranes}.

With Eq.~\eqref{eq:nfirstline}, we define $\mathcal{M}^{a}_{I}$ as the set $\{ t \}$ associated with the Sierpinski fractal structure generated by the first line $\tilde{S}^{I}_{1}(x)$. Similarly, $\mathcal{L}^{a}_{I}$ is defined as the straight line along the $z$-direction that touches only the respective $\mathcal{M}^{a}_{I}$. Consequently, the definitions of the symmetry operators in~\eqref{eq:wilson_lines} and~\eqref{eq:membrane} automatically respect the commutation in Eq.~\eqref{eq:commutation}.

This nontrivial algebra implies that the eigenstates of the Hamiltonian can be eigenstates of either $W^{a}_{I}$ or $M^{a}_{I}$,  but not both simultaneously. Accordingly, there is a $2^{2 \left( 2^{n} - 2^{m} \right)}$ degeneracy in the states, precisely matching  Eq.~\eqref{eq:gsd1}. This algebra can also be understood as a mixed 't Hooft anomaly between the symmetries, preventing the ground state from being trivially gapped.

For $L_x\neq L_y$, it seems to be very hard to construct explicitly a basis in which the unique points are brought to the first line, as in the previous case. Nevertheless, we have checked that the number of points in a given membrane which are not shared with the remaining ones is given by
\begin{equation}
\frac{2^{n-m}-1}{r} \ge1 .
\end{equation}
Therefore, there is at least one point which is not shared with the remaining membranes, so that we can always construct pair of operators satisfying \eqref{eq:commutation}.


\subsection{Non-degenerate Cases}\label{subsectionD}

For system sizes which are not of the form in the last row of the Table \ref{table:deg}, there are no topological constraints ($N_{c} = 0$) and the ground state is unique. In these cases, the fractal membranes do not close properly, meaning that they are not symmetry operators. On the contrary, the membrane operators create local excitations. In other words, the fractal symmetry is explicitly broken by the lattice size.

For these sizes it also occurs that the Wilson lines are trivial symmetries, as they reduce to a product of TD operators. Consider a product of TD operators in a set $\tilde{\mathcal{M}}^{a}_{k}$ along the $xy$-plane that reduces to identities at every site except for one, resulting in a $X$-operator localized at the site $k$,
\begin{equation} \label{eq:oneX}
	\prod_{p \, \in \, \tilde{\mathcal{M}}^{a}_{k} (z)} \mathcal{O}_{p} = X_{k}  \prod_{i \, \in \, \tilde{\mathcal{M}}^{a}_{k} (z + \frac{1}{2})} Z_{i} \prod_{i \, \in \,  \tilde{\mathcal{M}}^{a}_{k} (z - \frac{1}{2})} Z_{i}.
\end{equation}
Then, stacking this product along the $z$-direction results precisely in a Wilson line,
\begin{equation}
	\prod_{z} \prod_{p \, \in \,  \tilde{\mathcal{M}}^{a}_{k} (z)} \mathcal{O}_{p} = \prod_{i \, \in \, \mathcal{L}^{a}_{k}} X_{i} = W^{a}_{k}.
\end{equation}

\begin{figure}
	\includegraphics[angle = 90]{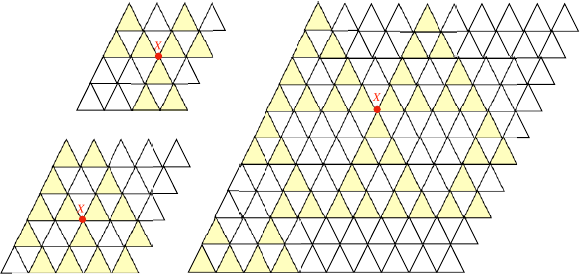}
	\caption{For non-degenerate system sizes, a single $X_{k}$ operator can be obtained from the product of TD operators over a plane. The figures above depict this product for $L=4$, $L=5$, and $L=10$.} \label{fig:nondegwilson}
\end{figure}

In order to find $\tilde{M}^{a}_{I}$ we need to discuss the configurations of $\{ t \}$ again. As discussed earlier for the constraints, given any $S_{1} (x)$ configuration, the Sierpinski evolution ensures that the product of $\mathcal{O}$'s reduces to identities in the interception of two subsequent lines, $S_{i} (x)$ and $S_{i+1} (x)$. Therefore, for a given $S_{1} (x)$ that satisfies
\begin{equation}
	S_{1} (x) + S_{1+L} (x) = x^{j}
\end{equation}
under the Sierpinski evolution, we can properly generate a set $ \tilde{\mathcal{M}}^{a}_{k}$ that results in a single $X_{k}$ on the plane. In Fig.~\ref{fig:nondegwilson}  we depict the product of $\mathcal{O}_{p}$ over the $xy$-plane for some of these sizes, such that the result is a single $X_{k}$. Stacking this structure along the $z$-direction results precisely in a Wilson line.

The first line configuration that gives rise to $\tilde{\mathcal{M}}^{a}_{k}$ is constructed by finding the coefficients $t_{1,i}$ that satisfy
\begin{equation}
	\begin{split}
		S_{1} (x) + S_{1+L} (x) & = \left[ 1 + \left( 1 + x\right)^{L} \right] S_{1} (x) \\
		& = \sum_{i = 1}^{L}  \sum_{j = 1}^{L} \binom{L}{j}  \, t_{1,i} \, x^{i+j}  = x^{k}.
	\end{split}
\end{equation}
For convenience, we introduce $j=j'-1$ and use the identification $x^{j' + L} = x^{j'}$ to rewrite the sum in the last line as
\begin{widetext}
	\begin{equation}
		\begin{split}
			\sum_{i = 1}^{L} \, \sum_{j' = i + 1}^{i + L} t_{1,i} \, \binom{L}{j' - i} \, x^{j'}  & = \sum_{i = 1}^{L} \, \left( \sum_{j' = i + 1}^{L} \binom{L}{j' - i} \, t_{1,i} \, x^{j'} + \sum_{j' = L + 1}^{i + L} \binom{L}{j' - i} \, t_{1,i} \, x^{j' - L} \right) \\
			& = \sum_{i = 1}^{L} \, \left( \sum_{j' = i + 1}^{L} \binom{L}{j' - i} \, t_{1,i} \, x^{j'} + \sum_{j' = 1}^{i} \binom{L}{j' + L - i} \, t_{1,i} \, x^{j'} \right) \\
			& = \sum_{i = 1}^{L}  \sum_{j = 1}^{L} \binom{L}{\left| i - j \right|} \, t_{1,i} \, x^{j},
		\end{split}
	\end{equation}
\end{widetext}
where we have used the binomial identity $\binom{L}{L-a} = \binom{L}{a}$. It results in
\begin{equation}
	\sum_{i = 1}^{L}  \sum_{j = 1}^{L} \binom{L}{\left| i - j \right|} \, t_{1,i} \, x^{j} = x^{k} \pmod{2}
\end{equation}
and the coefficients of $S_{1} (x)$ are the solutions of
\begin{equation}\label{eq:system}
	\sum_{i = 1}^{L} \binom{L}{\left| i - j \right|} \, t_{1,i} \, \bmod{2} = \begin{cases}
		1 \qquad \text{if} \quad  j = k, \\
		0 \qquad \text{otherwise}.
	\end{cases}
\end{equation}
This relation corresponds to a system of $L$ equations, with coefficients $\binom{L}{\left| i - j \right|}$ and variables $t_{1,i}$.  Expressing it in matrix form, where the elements are $M_{ij} = \binom{L}{\left| i - j \right|} \, \bmod{2}$, results in an $L \times L$ circulant matrix. For any system size $L$ such that $\det M \neq 0$, this system has a non-trivial solution, and it is straightforward to determine $S_{1} (x)$.

For these system sizes it also occurs that it is possible to create a localized excitation. Following the same strategy to construct membrane operators from topological constraints, the product of $Z$-operators in the sites belonging to $\tilde{\mathcal{M}}^{a}_{i}$ results in a membrane operator that excites a single TD operator,
\begin{equation}
	\prod_{j \, \in \, \tilde{\mathcal{M}}_{i}^{b} (z)} Z_{j} \equiv \tilde{M}_{i}^{b} (z).
\end{equation}
The fact that a single excitation can be created implies that there is only one global anyonic superselection sector.


\section{\label{sec:mutualst} Excitations and Mutual Statistics}

An excitation corresponds to a state whose eigenvalue of any $\mathcal{O}_{p}$ is $-1$. The action of an $X$-operator on the ground state creates a state with two excitations, while a $Z$-operator is responsible for creating three excitations. Open Wilson lines create pair of excitations, one at each end-point, and can also be used to move a single excitation along the $z$-direction. Analogously, three excitations can be separated through the action of open membrane operators, with some restrictions on the end positions due to the fractal structure. However, this \textit{mobility} in the  $xy$-plane occurs by crossing highly energetic virtual states, making the probability of tunneling very small. This is reminiscent of type-II fractonic physics, implying that the excitations are effectively immobile on the $xy$-plane.

Let us consider a pair of open operators $M^{a}_{I}$ and $W^{a}_{I}$. While the membrane transports charges of the sublattice $a$, the Wilson line transports charges of the other sublattice. Notice, however, that there is no way to move one excitation from one sublattice to the another, so that we can treat the excitations in the two sublattices as distinct\footnote{The Hamiltonian~\eqref{eq:hamiltonian} could be equivalently constructed in terms of TD operators defined as a product of only $X_{j}$ in one sublattice and as a product of only $Z_{j}$ in the other. In this notation, one could refer to them as electric or magnetic excitations depending on their sublattices, in analogy with the toric code.}. This leads naturally to the notion of mutual statistics between such excitations. 

\begin{figure}
	\includegraphics[width = 8cm]{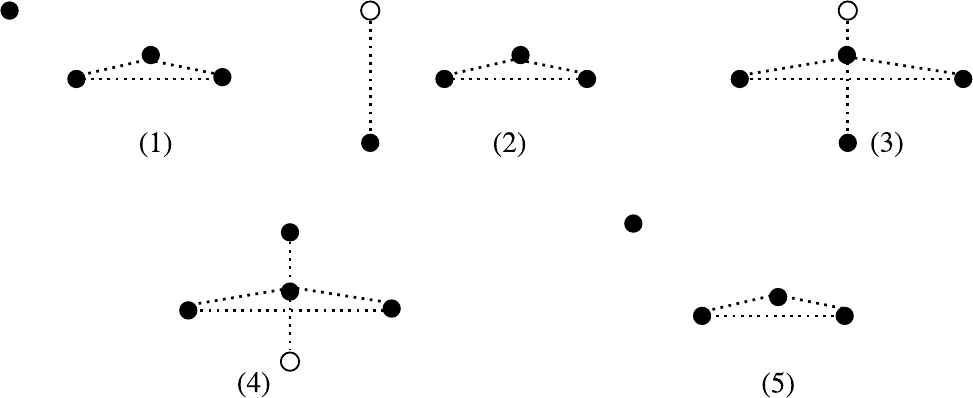}
	\caption{The mutual statistics process.} \label{fig:mutual}
\end{figure}

Consider a state represented by  $\ket{a_{3}, b_{ 2 }}$, containing three excitations in the sublattice $a$ and a pair of excitations in the sublattice $b$, with $b \neq a$. Then, we conceive the following transport of excitations: first, we move one excitation from the sublattice~$b$ using an open Wilson line~$W_{I}^{a}$; subsequently, we separate the three excitations in the sublattice~$a$ using an open membrane $M_{I}^{a}$; next, we return the excitation in the sublattice~$b$ to its original position with ${W^{a}_{I}}^{\dagger}$; and finally, we restore the initial positions of the three excitations in the sublattice~$a$ with ${M^{a}_{I}}^{\dagger}$. This process is  illustrated in Fig.~\ref{fig:mutual}.

Algebraically, the above procedure is expressed as
\begin{equation} \label{eq:stalgebra}
	{M^{a}_{I}}^{\dagger} \, {W^{a}_{I}}^{\dagger} \, M^{a}_{I}  \, W^{a}_{I} \, \ket{a_{3}, b_{ 2 }} = - \ket{a_{3}, b_{ 2 }},
\end{equation} 
where we use the fact that ${W^{a}_{I}}^{\dagger} \, M^{a}_{I} = - M^{a}_{I} \, {W^{a}_{I}}^{\dagger}$. The phase in the right-hand side implies a nontrivial mutual anyonic statistics between the excitations, which is an imprint of long-range entanglement inherent to topologically ordered phases.

We notice that the mutual statistics operation in \eqref{eq:stalgebra} involves only {\it open} membrane and line operators (not {\it symmetry} operators) and thus remains valid for the sizes with a unique ground state. Consequently, even when the ground state is non-degenerate and the fractal symmetry is explicitly broken by the lattice size, the phase is topologically ordered.

\begin{figure}
	\includegraphics[width = 0.9 \linewidth]{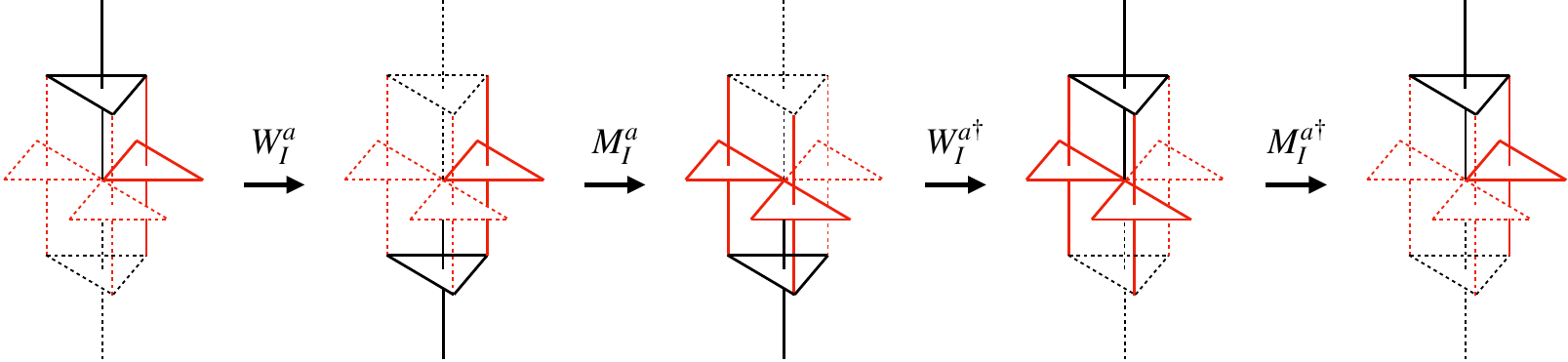}
	\caption{The self-statistics process.} \label{fig:self}
\end{figure}

An alternative way to characterize the topological order is by means of self-statistics of a composite excitation, which is defined through the windmill process discussed in \cite{Song:2023rml}. For example, regardless of the system size, the composite excitation of Fig.~\ref{fig:self} has fermionic self-statistics under the windmill. The windmill process in the HCP lattice starts with an excitation composed by an $\mathcal{O}_p$ with eigenvalue $-1$ in one sublattice and another in the opposite sublattice. Because the excitations are restricted to move along $z$-direction or on the $xy$-plane, the windmill process reduces to the sequence of operations shown in Fig.~\ref{fig:self}, returning to the initial configuration\footnote{Notice that both mutual and self-statistics follow a similar sequence of operations. Analogous to the toric code, mutual statistics corresponds to the braiding of two different excitations $e$ and $m$, while self-statistics involves only a single composite excitation, $f = e \times m$.}. This process involves the open operators $W^{a}_{I}$ and $M^{a}_{I}$ and their inverses, defining the universal statistics phase as $\theta = {M^{a}_{I}}^{\dagger} \,  {W^{a}_{I}}^{\dagger} \, M^{a}_{I} \, W^{a}_{I}$. Given that these operators act locally, the composite excitation has fermionic self-statistics, $\theta = Z_{\vec{r}} \, X_{\vec{r}} \, Z_{\vec{r}} \, X_{\vec{r}} = -1$.

\section{Spontaneous Breaking of Subsystem Symmetries \label{sec:SSB}}

The ground state degeneracy in topologically ordered phases is a reflection of spontaneous breaking of finite higher-form/subsystem symmetries. This can be understood by means of the characterization of topological order in terms of the local indistinguishability of degenerate ground states,
\begin{equation} \label{eq:topind}
	\bra{GS, \alpha} \Phi(x) \ket{GS, \beta} = C \delta_{\alpha \beta},
\end{equation}
where $C$ is a constant independent of the ground states $	\ket{GS, \alpha}$ and $\Phi(x)$ is a local operator. This relation encodes two important informations: i) the expectation value of a local operator cannot distinguish different ground states; ii) there is no {\it local} operator able to connect two distinct ground states. Rather, different ground states are connected by {\it extended} symmetry operators. This is precisely the notion of spontaneous symmetry breaking for an extended symmetry. In other words, the ground state is not invariant under such symmetries, but is mapped into another ground state. 

The existence of two extended symmetries (higher-form/subsystem) possessing a mixed 't Hooft anomaly is sufficient to ensure \eqref{eq:topind}. Let us consider the sizes whose ground states are degenerate.  We pick up a pair of operators $W$ and $M$ satisfying $WM=-MW$. Then we choose to diagonalize $W$ along with the Hamiltonian,
\begin{equation}
	H \ket{GS, \alpha} = E_0 \ket{GS, \alpha} \qquad \text{and} \qquad W\ket{GS, \alpha} = \alpha \ket{GS, \alpha}.
	\label{0101}
\end{equation}
Since $W^{2} = \openone$, the eigenvalues are constrained to $\alpha = \pm 1$. The action of the membrane operator on  $\ket{GS, \alpha}$ is $M \ket{GS, \alpha} = \ket{GS, - \alpha}$. With this, the condition \eqref{eq:topind} follows in a simple way. 

Consider the expectation value
\begin{equation}\label{eq:topind1}
	\begin{split}
		\bra{GS, \alpha} \Phi(x) \ket{GS, \alpha} & = \bra{GS, -\alpha} M^{\dagger} \Phi(x) M \ket{GS, -\alpha}. 
	\end{split}
\end{equation}
Now we can use the fact that the membrane $M$ is mobile along the $z$-direction to avoid eventual interception with the position $x$ of the local operator $\Phi$. This implies that $M$ commutes with $\Phi(x)$ in this expectation value, so that we obtain  
\begin{equation}\label{eq:topind12}
		\bra{GS, \alpha} \Phi(x) \ket{GS, \alpha}  = \bra{GS, -\alpha} \Phi(x) \ket{GS, -\alpha},
\end{equation}
which corresponds to \eqref{eq:topind} with $\alpha=\beta$.

\begin{figure}
	\includegraphics[width=6cm]{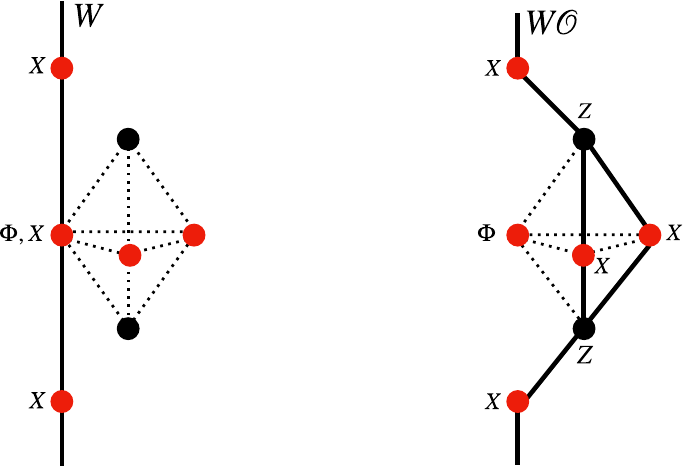}
	\caption{Splitability of the Wilson line.} \label{fig:split}
\end{figure}

To show \eqref{eq:topind} in the case $\alpha\neq \beta$, we consider the expectation value $\bra{GS, -\alpha} \Phi \ket{GS, \alpha}$. According to \eqref{0101}, it can be expressed as
\begin{equation}\label{eq:topind2}
	\begin{split}
		\bra{GS, -\alpha} \Phi(x) \ket{GS, \alpha} & = -\alpha \, \alpha \bra{GS, -\alpha} W^{\dagger} \, \Phi(x) \, W \ket{GS, \alpha}. 
	\end{split}
\end{equation}
To  avoid the position of the local operator, we would like to proceed as we did in the case of membrane operators, but we need to be careful since the Wilson lines are rigid. They enjoy a slightly different deformation property, which we refer to as splitability.  For a Wilson line intersecting the point $x$ of the local operator, we define a new object $\tilde{W}= W \mathcal{O}$, so that the $\tilde{W}$ avoids the point $x$, as illustrated in Fig. \ref{fig:split}.  Now, as all TD operators have eigenvalue $+1$ in the ground state, we can write \eqref{eq:topind2} as
\begin{equation}\label{eq:topind21}
	\begin{split}
		\bra{GS, -\alpha} \Phi(x) \ket{GS, \alpha} & = - \bra{GS, -\alpha} \mathcal{O} W^{\dagger} \, \Phi(x) \, W \mathcal{O}\ket{GS, \alpha} \\
		& = -\bra{GS, -\alpha} \tilde{W}^{\dagger} \Phi(x) \tilde{W}  \ket{GS, \alpha}\\
		&=-\bra{GS, -\alpha}  \Phi(x)  \ket{GS, \alpha}.
	\end{split}
\end{equation}
Therefore, $\bra{GS, -\alpha} \Phi \ket{GS, \alpha} = 0$, which corresponds to \eqref{eq:topind} with $\alpha\neq \beta$. To sum, the mixed 't Hooft anomaly between the two subsystem symmetries is responsible for ensuring the condition for topological ordering \eqref{eq:topind}.

In this context, it is interesting to discuss the lattice sizes for which the ground state is unique. Notice that the absence of extended symmetry operators does not precludes topological ordering, which can be detected through the mutual statistics. When the ground state is unique, the condition \eqref{eq:topind} is trivially satisfied and does not bring much for us. Another example of a topologically ordered system with a unique ground state has been reported in \cite{Watanabe:2022pgk}. 

It is enlightening to recall that in the early days, topological order was thought as lying outside the symmetry-breaking Landau paradigm. With the discovered of generalized (higher-form) symmetries, the topological order was reincorporated in the Landau framework, being understood as the spontaneous breaking of generalized higher-form symmetries, where the order parameter is the expectation value of an extended operator (for our case, the order parameter is $\langle W \rangle$).  In this concrete example, we see that the spontaneous breaking of fractal membrane symmetries is responsible for providing nontrivial degeneracy in the ground state, which is sufficient but not a necessary condition for ensuring topological order. Notably, the generalized symmetry  defined along the non-contractible fractal membrane is sensitive to system size and boundary conditions, making the degeneracy size-dependent. Nonetheless, the nontrivial braiding statistics indicate that the ground state is topological ordered.

\section{\label{sec:boundary} Boundary Theory}

\begin{figure}
	\includegraphics[width=8cm]{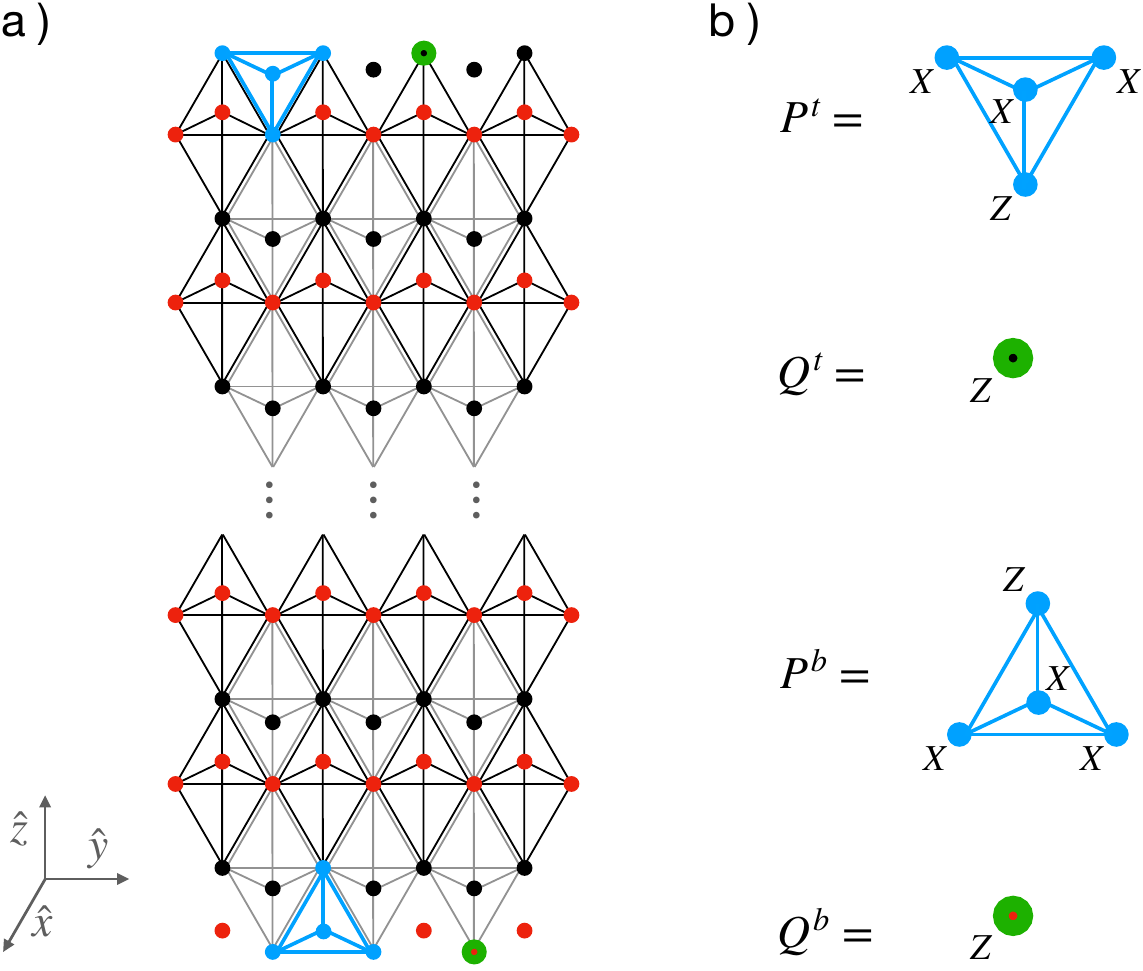}
	\caption{A top boundary of lattice open in $z$-direction and the representation of $P^{b}_{j}$ and $Q^{b}_{j}$ operators.}\label{fig:tboundary}
\end{figure}

We consider the open system with a physical boundary along the plane $xy$, as depicted in Fig.  \ref{fig:tboundary} a). The Hamiltonian is still given by \eqref{eq:hamiltonian}, but now we have new symmetry operators that live at the boundary,
\begin{equation}
	P_{p} \equiv \left( XXX \right)_{p} \, Z_{p - \frac{1}{2} \hat{z}}  \qquad \text{and} \qquad Q_{j}\equiv Z_{j},
\end{equation}
as in Fig. \ref{fig:tboundary} b). Note that $P_{j}$ and $Q_{j}$ correspond to cut TD operators of different sublattices. Both operators are symmetries of the Hamiltonian as they commute with all TD operators. However, in general they do not commute between themselves, namely, $\left[ P_p, Q_{j} \right] \neq 0$ when $j$ coincides with one of the sites whose center is $p$. This leads to anomalies at the boundary, and consequently the boundary theory cannot be trivially gapped. 

We can construct a boundary Hamiltonian in terms of these operators
\begin{equation}
H_{\text{boundary}}= - J_P\sum_p P_p - J_Q\sum_j Q_j.
\end{equation}
As the two terms in the Hamiltonian do not commute, we have the following phase structure. When $J_P>>J_Q$, it corresponds to a gapped phase with condensation of excitations associated with the sublattice of $Q_j$. In the opposite limit, $J_P<< J_Q$, we have another gapped phase with condensation of excitations associated with the sublattice of $P_p$. Finally, it must be a phase transition with gap closing presumably at $J_P \sim J_Q$.  

In the following we will characterize more precisely the boundary anomalies. 

\subsection{Boundary Anomalies \label{sec:boundanom}}

\subsubsection{Cases with a Unique Ground State}

Let us start with the sizes whose ground states are unique. In these cases, there are no anomalies in the bulk. Keeping in mind the discussion in subsection \ref{subsectionD}, it is possible to construct an operator constituted of the product of $P$'s that results in an isolated $X$ in the last layer of the boundary and a several $Z$'s in the penultimate layer of the boundary,
\begin{equation}
 \tilde{P}_j\equiv\prod_{p \, \in \,  \tilde{\mathcal{M}}^{a}_{j} (z)} P_{p}. 
\end{equation}
The operators constructed in this way are in one-to-one correspondence with the $Q$-operators and satisfy 
\begin{equation}
 \tilde{P}_j Q_i = (-1)^{\delta_{ij}} Q_i  \tilde{P}_j .
\end{equation}
This implies that there are $L^2$ anomalies at the boundary. Taking into account the presence of two boundaries, we have simply a total of $2L^2$ anomalies.

\subsubsection{Cases with Degenerate Ground States}

In these cases the analysis is a little more involved as there are anomalies in the bulk. To be careful with the counting of anomalies we treat the system with two boundaries. We find the following symmetry operators
\begin{equation}
W^1,\, W^2,\, M^1,\, M^2,\, P^t,\, Q^t,\, P^b,\, Q^b,  
\label{list}
\end{equation}
where we are omitting the position indices. The superscripts $t$ and $b$ in the boundary operators stand for the top and bottom boundaries, respectively. We need to count the number of independent operators. 

The constraints  \eqref{cmo} of the case with periodic boundary conditions translate now into 
\begin{equation}
\prod_z \prod_{p\, \in\, \mathcal{M}_I^1(z)} \mathcal{O}_p =  \prod_{p\, \in\, \mathcal{M}_I^1} \left( Q_p^t Q_p^b \right)
\label{c11}
\end{equation}
and 
\begin{equation}
	\prod_z \prod_{p\, \in\, \mathcal{M}_I^2(z)} \mathcal{O}_p =  \prod_{p\, \in\, \mathcal{M}_I^2} \left( P_p^t P_p^b \right).
	\label{c12}
\end{equation}
Among the $L^2$ pairs $Q_p^t Q_p^b$ of boundary operators, only $L^2-N_c$ are independent in view of the $N_c$ constraints \eqref{c11}. For the same reason, there are only  $L^2-N_c$ independent pairs of operators $P_p^t P_p^b$.

Now let us discuss the constraints among the Wilson lines, which follow from
\begin{equation}
Q_p^t \left(\prod_z \mathcal{O}_p\right) Q_p^b = W_i^1 W_j^1 W_k^1 
\end{equation}
and 
\begin{equation}
	P_p^t \left(\prod_z \mathcal{O}_p\right) P_p^b = W_i^2 W_j^2 W_k^2.
\end{equation}
These relations imply a total of $L^2- (L^2-N_c)=N_c$ independent Wilson lines for each sublattice, exactly as in the case of PBC.  

The counting of independent membrane operators is not affected by the presence of boundaries along the $xy$-plane, so that we also have $N_c$ independent membrane operators for each sublattice.  

The membrane operators can be written in terms of boundary operators
\begin{equation}
M^1_I = \prod_{p \, \in \, \mathcal{M}_I^1} Q^t_p.
\label{112}
\end{equation} 
These membranes can be transported from the top boundary to the bottom one through the application of TD operators. Thus, we can write
\begin{equation}
	M^1_I = \prod_{p \, \in \, \mathcal{M}_I^1} Q^b_p.
	\label{113}
\end{equation} 
The above relations mean that from the $L^2$  $Q^t_p$-operators of the top boundary, a number $N_c$ of them can be obtained from $N_c$ membranes, i.e., only $(L^2-N_c)$~ $Q^t_p$-operators are independent. The same reasoning follows for the $Q$-operators of the bottom boundary and also for the $P$-operators in both boundaries. Let us label the independent boundary operators with indices $Q_{\bar{I}}$ and $P_{\bar{I}}$, with $\bar{I},\bar{J}=1,2,\ldots,L^2-N_c$. Thus, the list of independent operators is
\begin{equation}
	W_I^1, W_I^2, M_I^1, M_I^2, P_{\bar{I}}^t, Q_{\bar{I}}^t, P_{\bar{I}}^b, Q_{\bar{I}}^b.
\end{equation}
The final step is to change the basis $Q_{\bar{I}}^t\rightarrow \tilde{Q}_{\bar{I}}^t$ and $Q_{\bar{I}}^b\rightarrow \tilde{Q}_{\bar{I}}^b$ (we discuss in the Appendix \ref{sec:MS2} how this is achieved) so that in the new basis the operators satisfy 
\begin{eqnarray}
	W_{I}^{1}M_{J}^{1}& =& \left( -1 \right)^{\delta_{IJ} } M_{J}^{1} W_{I}^{1}~~~\rightarrow~~~N_c ~~\text{anomalies}\nonumber\\
	W_{I}^{2}M_{J}^{2}& =& \left( -1 \right)^{\delta_{IJ} } M_{J}^{2} W_{I}^{2}~~~\rightarrow~~~N_c ~~\text{anomalies}\nonumber\\
	\tilde{Q}_{\bar{I}}^t P_{\bar{J}}^t  &=& (-1)^{\delta_{\bar{I},\bar{J}}}  P_{\bar{J}}^t 	\tilde{Q}_{\bar{I}}^t~~~~~\rightarrow~~~L^2-N_c ~~\text{anomalies}\nonumber\\
	\tilde{Q}_{\bar{I}}^b P_{\bar{J}}^b  &=& (-1)^{\delta_{\bar{I},\bar{J}}}  P_{\bar{J}}^b 	\tilde{Q}_{\bar{I}}^b~~~~~\rightarrow~~~L^2-N_c ~~\text{anomalies},
	\label{anomalies}
\end{eqnarray}
with all other commutators among these operators vanishing. These relations allow us to understand nicely the origin of the anomalies:
\begin{equation}
 \underbrace{2N_c}_{\text{bulk}}+\underbrace{(L^2-N_c)}_{\text{top boundary}}+\underbrace{(L^2-N_c)}_{\text{bottom boundary}}=2 L^2.
 \label{nice}
\end{equation}
It is interesting to note that the total number $2L^2$ of anomalies (bulk + boundaries) is the same as in the case of sizes with nondegenerate ground states. We see from \eqref{nice}   the amount of the anomalies that arise from the bulk and from the boundary. This expression also recovers the cases with unique ground states for $N_c=0$ (no topological constraints), where the anomalies arise exclusively from the boundaries.


\subsection{\label{sec:anomaly} Anomaly Inflow}

The anomaly inflow is a correspondence between anomalies of systems in different dimensions, namely, the anomalies of a $d+1$-dimensional system match the anomalies 
of a $d$-dimensional one, which can be taken as the boundary of the former. In an equivalent way, we can understand this matching of anomalies in terms of the transport of  anomalies from the boundary to the bulk and vice-versa\footnote{Anomaly inflow in other systems with subsystem symmetries has been studied recently in \cite{Burnell2022,Ebisu2024}.}. In the present case, this mechanism shows up in a quite explicit way. 

\begin{figure}
	\includegraphics[width=.75\linewidth]{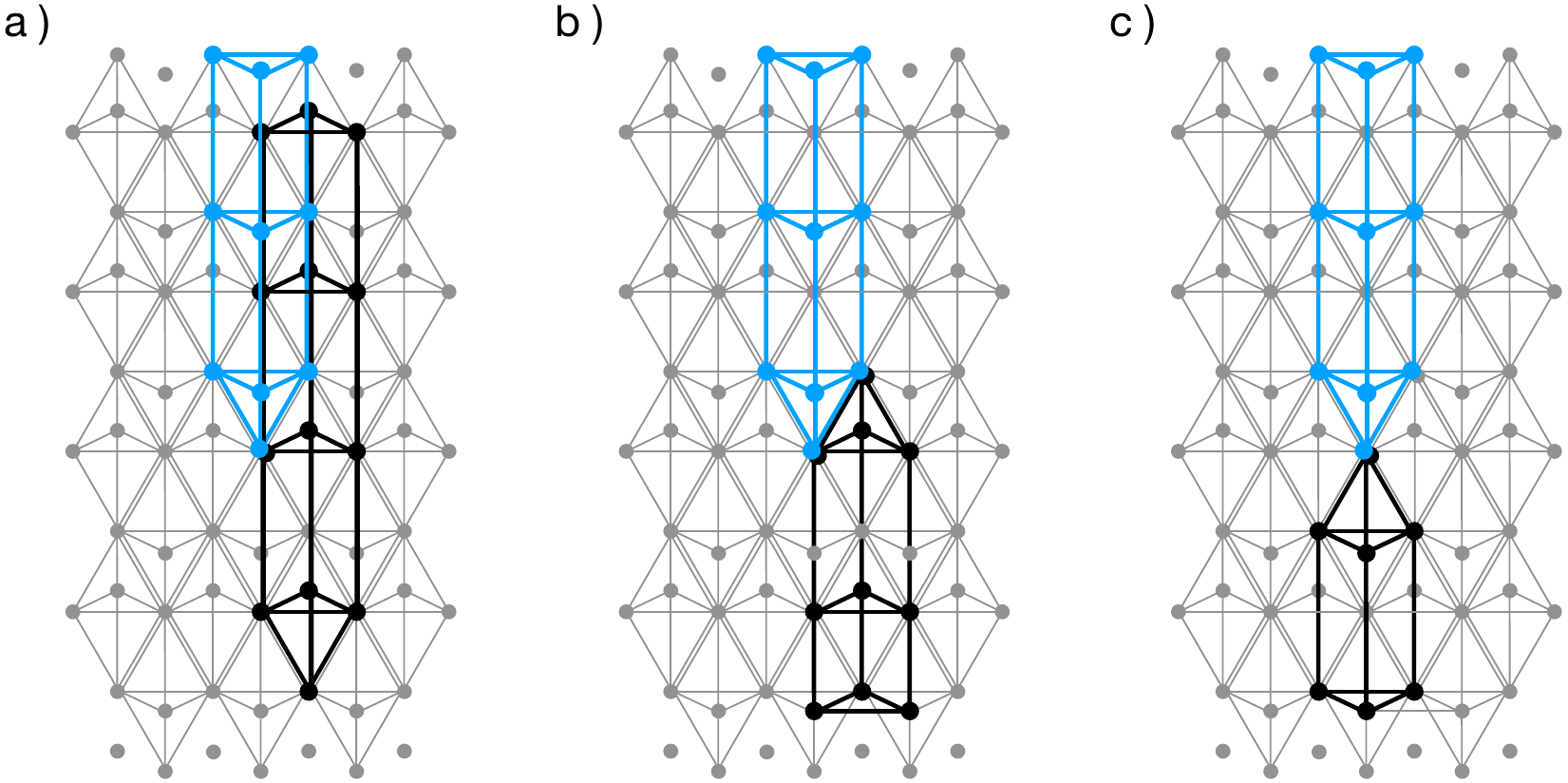}
	\caption{a) Mutually anomalous boundary operators extended into the bulk.  b) and c) Operators of different boundaries extended into the bulk, in which cases there are no anomalies. }\label{fig:ainflow2}
\end{figure}

Let us start with a pair of boundary operators of  \eqref{anomalies}, for example,
\begin{equation}	
	\tilde{Q}_{\bar{I}}^t P_{\bar{J}}^t  = (-1)^{\delta_{\bar{I},\bar{J}}}  P_{\bar{J}}^t 	\tilde{Q}_{\bar{I}}^t.
	\label{111}
\end{equation}
From these operators, we can define new ones that extend to a point $\vec{z}$ in the bulk
\begin{equation}
\bar{Q}_{\bar{I}}^{t,\text{bulk}}\equiv \tilde{Q}_{\bar{I}}^t \prod_{j=\bar{I}-\frac12\hat{z}}^{\bar{I}-\vec{z}} \mathcal{O}_j~~~\text{and}~~~ \bar{P}_{\bar{I}}^{t,\text{bulk}}\equiv {P}_{\bar{I}}^t  \prod_{j=\bar{I}-\hat{z}}^{\bar{I}- \vec{z}} \mathcal{O}_j,
\end{equation}
as shown in Fig. \ref{fig:ainflow2} a). Relation \eqref{111}  implies that these operators satisfy
\begin{equation}
\bar{Q}_{\bar{I}}^{t,\text{bulk}} \bar{P}_{\bar{J}}^{t,\text{bulk}} =  (-1)^{\delta_{\bar{I},\bar{J}}}  \bar{P}_{\bar{J}}^{t,\text{bulk}}\bar{Q}_{\bar{I}}^{t,\text{bulk}},
\end{equation}
showing that the boundary anomaly is realized now in the bulk. Similarly, the anomalies of the bottom boundary can also be transported to the bulk.  Notice, however, that there is no mixing between anomalies from opposite boundaries,
\begin{equation}
[\bar{Q}_{\bar{I}}^{t,\text{bulk}}, \bar{P}_{\bar{J}}^{b,\text{bulk}} ]=[\bar{Q}_{\bar{I}}^{b,\text{bulk}}, \bar{P}_{\bar{J}}^{t,\text{bulk}} ]=0.
\end{equation}
These relations are illustrated in Figs. \ref{fig:ainflow2} b) and  \ref{fig:ainflow2} c).

The anomalies between Wilson lines and membrane operators that take place in the bulk can also be realized in the boundary. We simply transport the membrane operators to the boundary, where they are expressed in terms of $P$-operators as in \eqref{112} and \eqref{113}.

\begin{figure}
	\includegraphics[width=.75\linewidth]{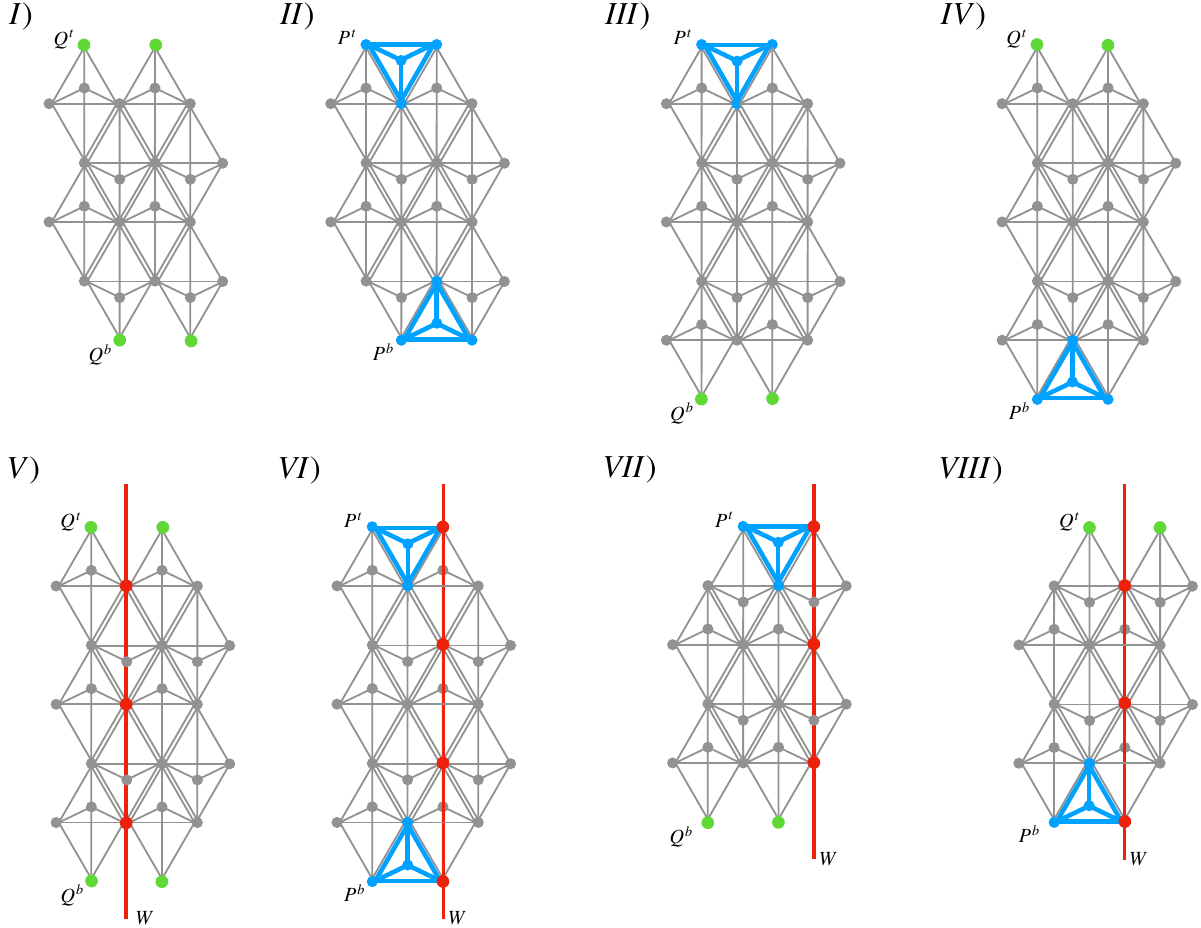}
	\caption{Boundary operators producing different two-dimensional phases.} \label{fig:boundop}
\end{figure}

\section{\label{sec:comparison} Connection with Two-Dimensional Fractal Models}

The Hamiltonian \eqref{eq:hamiltonian} shares its fractal structure  with certain two-dimensional models, namely, a SPT phase studied in \cite{Devakul_2019}
and a SSB system studied in \cite{Newman_1999,Zhou:2021wsv,Sfairopoulos2023}. Given that the fractal structure strongly dictates many physical properties, like the ground state degeneracy, we expect a direct comparison between properties of the models in $2d$ and the three-dimensional model \eqref{eq:hamiltonian}. It is enlightening to compare the physics of these models in $2d$ and $3d$, making clear the role of the  third dimension. It turns out that the boundary theory discussed previously can be used for this purpose. 

The idea underlying the connection between models in $2d$ and $3d$ is the following. We start with the model  \eqref{eq:hamiltonian} in the presence of two boundaries along the $xy$-plane, as in the previous section. Then we consider that each boundary condenses a single type of excitation, which means that we take the Hamiltonian as
\begin{equation}
H=H_{\text{bulk}}+H_{\text{boundary}},
\end{equation}
with $H_{\text{boundary}}$ involving only simultaneously commuting boundary operators. By taking the limit $L_x,L_y \gg L_z$, we obtain effectively a two-dimensional model. As we shall discuss, depending whether the two boundaries condense the same or different types of excitations, the resulting $2d$ model comprehends a SSB or a SPT phase.

Let us consider initially sizes $L_x=L_y=L$  so that the corresponding ground states are degenerate. The symmetries are $M^{a}_{I}$, $W^{a}_{I}$, and the boundary operators $P$'s and $Q$'s. There are several possibilities for introducing boundary operators producing different two-dimensional phases depending on which sublattice the operators belong, according to Fig.~\ref{fig:boundop}. If the operators introduced at the top and at the bottom boundaries belong to different sublattices, the Wilson line is not a symmetry and all $M^{a}_{I}$ reduce to a product of the operators of the Hamiltonian. Conversely, if the operators in the two boundaries belong to the same sublattice, $M^{a}_{I}$ and $W^{a}_{I}$ are still symmetries.

\begin{figure}
	\includegraphics[width=.50\linewidth]{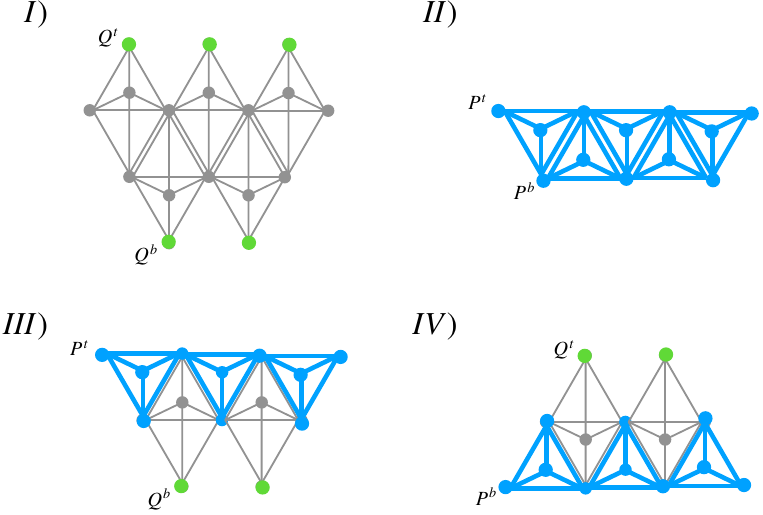}
	\caption{Two-dimensional SPT phase obtained in the limit $L\gg L_z$ from different boundaries.} \label{SPT}
\end{figure}

Cases $I,\ldots, IV$ in Fig.~\ref{fig:boundop} give rise to a SPT phase after we take limit $L\gg L_z$, shown in Fig. \ref{SPT}. In other words, all the possibilities $I,\ldots, IV$ are simply different representations of the same SPT phase. Let us discuss explicitly some cases to clarify this point. 

\begin{figure}
	\includegraphics[width=.3\linewidth]{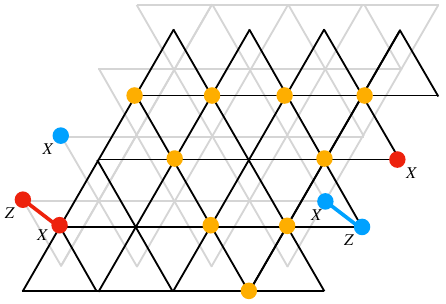}
	\caption{Top view of the two-dimensional SPT phase with the corresponding boundary operators.} \label{SPT2}
\end{figure}

We start with $I$, whose Hamiltonian, after the limit $L\gg L_z$, reads
\begin{equation}
H_{\text{SPT}}= - J \sum_{p\, \in \, (xy)_1} \mathcal{O}_p   - J \sum_{p\, \in \, (xy)_2} \mathcal{O}_p -J_Q^t \sum_{p\, \in \, (xy)_1} Q_p^t - J_Q^b \sum_{p\, \in \, (xy)_2} Q_p^b,
\label{114}
\end{equation}
where the notation $(xy)_{1,2}$ means that the sum runs over the whole plane $xy$ (not in $z$) associated with the sublattice 1,2 displaced from each other along $z$ by $\hat{z}/2$, as in Fig. \ref{SPT} $I$. There are no nontrivial symmetries since the membrane operators, which would be symmetries, can be expressed in terms of the operators of the Hamiltonian \eqref{114}. Thus, the ground state is unique. However, this is not a trivially gapped phase due to the presence of anomalous edge symmetries when we consider a boundary in the plane $xy$. As shown in Fig. \ref{SPT2}, there are boundary operators that commute with \eqref{114} but do not commute among themselves, implying that the boundary is not trivially gapped. Therefore, this phase corresponds to a SPT one, which has been discussed in \cite{Devakul_2019}.

Another representative of the SPT class is the case $II$. After taking the limit $L\gg L_z$, the Hamiltonian becomes
\begin{equation}
H_{\text{SPT}}=-J_P^t \sum_{p\, \in \, (xy)_1} P_p^t - J_P^b \sum_{p\, \in \, (xy)_2} P_p^b, 
\label{115}
\end{equation}
as shown in Fig. \ref{SPT} $II$. Note that, in contrast to the previous case, here the boundary operators are accommodated without the need to maintain the TD operators.  In this case, the membrane operators also can be written in terms of the operators of the Hamiltonian \eqref{115} and then the ground state is unique. Introducing a boundary in the $xy$-plane we find again anomalous edge symmetries, implying that this is a SPT phase. It is just a different parametrization of the SPT phase of $I$. The same reasoning follows for the cases $III$ and $IV$.

\begin{figure}
	\includegraphics[width=.5\linewidth]{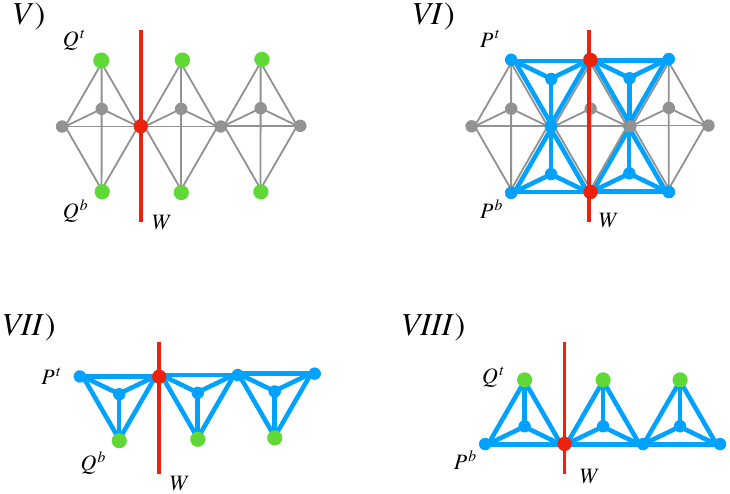}
	\caption{Two-dimensional SSB phase obtained in the limit $L\gg L_z$ from different boundaries.} \label{fig:approx2d}
\end{figure}

Now we discuss the SSB class. Let us take the representative $V$ of Fig. \ref{fig:approx2d}, whose Hamiltonian in the limit $L\gg L_z$ is
\begin{equation}
H_{\text{SSB}}=- J \sum_{p\, \in \, (xy)_1} \mathcal{O}_p  -J_Q^t \sum_{p\, \in \, (xy)_1} Q_p^t - J_Q^b \sum_{p\, \in \, (xy)_1} Q_p^b.
\end{equation}
Now, the membrane operators are nontrivial symmetries just like the Wilson lines, in which case become local operators. The nontrivial commutation between these symmetry operators in the bulk leads to the ground state degeneracy $2^{N_c}$, with the ground states being distinguished by a local operator. Therefore, this corresponds to a SSB phase. It is known as the Newman-Moore phase, and has been discussed in  \cite{Newman_1999,Zhou:2021wsv,Sfairopoulos2023}. All the other cases $VI, VII,$ and $VIII$ in Fig. \ref{fig:approx2d} exhibit the same physics.

As discussed in Sec. \ref{sec:mutualst}, the model \eqref{eq:hamiltonian} is topologically ordered because the long-range entanglement makes impossible to deform the ground state into a product state at finite time. The nontrivial mutual statistics depends on the mobility along the $z$-direction, and thus the two dimensional models cannot exhibit the long-range entanglement. Additionally, in the cases where a two-dimensional fractal model presents nontrivial ground state degeneracy, there is a local operator distinguishing different ground states. Consequently, the two-dimensional fractal models are not topologically ordered. This is the reason why they lie out the Haah's bound \cite{Haah_2021} for the ground state degeneracy.

\section{Effective Field Theory \label{sec:eft}}

The effective field theory of a physical system is a description that captures fundamental information of its physics in the long wavelength  limit, neglecting non-universal details. It is important to mention, however, that due to the strong dependence of fractal subsystem symmetries on the lattice details we cannot consider the strict continuous limit of the model \eqref{eq:hamiltonian}. With this in mind, what we shall refer to effective theory is the embedding of the lattice model into a $U(1)$ gauge theory which contains all the universal data of the lowest excited states. Such a description, as we will argue, is a  powerful tool for examining various aspects of the system, such as the presence of a mixed `t Hooft anomaly and the physics at the boundary.


\subsection{Effective Field Theory}

We start with a bosonic lattice Hamiltonian that is invariant under $U(1)$ fractal subsystem symmetries and proceed to gauge such symmetries. We then argue that  the deep IR of the Higgs phase, where $U(1)$ is effectively reduced to $\mathbb{Z}_2$, corresponds to the ground state manifold of the model in Eq. \eqref{eq:hamiltonian}. 

 Consider the following Hamiltonian
\begin{equation}\label{eq:btheory}
	H _{\text{matter}}= \dfrac{1}{2}\sum_{\vec{r}\,\in\, \Lambda_1}  \Pi_{\vec{r}}^{\dagger}\,\Pi_{\vec{r}} -J  \sum_{\vec{r}\, \in \,  \Lambda_1}  \left (\Phi_{\vec{r}} \, \Phi_{\vec{r}+\hat\epsilon_{1}}\, \Phi_{\vec{r}+\hat\epsilon_{2}} + \sigma_0 \, \Phi_{\vec{r}}^\dagger \, \Phi_{\vec{r}+\hat\epsilon_3} +\text{H. c.} \right )+ \dots ~ ,
\end{equation}
where $\Phi_{\vec r}$ is a complex bosonic field and $\Pi_{\vec{r}}$ its conjugate momentum. In this Hamiltonian, $\hat\epsilon_{1}$ and $\hat\epsilon_2$ represent triangular lattice unit vectors,  $\hat\epsilon_3$ corresponds to a unit vector along the vertical direction, which span the whole lattice $\Lambda_1$ consisting of vertical stacks of 2D planar triangular lattices (one of the sublattices of \eqref{eq:hamiltonian}). We set the dimension of the scalar field as $[\Phi]=1/2$ in mass units. The dimensions of the parameters $J$ and $\sigma_0$ are $[J]=3/2$ and $[\sigma_0]=1/2$.

The Hamiltonian in Eq.~\eqref{eq:btheory} allows for three and two particle creation processes in the $xy$-plane and in the $z$-axis, respectively, which mimics the action of $Z_{i}$ and $X_{i}$ operators in the model \eqref{eq:hamiltonian}. 

\begin{figure}
	\includegraphics[width=.6\linewidth]{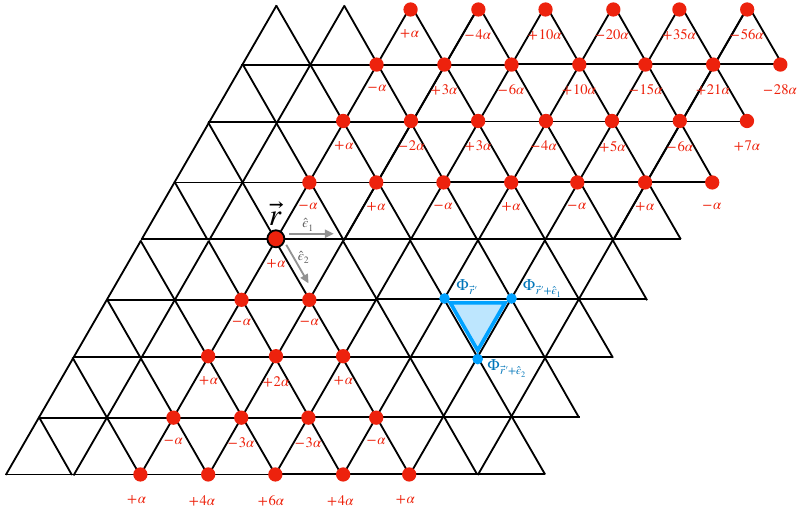}
	\caption{Action of the $U(1)$ symmetry in the $xy$-plane. The fields $\Phi$'s in the vertices of the lattice transform with a phase, according to Eq.~\eqref{symmetries}. All the products $\Phi_{\vec{r}'} \, \Phi_{\vec{r}'+\hat\epsilon_{1}}\, \Phi_{\vec{r}'+\hat\epsilon_{2}}$, as depicted in blue, are invariant under the action of this symmetry.} \label{fig:u1sym}
\end{figure}

For infinite size, the model \eqref{eq:btheory} possesses an infinite number of nonuniform global $U(1)$ symmetries,  
\begin{equation}
\Phi_{\vec{r}} \rightarrow e^{i \alpha_{\vec{r}_0} f_{\vec{r}-\vec{r}_0}} \Phi_{\vec{r}},
\label{symmetries}
\end{equation}
where $\alpha_{\vec{r}_0}$ are the $U(1)$ global parameters and $f_{\vec{r}}$ is defined as
\begin{equation}
\quad f_{\vec{r}} \equiv(-1)^{i_2}\, \Theta(i_{1} + i_{2} + 1/2) \binom{i_{2}}{i_{2} + i_{1}},
\label{newton}
\end{equation}
with $\Theta$ being the Heaviside step function and the site $\vec{r}$ specified by $\vec{r}= i_1 \hat{\epsilon}_1+ i_2 \hat{\epsilon}_1+ i_3 \hat{\epsilon}_3$. Note that $f_{\vec{r}}$ is an integer-valued function. The action of this symmetry is illustrated in Fig.~\ref{fig:u1sym}.

It is useful to see the fate of these symmetries for finite $L$. For $L$ multiple of 3 but $L\neq k(2^n - 2^m),$ with $n > 2$ (for example, $L=33$), the lattice can be partitioned into three sublattices, $A, B$, and $C$. Then the symmetries \eqref{symmetries} break down to
\begin{equation}
	U(1)_{A \cup C}: \Phi_{\vec r}\rightarrow e^{\mathrm i\alpha f_{\vec{r}}^{(1)}}\Phi_{\vec{r}}\quad \text{for}\quad \vec{r}\in A \label{ab'}
\end{equation}
and
\begin{equation}
	U(1)_{B \cup C}: \Phi_{\vec r}\rightarrow e^{\mathrm i \beta f_{\vec{r}}^{(2)} }\Phi_{\vec{r}}\quad \text{for}\quad \vec{r}\in A, \label{bc}
\end{equation}
with $f_{\vec{r}}^{(1)} $ and $f_{\vec{r}}^{(2)} $ being certain linear combinations of the integer-valued function in \eqref{newton},
\begin{equation}
	f_{\vec{r}}^{(1)} \equiv  \sum_{k=0}^{\frac{L}{3}-1}\left(f_{\vec{r}+3k\hat{\epsilon}_1} - f_{\vec{r}+(3k+2)\hat{\epsilon}_1}\right)~~~\text{and}~~~
		f_{\vec{r}}^{(2)} \equiv  \sum_{k=0}^{\frac{L}{3}-1}\left(f_{\vec{r}+(3k+1)\hat{\epsilon}_1} - f_{\vec{r}+(3k+2)\hat{\epsilon}_1}\right),
		\label{functions}
\end{equation}
which are also integer-valued and in addition are periodic in the $xy$-plane (periodicity in $z$ is trivial since $f_{\vec{r}}$ does not depend on $i_3$). Note that a similar transformation for $A\cup B$ sublattices  is also a symmetry, but it is not independent from the other two. 

 For $L=k(2^n-2^m)$, but not a multiple of 3, the symmetries \eqref{symmetries} break down to the discrete fractal symmetries $\mathbb{Z}_2$ of the model \eqref{eq:hamiltonian} (if $\alpha_{\vec{r}_0}$ are integers defined mod 2). 
If $L$ is both a multiple of 3 and of the form $L=k(2^n-2^m)$, we have a mix of the two previous cases. If $L$ does not fit into any of the cases, then there is no remaining symmetry.

We proceed with the gauging of the $U(1)$ symmetries, which is more naturally implemented in the Lagrangian description. For this, we introduce dynamical gauge fields  $a_t, a_{xy}, a_z$ governed by a Maxwell-like Lagrangian,
\begin{equation}
	\mathcal{L}[a] = \frac{1}{2h_{xy}} \left(\Delta_{xy} a_{t} - \partial_t a_{xy} \right)^{2} +  \frac{1}{2h_{z}} \left(\Delta_{z} a_{t} - \partial_t a_{z} \right)^{2}  - \dfrac{1}{2g}\left (\Delta_{xy}a_z-\Delta_z a_{xy}\right )^2.
	\label{ML}
\end{equation}
The dimensions of the gauge fields and of the coupling constants are $[a_t]=[a_{xy}]=[a_z]=1$ and $[h_{xy}]=[h_z]=[g]=1$.

The Lagrangian \eqref{ML} is invariant under the gauge transformations
\begin{equation}
a_t \rightarrow a_t + \partial_t \xi,~~~ a_z \rightarrow a_z + \Delta_z \xi,~~~ a_{xy}\rightarrow a_{xy}+\Delta_{xy}\xi,
\end{equation}
where  
\begin{equation}
	\Delta_{xy} \, \xi_{\vec{r}} \equiv \frac{1}{\ell}( \xi_{\vec{r}} +\xi_{\vec{r}+\hat\epsilon_{1}} +\xi_{\vec{r}+\hat\epsilon_{2}})\qquad
	\text{and} \qquad \Delta_z \, \xi_{\vec{r}} \equiv \frac{1}{\ell_z} (\xi_{\vec{r} +\hat\epsilon_3} - \xi_{\vec{r}}),
\end{equation} 
with $\ell$ and $\ell_z$ corresponding to the lattice spacing in the $xy$-plane and in the $z$-direction, respectively. Notice that the operator $\Delta_{xy}$ is not well-defined in the limit $\ell\rightarrow 0$. 
In the absence of boundaries, the integration by parts involving the operator $\Delta_{xy}$ is
\begin{equation}
	\begin{aligned}
		\sum_{\vec{r}} \xi_{\vec{r}} \, \Delta_{xy} \eta_{\vec{r}} &= \sum_{\vec{r}} \xi_{\vec{r}} \frac{1}{\ell} \left(\eta_{\vec{r}} + \eta_{\vec{r} + \hat\epsilon_{1}} + \eta_{\vec{r}+\hat\epsilon_{2}}\right) \\
		&=\sum_{\vec{r}} \eta_{\vec{r}}\frac{1}{\ell} \left(\xi_{\vec{r}} + \xi_{\vec{r} - \hat\epsilon_{1}} + \xi_{\vec{r} - \hat\epsilon_{2}}\right)\\
		&= \sum_{\vec{r}} \eta_{\vec{r}} \, \Delta_{xy}' \xi_{\vec{r}}, \label{ibp}
	\end{aligned}
\end{equation}
where we have defined 
\begin{equation}
	\Delta_{xy}' \xi_{\vec{r}} \equiv \frac{1}{\ell}( \xi_{\vec{r}} + \xi_{\vec{r} - \hat\epsilon_{1}} + \xi_{\vec{r} - \hat\epsilon_{2}}).
\end{equation}
The operators $\Delta_{xy}$ and $\Delta_{xy}'$ can be taken as acting on opposite sublattices, so that the integration by parts, in effect, switches the sublattice. They satisfy the respective properties
\begin{equation}
\Delta_{xy}'f_{\vec{r}}=0~~~\text{and}~~~ \Delta_{xy}g_{\vec{r}}=0,
\label{annihilation}
\end{equation}
with $f_{\vec{r}}$ given in \eqref{newton} and $g_{\vec{r}}$ being the counterpart of the opposite sublattice,
\begin{equation}
	\quad g_{\vec{r}} \equiv(-1)^{i_2}\, \Theta(-i_{1} - i_{2} + 1/2) \binom{-i_{2}}{- i_{2} - i_{1}}.
	\label{newton'}
\end{equation}
As we shall see, the properties \eqref{annihilation} are important for the construction of gauge-invariant objects. In the case of a finite system along the $xy$-plane with $L$ being a multiple of 3, we can construct $g_{\vec{r}}^{(1)}$ and $g_{\vec{r}}^{(2)}$ analogous to  \eqref{functions}, 
\begin{equation}
	g_{\vec{r}}^{(1)} \equiv  \sum_{k=0}^{\frac{L}{3}-1}\left(g_{\vec{r}+3k\hat{\epsilon}_1} - g_{\vec{r}+(3k+2)\hat{\epsilon}_1}\right)~~~\text{and}~~~
g_{\vec{r}}^{(2)} \equiv  \sum_{k=0}^{\frac{L}{3}-1}\left(g_{\vec{r}+(3k+1)\hat{\epsilon}_1} - g_{\vec{r}+(3k+2)\hat{\epsilon}_1}\right).
\label{functions'}
\end{equation}

The coupled system of matter and gauge fields is then incorporated in the partition function as
\begin{equation}
	\mathcal Z = \int [\mathcal Da] \mathcal D\Phi \mathcal D \Phi^\dagger \, e^{\mathrm i \int dt \,  \ell \ell_z\sum_r\, \left (\mathcal L[\Phi,a]+\mathcal{L}[a]\right )},
\end{equation}	
where	
\begin{equation}
	\mathcal L[\Phi,a] = \left |(\partial_t -\mathrm i q a_t)\Phi\right |^2 +J e^{-i q \ell a_{xy,\vec{r}+\vec{\delta}}} \Phi_{\vec{r}} \, \Phi_{\vec{r}+\hat\epsilon_{1}}\, \Phi_{\vec{r}+\hat\epsilon_{2}} + J\sigma_0 \, e^{-i q \ell_z a_{z, \vec{r}+\frac12\hat\epsilon_3}} \Phi^\dagger_{\vec{r}} \, \Phi_{\vec{r}+\hat\epsilon_3} + \text{H. c.},
	\label{ScalarLag}
\end{equation}
for an integer $q$, which sets the charge of $\Phi$ under the gauge group, and $\vec{\delta} \equiv \frac13 ( \hat\epsilon_1- \hat\epsilon_2)$ localizes the center of the triangle.

Aiming to isolate the physics of the deep IR of the Higgs phase, we parametrize the complex field in terms of a phase as $\Phi_{\vec{r}} = \rho_0  e^{i \phi_{\vec{r}}}$, which allows us to rewrite the theory as
\begin{eqnarray}
	\mathcal Z= \int [\mathcal D a] \, \mathcal D\phi \, e^{\mathrm i \int dt\, \ell \, \ell_z\, \sum_r\, \left (\mathcal L [\phi,a]+\mathcal{L} [a]\right )}, 
\end{eqnarray}	
with
\begin{eqnarray}
	\mathcal L[\phi,a] =\rho_0^2(\partial_t\phi - q a_t)^2+2 J \rho_0^3 \cos[\ell\left (\Delta_{xy}\phi - q a_{xy}\right )] + 2 J \rho_0^2 \sigma_0  \cos[\ell_z\left (\Delta_{z}\phi - q a_{z}\right)].
\end{eqnarray}

We want to express such a theory in terms of a pure gauge theory, where the extended symmetries become evident. For this, we introduce fields  $c_{t} \equiv  \partial_{t} \phi$,  $c_{xy} \equiv  \, \Delta_{xy} \phi$, and $c_{z} \equiv  \, \Delta_{z} \phi$, so that we rewrite the theory in the form
\begin{equation}
	\mathcal Z = \int [\mathcal D a][\mathcal D b] [\mathcal D c] \, e^{\mathrm i \int dt\, \ell \, \ell_z \, \sum_r\, \left (\mathcal L [a,c] + \mathcal L[b,c]+\mathcal{L} [a]\right )}, 
\end{equation}
with
\begin{eqnarray}
		\mathcal L[a,c] &=& \rho_0^2(c_t- q a_t)^2+2 J \rho_0^3 \cos[\ell\left (c_{xy}- q a_{xy}\right )] + 2 J \rho_0^2 \sigma_0  \cos[\ell_z\left (c_z - q a_{z}\right)]\nonumber\\
		&=& \rho_0^2 \left(c_t-q a_{t} \right )^2- J \rho_0^3 \ell^2 \left(c_{xy}-q a_{xy}  \right)^{2} - J \rho_0^2 \sigma_0 \ell_z^2  \left(c_{z}-q a_{z}\right)^{2} + \cdots
\end{eqnarray}	
and
\begin{equation}	
		\mathcal{L}[b,c]=  \frac{1}{2 \pi} \left[ b_{t} \left( \Delta_{xy} c_{z} - \Delta_{z} c_{xy}\right) + b_{xy}\left(\Delta_{z} c_{t} - \partial_{t} c_{z}\right) - b_{z}\left(\Delta_{xy} c_{t} - \partial_{t} c_{xy}\right) \right].
		\label{01234}
\end{equation}
The Lagrangian \eqref{01234} is introduced to enforce dynamically the constrains $\Delta_{xy}c_{t} = \partial_{t} c_{xy}$, $\Delta_{z}c_{t} = \partial_{t} c_{z}$, and $\Delta_{xy} c_{z} = \Delta_{z} c_{xy}$. If we integrate out the Lagrange multiplier fields $b_t$, $b_{xy}$, and $b_z$, we return to the original action. On the other hand, integrating out the $c$-fields results in 
\begin{eqnarray}
	\mathcal Z = \int [\mathcal D a][\mathcal D b] \, e^{\mathrm i \int dt \, \ell \, \ell_z\, \sum_r\, \left (\mathcal L [a,b] + \mathcal L[b]+\mathcal{L} [a]\right )},
	\label{BF}
\end{eqnarray}
with
\begin{eqnarray}
	\mathcal{L}[a,b] &=& \frac{q}{2 \pi } \left[b_{t} \left(\Delta_{xy} a_{z} - \Delta_{z} a_{xy} \right) + b_{xy} \left( \Delta_{z} a_{t} - \partial_{t} a_{z}\right) + b_{z} \left( \partial_{t} a_{xy} - \Delta_{xy} a_{t} \right)\right]\nonumber\\
	&\equiv&  \frac{q}{2 \pi } b\,\wedge  Da
	\label{ab}
\end{eqnarray}	
and
\begin{eqnarray}	
	\mathcal L[b] =  &-&\frac{1}{16 \pi ^2 \rho_0^2} \left(\Delta_{z} b_{xy} + \Delta_{xy}' b_{z}\right)^{2}+ \frac{1}{16 \pi ^2 J \rho_0^2 \sigma_0 \ell_z^2} \left(\partial_t b_{xy} + \Delta_{xy}' b_{t} \right)^{2} \nonumber\\&+& \frac{1}{16 \pi ^2 J \rho_0^3 \ell^2} \left(\partial_t b_{z} - \Delta_{z} b_{t} \right)^{2}.
\end{eqnarray}

From \eqref{BF}, we are ready to take the low-energy limit but not long-distances, which means that all constants with positive dimension in mass units, $h_{xy}, h_z, g, J, \rho_0$, and  $\sigma_0$ are taken to infinite while keeping fixed the length scales $\ell$ and $\ell_z$. In this limit all gapped excitations become inaccessible and the effective description is governed by a BF-like theory \eqref{ab},
\begin{eqnarray}\label{IR}
	\mathcal Z_{\text{IR}} = \int [\mathcal D a][\mathcal D b] \, e^{\frac{\mathrm i \,q }{2\pi}\int b\, \wedge\, Da},
\end{eqnarray}
where we have introduced the short notation $\int \equiv  \int dt \, \ell \, \ell_z\, \sum_r\,$. 

In the absence of boundaries, this theory is invariant under the gauge transformations
\begin{eqnarray}
	a_{t} \to a_{t} + \partial_{t} \xi, \quad &a_{xy} \to a_{xy} + \Delta_{xy} \xi,   \quad &a_{z} \to a_{z} + \Delta_{z} \xi, \label{eq:Agauge}\\
	b_{t} \to b_{t} + \partial_{t} \eta, \quad &b_{xy} \to b_{xy} - \Delta_{xy}' \eta,  \quad &b_{z} \to b_{z} + \Delta_{z} \eta.\label{eq:Bgauge}
\end{eqnarray}


\subsection{Extended Defects and Operators}

We would like to construct extended gauge-invariant defects and operators. Let start with the defects, which are extended objects along the time direction. They are simply 
\begin{equation}
\exp \left( i u_e \oint dt \, a_t \right) ~~~\text{and}~~~  \exp \left( i u_m \oint dt \, b_t \right),
\end{equation}
and are interpreted as describing the world-lines of single probe excitations at rest. If we take the time direction as a circle of length $\tilde{L}_t$ (dimensional), then we can construct two types of large gauge transformation that wind around it. The first type are the uniform (usual) ones 
\begin{equation}
\xi= 2\pi \frac{t}{\tilde{L}_t} n_t, ~~~ n_t\,\in \mathbb{Z},
\end{equation}
which are the same for all points in the $xy$-plane, inducing  
\begin{equation}
a_t \rightarrow a_t + \frac{2\pi }{\tilde{L}_t}n_t,~~~a_z\rightarrow a_z,~~~ a_{xy}\rightarrow a_{xy} + 6\pi \frac{t}{\ell\tilde{L}_t}n_t.
\label{trans1}
\end{equation}
The second type of large gauge transformations are the nonuniform ones 
\begin{equation}
	\xi= 2\pi \frac{t}{\tilde{L}_t} g_{\vec{r}}^{(\sigma)},~~~\sigma=1,2,
\end{equation}
where $g_{\vec{r}}^{(1)}$ and $g_{\vec{r}}^{(2)}$ are the integer-valued functions defined in \eqref{functions'}. In this case, the winding depends on the position in the $xy$-plane. The corresponding field transformations are
\begin{equation}
	a_t \rightarrow a_t + \frac{2\pi }{\tilde{L}_t}g_{\vec{r}}^{(\sigma)} ,~~~a_z\rightarrow a_z,~~~ a_{xy}\rightarrow a_{xy}.
	\label{trans1'}
\end{equation}
The defect $\exp \left( i u_e \oint dt \, a_t \right) $ is invariant under both types of gauge transformations provided that $u_e$ is an integer. A similar reasoning for the $b$-field  implies that $u_m$ is also an integer. 

Now we construct extended operators. The first ones are the line operators along the $z$-direction
\begin{equation}
\exp \left(i v_e\, \ell_z \osum_z a_z \right) ~~~\text{and}~~~\exp \left(i v_m\, \ell_z \osum_z b_z \right). 
\label{line}
\end{equation}
As in the previous case, we can consider two types of large gauge trasformations. A uniform large gauge transformation that winds around the $z$-direction is 
\begin{equation}
\xi = 2\pi \frac{i_3}{L_z} n_3,~~~n_3\,\in\,\mathbb{Z},
\end{equation}
under which the $a$-fields transform as 
\begin{equation}
a_t\rightarrow a_t,~~~a_z \rightarrow a_z + \frac{2\pi}{\ell_z L_z}n_3, ~~~ a_{xy}\rightarrow a_{xy} + 6\pi \frac{i_3}{\ell \,L_z}n_3.
\label{trans2}
\end{equation}
The nonuniform large gauge trasformations are
\begin{equation}
	\xi = 2\pi \frac{i_3}{L_z} g_{\vec{r}}^{(\sigma)},~~~\sigma=1,2,
\end{equation}
and the corresponding field transformations are 
\begin{equation}
	a_t\rightarrow a_t,~~~a_z \rightarrow a_z + \frac{2\pi}{\ell_z L_z} g_{\vec{r}}^{(\sigma)}, ~~~ a_{xy}\rightarrow a_{xy} .
	\label{trans2'}
\end{equation}
Invariance of $\exp \left(i v_e\, \ell_z \displaystyle\osum_z a_z \right)$ under both types of gauge transformations implies that $v_e$ is an integer. A similar reasoning for the $b$-fields also implies that $v_m$ is an integer.

Finally, we can construct extended operators in the $xy$-plan. We need to be careful here because such operators are very sensitive to the lattice size along the $x$ and $y$ directions. For infinite size, they are 
\begin{equation}
\exp  \left(i w_e \ell \sum_{xy} f_{\vec{r}-\vec{r}_0}\, a_{xy}  \right)~~~\text{and}~~~ \exp  \left(i w_m \ell \sum_{xy} g_{\vec{r}-\vec{r}_0} \, b_{xy}  \right).
\label{infinitesize}
\end{equation}
Under a small gauge transformation $a_{xy}\rightarrow a_{xy}+ \Delta_{xy}\xi$, the extended operator $\exp  \left(i w_e \ell \sum_{xy} f_{\vec{r}-\vec{r}_0}\, a_{xy}  \right)$ changes by a factor
\begin{equation}
	\exp  \left(i w_e \ell \sum_{xy} f_{\vec{r}-\vec{r}_0}\,  \Delta_{xy}\xi \right).
\end{equation}
Using \eqref{ibp} and \eqref{annihilation}, we see that 
\begin{equation}
	\exp  \left(i w_e \ell \sum_{xy} f_{\vec{r}-\vec{r}_0}\,  \Delta_{xy}\xi \right) = 	\exp  \left(i w_e \ell \sum_{xy}  \xi \Delta_{xy}'f_{\vec{r}-\vec{r}_0} \right)=1,
\end{equation}
which shows that $\exp  \left(i w_e \ell \sum_{xy} f_{\vec{r}-\vec{r}_0}\, a_{xy}  \right)$  is gauge invariant. Actually, the nonexponentiated object $\sum_{xy} f_{\vec{r}}\, a_{xy} $ is gauge invariant itself. The need for the exponentiation comes when we consider the system at finite size along the $xy$-plane.

If the system is finite with $L$ being a multiple of 3, then it is possible to construct a gauge-invariant objects as
\begin{equation}
\exp  \left(i w_e \ell \osum_{xy} f_{\vec{r}}^{(\sigma)}\, a_{xy}  \right) ~~~\text{and}~~~\exp  \left(i w_m \ell \osum_{xy} g_{\vec{r}}^{(\sigma)}\, b_{xy}  \right), 
\label{extendeda}
\end{equation}
where $f_{\vec{r}}^{(\sigma)}$ and $g_{\vec{r}}^{(\sigma)}$, with $\sigma=1,2$, are defined in \eqref{functions} and \eqref{functions'}. In addition, $f_{\vec{r}}^{(\sigma)}$ and  $g_{\vec{r}}^{(\sigma)}$, satisfy 
\begin{equation}
	\osum_{xy} f_{\vec{r}}^{(\sigma)}=0~~~\text{and}~~~ 	\osum_{xy} g_{\vec{r}}^{(\sigma)}=0,
\end{equation}
ensuring that \eqref{extendeda} is also invariant under the transformations in \eqref{trans1} and \eqref{trans2}.

A large gauge transformation that winds around directions $\hat{\epsilon}_{1}$ and $\hat{\epsilon}_{2}$ is
\begin{equation}
\xi= 2\pi \left(\frac{i_1}{L}n_1+ \frac{i_2}{L}n_2\right), ~~~ n_1, n_2 \in \mathbb{Z}.
\end{equation}
In this case, there are no nonuniform transformations because there is no $t$ or $z$-dependent quantity annihilated by the operators $\partial_t$ and $\Delta_z$. 
The corresponding field transformations are
\begin{equation}
	a_t\rightarrow a_t,~~~a_z\rightarrow a_z,~~~a_{xy}\rightarrow a_{xy} + 2\pi \left(\frac{3 i_1+1}{\ell L}n_1+ \frac{3 i_2+1}{\ell L}n_2\right).
\end{equation}
The quantity $\displaystyle\osum_{xy}f_{\vec{r}}^{(\sigma)}\, a_{xy} $ in \eqref{extendeda} is invariant under these transformations, still without the need for exponentiation. But we can also conceive the following gauge transformation
\begin{equation}
\xi = 2\pi \frac{i_1 i_2}{L^2} n_{12}, ~~~n_{12}\,\in\,\mathbb{Z}. 
\end{equation}
The field transformations are
\begin{equation}
	a_t\rightarrow a_t,~~~a_z\rightarrow a_z,~~~a_{xy}\rightarrow a_{xy} + 2\pi \left(\frac{3 i_1 i_2+i_1+i_2}{\ell L^2 }n_{12}\right),
	\label{trans3'}
\end{equation}
according to which
\begin{equation}
\osum_{xy}  f_{\vec{r}}^{(\sigma)}\, a_{xy} \rightarrow \osum_{xy} f_{\vec{r}}^{(\sigma)}\, a_{xy} + 2\pi \frac{n_{12}}{\ell}.	
\end{equation}
Invariance under this gauge transformation implies that this object needs to be exponentiated as in \eqref{extendeda}, with $w_e$ being an integer. All these discussions follow for the operators involving the $b$-fields and also imply that $w_m$ is an integer.

Now we consider the algebra between such extended operators. The commutation relations following from \eqref{ab} are
\begin{equation}
[a_z(\vec{r}), b_{xy}(\vec{r}')]= - \frac{2\pi i}{q} \frac{\delta_{\vec r, \vec r'}}{\ell  \ell_z}~~~\text{and}~~~ [a_{xy}(\vec{r}), b_{z}(\vec{r}')]=  \frac{2\pi i}{q} \frac{\delta_{\vec r, \vec r'}}{\ell \ell_z}.
\label{commut}
\end{equation}
These commutators lead to
\begin{equation}
\exp \left(i v_e\, \ell_z \osum_z a_z \right) \exp  \left(i w_m \ell \osum_{xy}g_{\vec{r}}^{(\sigma)}\, b_{xy}  \right)=   e^{-\frac{2\pi i}{q} v_e w_m g_{\vec{r}}^{(\sigma)}}\exp  \left(i w_m \ell \osum_{xy} g_{\vec{r}}^{(\sigma)}\, b_{xy}  \right)\exp \left(i v_e\, \ell_z \osum_z a_z \right) 
\label{al1}
\end{equation}
and
\begin{equation}
	\exp  \left(i w_e \ell \osum_{xy}f_{\vec{r}}^{(\sigma)}\, a_{xy}  \right) \exp \left(i v_m\, \ell_z \osum_z b_z \right)=  e^{\frac{2\pi i}{q} w_e v_m f_{\vec{r}}^{(\sigma)}}	\exp \left(i v_m\, \ell_z \osum_z b_z \right)\exp  \left(i w_e \ell \osum_{xy}f_{\vec{r}}^{(\sigma)}\, a_{xy}  \right).
	\label{al2}
\end{equation}
These relations make clear that the integers $v_e, v_m, w_e$, and $w_m$ are all defined mod $q$, since the corresponding operators behave as the identity whenever any one of these integers is a multiple of $q$. Naturally, the same conclusion extends to the integers $u_e$ and $u_m$, since the defects are simply rotated versions of the line operators along the $z$-direction. The integer-valued functions $f_{\vec{r}}^{(\sigma)}$ and $g_{\vec{r}}^{(\sigma)}$ are also defined mod $q$. In sum, the algebra of the extended operators corresponds to a $\mathbb{Z}_q$ algebra. The phase factors in \eqref{al1} and \eqref{al2} imply that the corresponding symmetries possess mix 't Hooft anomalies, which will be discussed shortly from the functional perspective.

\subsection{Generalized Global  Symmetries}

Now that we have constructed all the extended objects, let us discuss the generalized global symmetries acting on them. Consider the transformations
\begin{equation}
a_t \rightarrow a_t + \frac{2\pi }{q} \frac{g_{\vec{r}}^{(\sigma)}}{\tilde{L}_t},~~~ a_z\rightarrow a_z,~~~a_{xy}\rightarrow a_{xy}.
\label{global0}
\end{equation}
They look like a gauge transformation as in \eqref{trans1'} but this is not the case because of the factor of $q$ (they cannot be undone by a gauge transformation). They correspond to a global symmetry, which acts nontrivially on the defect $\exp \left( i u_e \oint dt \, a_t \right)$, 
\begin{equation}
\exp \left( i u_e \oint dt \, a_t \right)  \rightarrow  \exp \left( i u_e \frac{2\pi}{q} g_{\vec{r}}^{(\sigma)} \right)  \exp \left( i u_e \oint dt \, a_t \right). 
\end{equation}
The factor $\exp \left( i u_e \frac{2\pi}{q} g_{\vec{r}}^{(\sigma)} \right)$ is the charge of the defect under the global symmetry. As it depends on the position in the $xy$-plane through $g_{\vec{r}}^{(\sigma)}$, this implies that a single excitation cannot move along the $xy$-plane, otherwise it will violate the conservation of the charge. On the other hand, there is no violation in moving along $z$-direction, which means that a single excitation is mobile in that direction. This is how the mobility of excitations can be understood in terms of generalized global symmetries. 
 
The global symmetries acting on the extended operators are 
\begin{equation}
a_t\rightarrow a_t,~~~a_z \rightarrow a_z + \frac{2\pi}{q}\frac{g_{\vec{r}}^{(\sigma)}}{\ell_z L_z} , ~~~ a_{xy}\rightarrow a_{xy} 
\label{globalz}
\end{equation} 
and
\begin{equation}
a_t\rightarrow a_t,~~~a_z\rightarrow a_z,~~~a_{xy}\rightarrow a_{xy} + \frac{2\pi}{q} \left(\frac{3 i_1 i_2+i_1+i_2}{\ell L^2 }n_{12}\right).
\label{globalxy}
\end{equation}
As in the previous case, these transformations look like the gauge transformations in \eqref{trans2'} and \eqref{trans3'} but are global symmetries. The extended operators transform respectively as 
\begin{equation}
\exp \left(i v_e\, \ell_z \osum_z a_z \right) \rightarrow \exp\left(i v_e \frac{2\pi}{q}g_{\vec{r}}^{(\sigma)} \right) \exp \left(i v_e\, \ell_z \osum_z a_z \right) 
\end{equation}
and
\begin{equation}
\exp  \left(i w_e \ell \osum_{xy} f_{\vec{r}}^{(\sigma)}\, a_{xy}  \right)   \rightarrow\exp \left(i w_e \frac{2\pi}{q} n_{12}\right) \exp  \left(i w_e \ell \osum_{xy} f_{\vec{r}}^{(\sigma)}\, a_{xy}  \right). 
\end{equation}


\subsection{Fractal Membrane Operators}

So far the constraints that all the charges of the extended objects are defined mod $q$ have been obtained for sizes $L$ being a multiple of 3. Nevertheless, we shall assume these constraints for any $L$. In particular, for sizes $L=k (q^n-q^m)$, with $q$ being a prime, the operators in \eqref{infinitesize} turn out to be compatible with periodic boundary conditions due to the fact that $w_e$, $w_m$, $ f_{\vec{r}-\vec{r}_0}$, and $g_{\vec{r}-\vec{r}_0}$ are defined mod $q$ \footnote{More precisely, certain compositions of these operators  or, equivalently, certain linear combinations of $f$'s and $g$'s are compatible with periodic boundary conditions.}. However, not all $L^2$ operators $	\exp  \left(i w_e \ell \sum_{xy} f_{\vec{r}-\vec{r}_0}\, a_{xy}  \right)$ specified by $\vec{r}_0$ are independent, but only a number $q^n - q^{m+1}$. Therefore, we label the independent operators compatible with the periodic boundary conditions by $ f_{\vec{r}-\vec{r}_0} \rightarrow f_{\vec{r}}^I$, i.e., 
\begin{equation}
	\exp  \left(i w_e \ell \sum_{xy} f_{\vec{r}}^I\, a_{xy}  \right)~~~\text{and}~~~ \exp  \left(i w_m \ell \sum_{xy} g_{\vec{r}}^I \, b_{xy}  \right),
	\label{M}
\end{equation}
with $I=1,2,\ldots, q^n - q^{m+1}$. These operators are nothing else but the $\mathbb{Z}_q$ generalization of the membrane operators of the model \eqref{eq:hamiltonian} for any prime $q$. For the case $q=2$, they correspond precisely to the membrane operators in  \eqref{eq:membrane}. This correspondence becomes more evident rewriting \eqref{M} as 
\begin{equation}
	\exp  \left(i \ell \sum_{xy \,\in\, \mathcal{M}_I^a}  a_{xy}  \right)~~~\text{and}~~~ \exp  \left(i \ell \sum_{xy\,\in\, \mathcal{M}_I^b}  b_{xy}  \right),
\end{equation}
where we have used that $f_{\vec{r}}^I$ and $g_{\vec{r}}^I$ are defined mod 2.


\subsection{Mixed 't Hooft Anomalies from the Partition Function}

To discuss the mixed 't Hooft anomalies from the functional perspective it is very convenient to resort to the notation of \eqref{ab}, which we write explicitly as
\begin{equation}
b\wedge Da \equiv \epsilon_{\mu\nu\rho} b_{\mu}D_{\nu} a_{\rho}, ~~~\mu,\nu,\rho=0,1,2,
\end{equation}
with 
\begin{equation}
a_{\mu}\equiv(a_t, a_{xy}, a_z),~~~ b_{\mu}\equiv(b_t, b_{xy}, b_z),~~~D_{\mu}\equiv(\partial_{t}, \Delta_{xy}, \Delta_z).
\end{equation}
In this notation, integration by parts corresponds to
\begin{equation}
\int b\wedge Da = \int a \wedge D'b,
\end{equation}
where 
\begin{equation}
D'_{\mu}\equiv(\partial_{t}, -\Delta_{xy}', \Delta_z).
\end{equation}

The generalized symmetries in \eqref{global0}, \eqref{globalz}, and  \eqref{globalxy} can be unified as
\begin{equation}
a_{\mu}\rightarrow a_{\mu} +\alpha_{\mu}. 
\end{equation}
Similarly, we have the global symmetries for the $b$-fields
\begin{equation}
	b_{\mu}\rightarrow b_{\mu} +\beta_{\mu}. 
\end{equation}

We can now see that $\alpha$ and $\beta$ symmetries have a mixed 't Hooft anomaly, as they cannot be simultaneously gauged. To see this, let us first gauge the $\alpha$-symmetries, by promoting $\alpha$ to arbitrary functions, which are no longer given by  \eqref{global0}, \eqref{globalz}, and  \eqref{globalxy}. This needs to be accompanied by  background gauge fields $A$ in the partition function
\begin{eqnarray}
	\mathcal Z_{\text{IR}}[\mathcal A]= \int [\mathcal D a][\mathcal D b] \, e^{\frac{\mathrm i q}{2\pi}\int b\, \wedge \,(Da- A) },
\end{eqnarray}
with the background fields $A$ transforming as
\begin{eqnarray}
 A\rightarrow  A+ D\alpha.
\end{eqnarray}
It follows immediately that the partition function is invariant, i.e., $\mathcal Z_{\text{IR}}[A] = \mathcal Z_{\text{IR}}[A+D\alpha]$.

Similarly, introducing background gauge fields $B$ to cancel the additional contributions from gauging the $\beta$-symmetries, we obtain
\begin{eqnarray}
	\mathcal Z_{\text{IR}}[B]= \int [\mathcal D a][\mathcal D b] \, e^{\frac{\mathrm i q}{2\pi}\int b\, \wedge\, Da- a \,\wedge\, B }.
\end{eqnarray}
This partition function is invariant under the gauge transformations
\begin{eqnarray}
	b\rightarrow b+\beta\quad \text{and}\quad  B \rightarrow  B+D'\beta,
\end{eqnarray}
namely, $ \mathcal Z_{\text{IR}}[ B] = \mathcal Z_{\text{IR}}[B+D'\beta]$.

However, there is no way to gauging both $\alpha$ and $\beta$ symmetries simultaneously. Consider the partition function with both background fields
\begin{eqnarray}
	\mathcal Z_{\text{IR}}[ A,  B]= \int [\mathcal D a][\mathcal D b] \, e^{\frac{\mathrm i q}{2\pi}\int b\, \wedge \, (Da- A)-a \,\wedge\, B}.
\end{eqnarray}
Under a gauge transformation, say, involving the $\alpha$-symmetries, the partition function changes as 
\begin{eqnarray}
	\mathcal Z_{\text{IR}}[A+D \alpha, B] =e^{-\frac{\mathrm{iq}}{2\pi} \int  \alpha\, \wedge \, B} \mathcal Z_{\text{IR}}[ A,  B].
\end{eqnarray}
The 't Hooft anomaly manifests through the phase $e^{-\frac{\mathrm{iq}}{2\pi} \int  \alpha\, \wedge \, B}$, which cannot be canceled by any local counterterm involving only the background fields $A$ and $B$. As in the case of usual BF theory, such anomaly can be canceled if we consider the system as the boundary of a system in $(4+1)D$. In this case, it is possible to construct a local counterterm in $(4+1)D$ whose variation under a gauge transformation is a boundary term that precisely cancels the phase $e^{-\frac{\mathrm{iq}}{2\pi} \int  \alpha\, \wedge \, B}$.


\subsection{Boundary Theory}

We place the theory \eqref{IR} on a manifold periodic in $x$ and $y$ directions but with an open boundary in the $z$-direction, at $z=0$. In this situation, the bulk action is no longer invariant under the gauge transformation~\eqref{eq:Agauge} and~\eqref{eq:Bgauge}, but changes by boundary terms. This implies that gauge degrees of freedom become physical ones at the edge. To explore this point, let us consider the bulk equations of motion for $a_t$ and $b_t$:
\begin{equation}
\Delta_{xy}a_z- \Delta_{z}a_{xy}=0~~~\text{and}~~~ \Delta_{xy}'b_z+\Delta_{z}b_{xy}=0.
\end{equation}
The solutions of these conditions can be parametrized in terms of two scalars fields $\varphi$ and $\theta$ as
\begin{equation}
a_z=\Delta_{z}\varphi,~~~ a_{xy}=\Delta_{xy}\varphi~~~\text{and}~~~ b_z=-\Delta_{z}\theta,~~~ b_{xy}=\Delta_{xy}'\theta.
\label{par}
\end{equation}
Now we use these parametrizations in the line operators \eqref{line}, which become local operators at the boundary
\begin{equation}
	\exp \left(i \, \ell_z \sum_{i_3=-\infty}^0 a_z \right) = 	\exp \left(i \, \ell_z \sum_{i_3=-\infty}^0 \Delta_{z}\varphi \right) = e^{i \varphi(i_1,i_2)}
	\label{local1}
\end{equation}
and
\begin{equation}
\exp \left(i \, \ell_z \sum_{i_3=-\infty}^0  b_z \right)=\exp \left(-i \, \ell_z \sum_{i_3=-\infty}^0  \Delta_z\theta \right)= e^{-i\theta(i_1,i_2)}.
\label{local2}
\end{equation}

The commutation relations in \eqref{commut} imply nontrivial commutations at the boundary. From the first one, we can write
\begin{equation}
	\sum_{i_3=-\infty}^0[a_z(i_i,i_2,i_3), b_{xy}(j_1,j_2,0)]= - \frac{2\pi i}{q}\sum_{i_3=-\infty}^0  \frac{\delta_{i_1,j_1} \delta_{i_1,j_2}\delta_{i_3,0}}{\ell \ell_z}.
\end{equation}
Using the parametrizations in \eqref{par}, we get a commutation rule at the boundary
\begin{equation}
[\varphi(i_i,i_2), \Delta_{xy}'\theta(j_1,j_2)]=  - \frac{2\pi i}{q}  \frac{\delta_{i_1,j_1} \delta_{i_1,j_2}}{\ell}.
\end{equation}
Proceeding in a similar way, the second commutation rule in \eqref{commut} leads to
\begin{equation}
	[\theta(i_i,i_2), \Delta_{xy}\varphi(j_1,j_2)]=   \frac{2\pi i}{q}  \frac{\delta_{i_1,j_1} \delta_{i_1,j_2}}{\ell}.
\end{equation}

Now we can construct an edge theory by requiring that it produces the above commutations relations. This fixes the term in the edge theory involving time derivative
\begin{equation}
S_{\text{edge}} = \int dt \, \ell \sum_{xy} \frac{q}{2\pi} \Delta_{xy}\varphi \, \partial_{t}\theta + \cdots.
\end{equation} 
The next step in constructing the edge theory is to include terms in this action which are compatible with the symmetries. 

From the local operators \eqref{local1} and \eqref{local2} at the boundary, we can construct the operators 
\begin{equation}
e^{i \ell \Delta_{xy}\varphi}~~~\text{and}~~~ e^{-i \ell \Delta_{xy}'\theta}.
\end{equation}
These operators generate the following transformations
\begin{equation}
e^{-i \ell \Delta_{xy}'\theta} \,e^{i \varphi}\,  e^{i \ell\Delta_{xy}'\theta} =  e^{i \left(\varphi +\frac{2\pi}{q}\right) }
\end{equation}
and
\begin{equation}
e^{i \ell \Delta_{xy}\varphi}  \,e^{-i \theta}\,e^{-i \ell \Delta_{xy}\varphi} = e^{-i \left(\theta +\frac{2\pi}{q}\right) },
\end{equation}
which correspond to the shifts in the scalar fields
\begin{equation}
\varphi \rightarrow \varphi +\frac{2\pi}{q}~~~\text{and}~~~\theta \rightarrow \theta +\frac{2\pi}{q}. 
\label{shift1}
\end{equation}

For sizes $L=k(q^n-q^m)$, we also have the fractal operators, that can be placed at the boundary $i_3=0$:
\begin{equation}
	\exp  \left(i  \ell \sum_{xy} f_{\vec{r}}^I\, \Delta_{xy}\varphi  \right)~~~\text{and}~~~ \exp  \left(i  \ell \sum_{xy} g_{\vec{r}}^I \, \Delta_{xy}'\theta  \right).
	\label{Medge}
\end{equation}
They generate the following shifts in the fields
\begin{equation}
\varphi \rightarrow \varphi +\frac{2\pi}{q}g_{\vec{r}}^I~~~\text{and}~~~\theta \rightarrow \theta +\frac{2\pi}{q}f_{\vec{r}}^I.
\label{shift2}
\end{equation}

Then, the edge theory must involve only operators which are invariant under the shift symmetries \eqref{shift1} and \eqref{shift2}. With this requirement, the edge theory containing the simplest operators is 
\begin{equation}
	\begin{aligned}
		S_{\text{edge}} = \int \!\!\! dt \, \ell \sum_{xy} & \left[ \frac{q}{2\pi} \Delta_{xy}\varphi \, \partial_{t}\theta - K_1 (\Delta_x \Delta_{xy}\varphi)^2 - K_2 (\Delta_y \Delta_{xy}\varphi)^2\right. \\
		& -
		K_{12} (\Delta_x \Delta_{xy}\varphi) (\Delta_y \Delta_{xy}\varphi) -  K_1' (\Delta_x \Delta_{xy}'\theta)^2 - K_2' (\Delta_y \Delta_{xy}'\theta)^2\\
		&-
		K_{12}' (\Delta_x \Delta_{xy}'\theta) (\Delta_y \Delta_{xy}'\theta) +  \lambda \cos(q\,\varphi)+ \lambda' \cos(q\,\theta)\big{]},
	\end{aligned}
\end{equation} 
where $\Delta_x\varphi\equiv \frac{1}{\ell}(\varphi(\vec{r}-\hat\epsilon_1)-\varphi(\vec{r}))$ and  $\Delta_y\varphi\equiv \frac{1}{\ell}(\varphi(\vec{r}-\hat\epsilon_2)-\varphi(\vec{r}))$. 
As the two cosines involve operators that do not commute, they cannot be simultaneously relevant at the IR. Suppose that $ \cos(q\,\varphi)$ is relevant. Then, the field $\varphi$ is pinned at one of the values $\frac{2\pi }{q} n$, with $n$ an integer defined mod $q$. This implies that the $\mathbb{Z}_q$ shift symmetry of $\varphi$ is spontaneously broken. Similarly, if the $ \cos(q\,\theta)$ is relevant, the $\mathbb{Z}_q$ shift symmetry of $\theta$ is spontaneously broken. These two gapped phases are separated by a gapless critical point, where both symmetries are exact. This analysis leads to the conclusion that the edge theory is not trivially gapped.


\section{Final Remarks \label{sec:conclusion}}

The class of topological ordered phases that we have discussed in this work has very intriguing physical properties. They follow from the fractal subsystem symmetries which are extremely sensitive to the lattice sizes $L_x$ and $L_y$.  The result is a ground state degeneracy with an intricate dependence on the sizes $L_x$ and $L_y$. This form of UV/IR mixing leads to an interesting interplay between spontaneous and explicit breaking of fractal subsystem symmetries. 

Certain sizes can accommodate fractal membranes as symmetry operators, which have mixed 't Hooft anomalies with Wilson lines. Choosing the ground states as eigenstates of Wilson operators, the anomalies imply that the membrane operators act nontrivially on the ground states, meaning that these subsystem symmetries are spontaneously broken. For sizes in which fractal membranes fail to close properly, the fractal symmetry is explicitly broken and the ground state is unique.

We have also studied the boundary physics of these topological phases. This analysis is enlightening since it reveals an interesting notion of conservation of the amount of anomalies given in Eq. \eqref{nice}, which we repeat here for convenience
\begin{equation}
	\underbrace{2N_c}_{\text{bulk}}+\underbrace{(L^2-N_c)}_{\text{top boundary}}+\underbrace{(L^2-N_c)}_{\text{bottom boundary}}=2 L^2.
\end{equation}
For sizes in which the ground state is unique, $N_c=0$, only the boundaries contribute to the total amount of anomalies. On the other hand, for the cases where the ground state is degenerate, both bulk and boundaries give rise to anomalies. In either case, the total number of anomalies is $2 L^2$.

The discussion of the boundary theory is also quite helpful in the connection with two-dimensional models, which emerge in the limit $L_x=L_y \gg L_z$. Depending whether the two boundaries condense the same type of excitations or different types, the resulting two-dimensional system corresponds to a SSB or a SPT phase.

A delicate question that pervades all the fracton-like phases is concerning the effective field theory descriptions. The UV/IR mixing appears as an enormous obstacle to the derivation of  theories in the continuum and in some cases becomes practically insurmountable. 
These seem to be the cases of phases involving fractal symmetries, which are not much amenable to the continuum limit. The best we can do in such situations is to conceive an effective theory of low energies but not long distances. This is the strategy adopted in this work. 

Even without a theory in the continuum, the effective description we derived is rewarding in several aspects. It makes clear the content of generalized global symmetries of the system. 
In particular, the pattern of mobility of excitations is nicely described in terms of the position-dependent charges of the defects under the fractal subsystem symmetries. The mixed t' Hooft anomalies become very transparent in the path integral formulation supplied by the effective description. Finally, the effective theory incorporates in a very natural way the boundary physics.

	
\begin{acknowledgments}
	
We would like to thank Carlos Hernaski, Gustavo Yoshitome, and Julio Toledo for enlightening discussions. We also thank K. Sfairopoulos for useful comments on the manuscript.  G.D. is supported by DOE Grant No. DE-FG02-06ER46316.	
We also acknowledge the financial support from the Brazilian funding agencies CAPES and CNPq. 

\end{acknowledgments}
	

\appendix

\section{Fractal Configurations \label{sec:MS1}}

The configurations given by Eq.~\eqref{eq:basis} are the elements of a basis for the set of first lines,  $\left\{S_{1}(x)\right\}$, in the sense that they form a Module (a generalization of the notion of a vector space for fields defined in a ring\footnote{These elements can be added and multiplied by a $\mathbb{Z}_{2}$ scalar $\alpha$. The addition respect all axioms,  including associativity and commutativity, which hold trivially. The identity element is $S_{1}^{0} = 0$ (given by the ``repetition code'' $t_{1,i} = 0$ for all $i$) and each element is its own inverse.  Scalar multiplication by $\alpha = 0$ or $\alpha = 1$ respects the axioms trivially.}). As discussed in Sec.~\ref{sec:degeneracy}, for a size $L=2^{n}-2^{m}$ there are $2^{n}-2^{m+1}$ configurations $S_{1}^{i}(x)$ that span the set of all first lines that respect the PBC. Let us examine the details with some illustrative examples.

\begin{figure}
	\begin{centering}
		\includegraphics[width=0.2\linewidth]{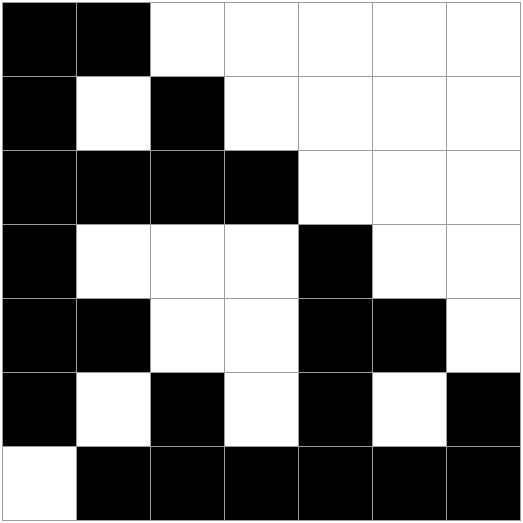} \hspace{0.02\linewidth} \includegraphics[width=0.2\linewidth]{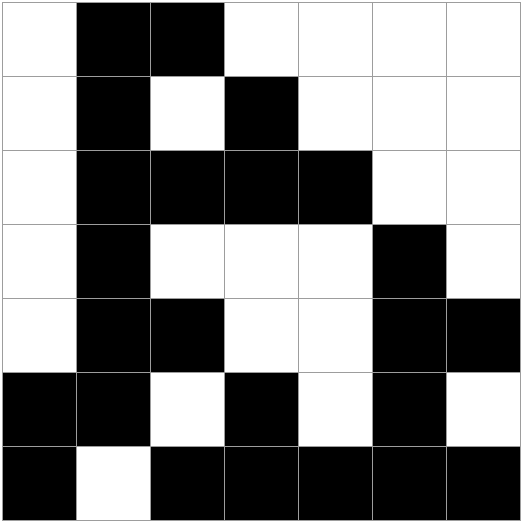} \hspace{0.02\linewidth} \includegraphics[width=0.2\linewidth]{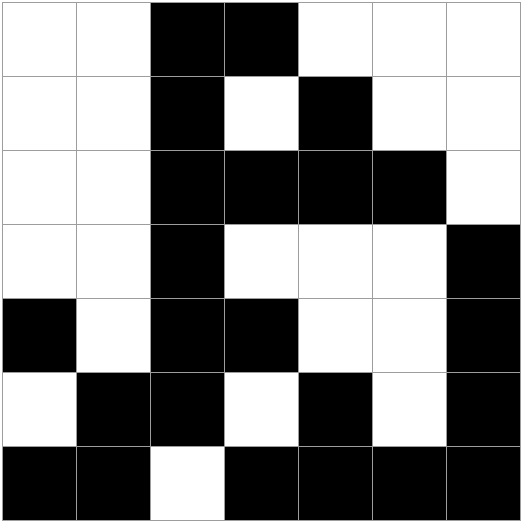} \\ \phantom{M} \\
		\includegraphics[width=0.2\linewidth]{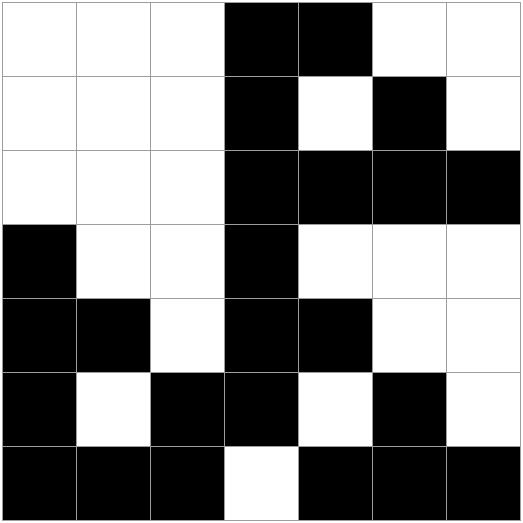} \hspace{0.02\linewidth} \includegraphics[width=0.2\linewidth]{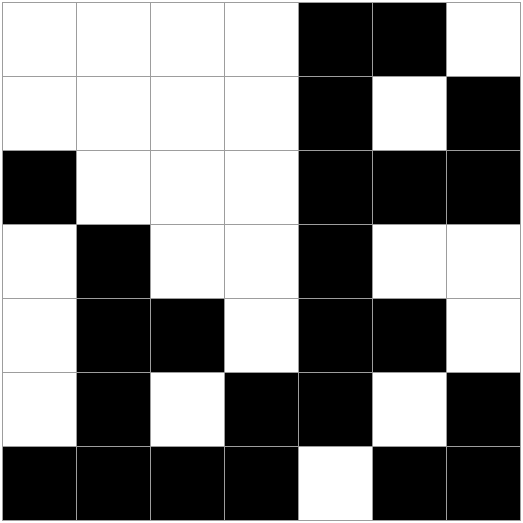} \hspace{0.02\linewidth} \includegraphics[width=0.2\linewidth]{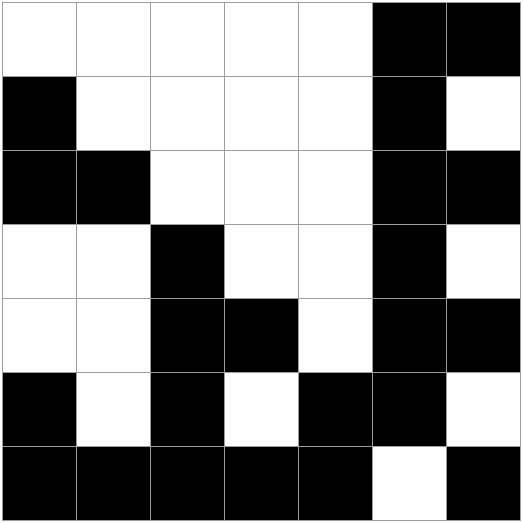} 
		\caption{Membranes generated by $S_{1}^{i}(x)$, where  $i=1,2, \, \dots \, ,6$, respectively, for a system size $L=7$.} \label{fig:7x7}
	\end{centering}
\end{figure}

\textit{Example 1:} Consider a system size $L=7$. This size can only be achieved with $n=3$ and $m=0$ (resulting in $L = 2^{3} - 2^{0} = 7$). Consequently, from Eq.~\eqref{eq:basis} we find the basis elements
\begin{equation}
	\begin{aligned}
		&S_{1}^{1}(x) = x+x^{2}
		&\hspace{0.1\linewidth} 
		&S_{1}^{2}(x) = x^{2}+x^{3} 
		& \hspace{0.1\linewidth} 
		& S_{1}^{3}(x) = x^{3}+x^{4} \\
		&S_{1}^{4}(x) = x^{4}+x^{5} 
		& \hspace{0.1\linewidth}
		&S_{1}^{5}(x) = x^{5}+x^{6} 
		&\hspace{0.1\linewidth} 
		&S_{1}^{6}(x) = x^{6}+x^{7},
	\end{aligned}  \label{eq:A1}
\end{equation}
represented in Fig.~\ref{fig:7x7}. These configurations are ``linearly independent'' elements in the sense that it is impossible to construct any of them as a combination of the other five. Note that $S_{1}^{7}(x)$ is excluded from \eqref{eq:A1} as it is not an independent configuration. It can be constructed as
\begin{equation}
	S_{1}^{7}(x) = x^{7}+x = \sum_{i=1}^{6}S_{1}^{i}(x).
\end{equation}

\begin{figure}
	\begin{centering}
		\includegraphics[width=0.2\linewidth]{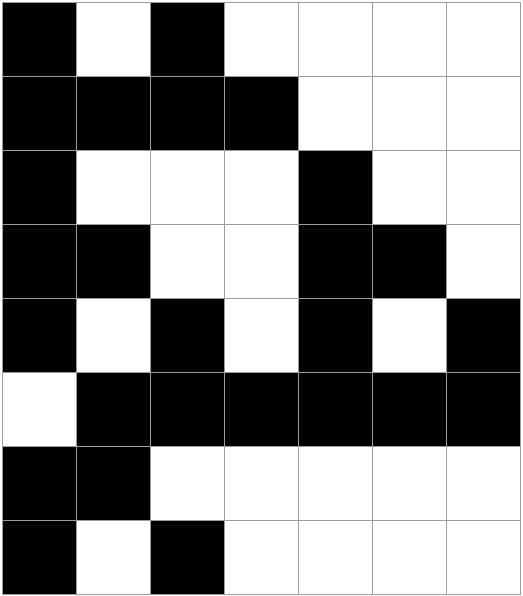} \hspace{0.02\linewidth} \includegraphics[width=0.2\linewidth]{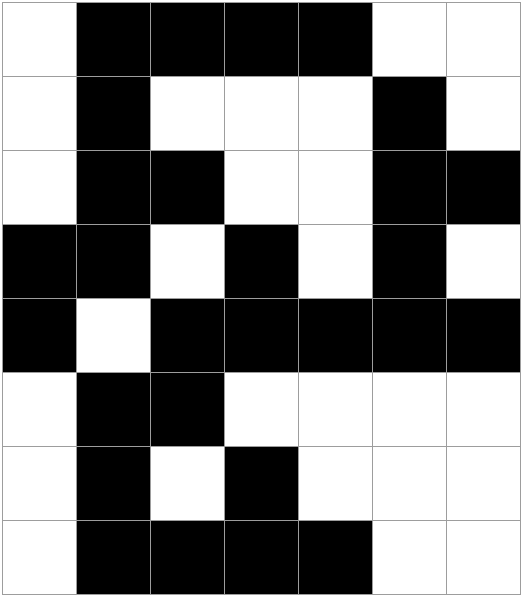} \hspace{0.02\linewidth} \includegraphics[width=0.2\linewidth]{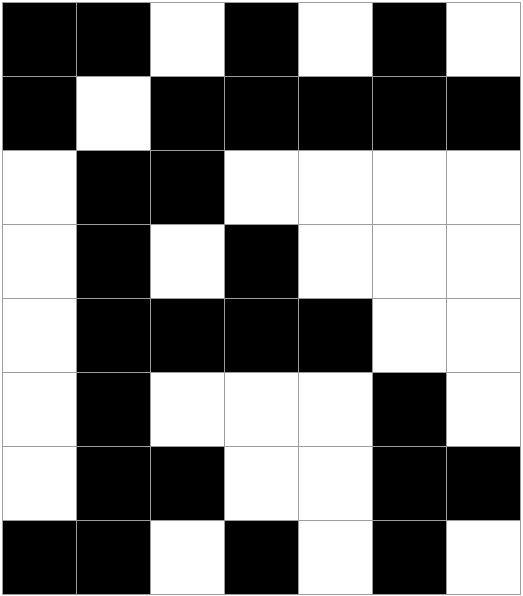} \hspace{0.02\linewidth} \includegraphics[width=0.2\linewidth]{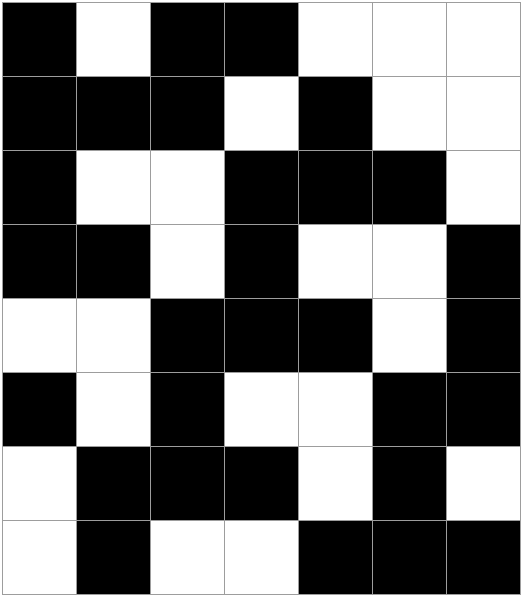}
		\caption{The first three configurations are generated by the  Eq.~\eqref{eq:A3} and the last one by Eq.~\eqref{eq:A72}. The configurations are depicted with eight rows and seven columns. The extra row included in the end is equivalent to the first one by periodic boundary conditions. We see that while the first three configurations are periodic in a size $7 \times 7$, the last one does not respect periodicity.} \label{fig:7x7others}
	\end{centering}
\end{figure}

Any other configuration that respects the periodicity can be constructed as combination of the $S_{1}^{i}(x)$ defined in \eqref{eq:A1}. For example, the configurations
\begin{equation}\label{eq:A3}
	\begin{aligned}
		x+x^{3} &= S_{1}^{1} + S_{1}^{2}, \\
		x + x^{2} + x^{4} + x^{6} &= S_{1}^{1} + S_{1}^{4} + S_{1}^{5}, \\
		x^{2} + x^{3}+x^{4}+x^{5} &= S_{1}^{2}+S_{1}^{4}
	\end{aligned}
\end{equation}
are depicted in Fig.~\ref{fig:7x7others}. Since the elements of \eqref{eq:A1} span the space of configurations, for $L=7$ the dimension of $\left\{S_{1}(x)\right\}$ is $6$. 

Configurations that cannot be constructed as combination of the elements of Eq.~\eqref{eq:A1} do not form closed structures under PBC. For example, consider the configuration $\tilde{S}_{1}(x) = x + x^{3} + x^{4}$, that cannot  be construct by combining the $S_{1}^{i}(x)$. This configuration do not satisfy $\tilde{S}_{1 + 7}(x) = \tilde{S}_{1}(x)$. Instead, 
\begin{eqnarray}\label{eq:A72}
	\tilde{S}_{1+7}(x) &=& \left(1+x\right)^{7} \tilde{S}_{1}(x) \nonumber \\
	&=& \left(x^{2} + x^{3} + x^{4} + x^{5} + x^{6}\right) \left(x+x^{3} + x^{4}\right) \nonumber \\
	&=& x^{2}+x^{5}+^{6}+x^{7} \neq \tilde{S}_{1}(x),
\end{eqnarray}
as depicted in Fig.~\ref{fig:7x7others}.

\begin{figure}
	\begin{centering}
		\includegraphics[width=0.2\linewidth]{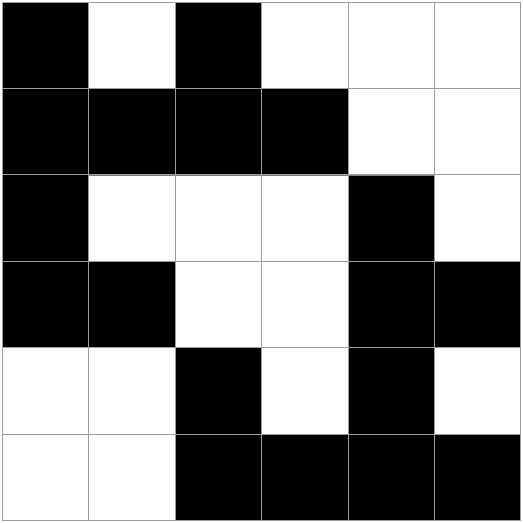} \hspace{0.02\linewidth} \includegraphics[width=0.2\linewidth]{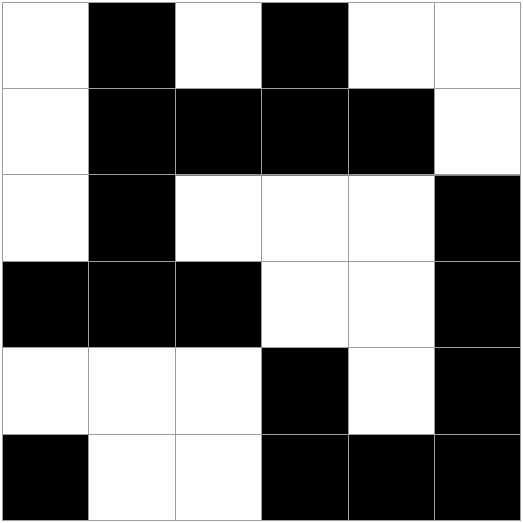} \hspace{0.02\linewidth} \includegraphics[width=0.2\linewidth]{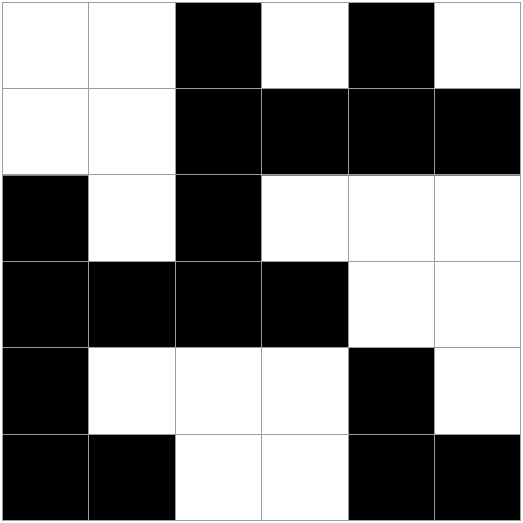}\hspace{0.02\linewidth} \includegraphics[width=0.2\linewidth]{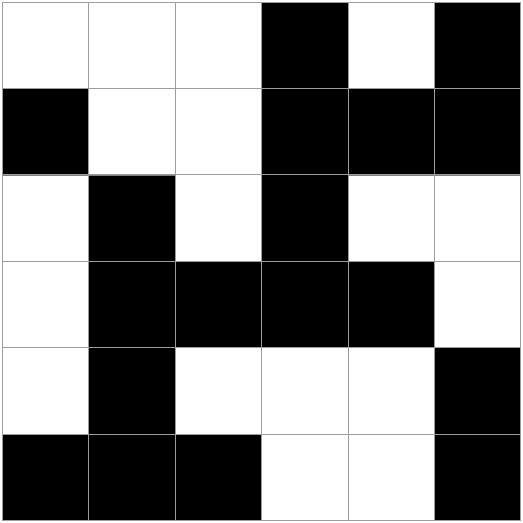}
		\caption{Membranes generated by $S_{1}^{i}(x)$, where  $i=1$, $2$, $3$, $4$, respectively, for a system size $L=6$.} \label{fig:6x6}
	\end{centering}
\end{figure}

\begin{figure}
	\begin{centering}
		\includegraphics[width=0.2\linewidth]{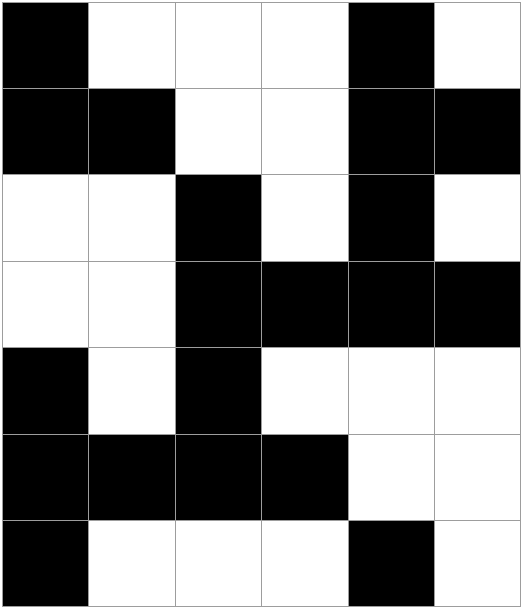} \hspace{0.02\linewidth} \includegraphics[width=0.2\linewidth]{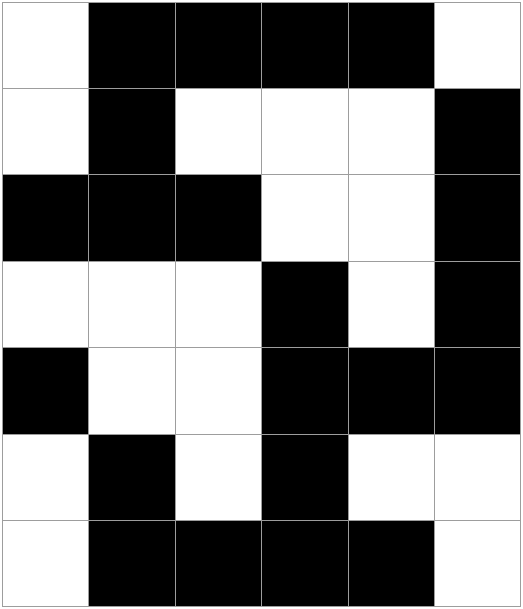} \hspace{0.02\linewidth} \includegraphics[width=0.2\linewidth]{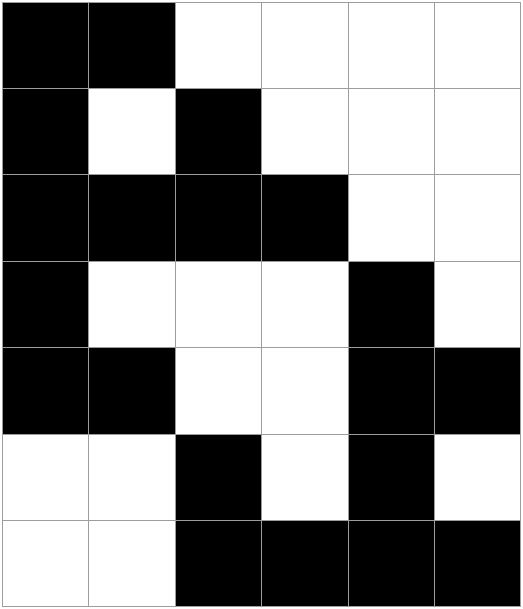}
		\caption{The first and second configurations are generated by Eq.~\eqref{eq:A6} and the last one is generated by Eq.\eqref{eq:A7}. The configurations are depicted with seven rows and six columns. The extra row included in the end is equivalent to the first one by periodic boundary conditions. We see that while the first two configurations are periodic in a size $6 \times 6$, the last one does not respect periodicity.} \label{fig:6x6others}
	\end{centering}
\end{figure}

\textit{Example 2:} Consider a system size $L=6$. The basis for this size is defined Eq.~\eqref{eq:basis} with $n=3$ and $m=1$. Consequently, the basis elements are
\begin{equation}
	\begin{aligned}
		&S_{1}^{1}(x) = x+x^{3} 
		&\hspace{0.13\linewidth}
		&S_{1}^{2}(x) = x^{2}+x^{4}\\
		&S_{1}^{3}(x) = x^{3}+x^{5}
		&\hspace{0.13\linewidth}
		& S_{1}^{4}(x) = x^{4}+x^{6},
	\end{aligned} \label{eq:A5}
\end{equation}
and are represented in Fig.~\ref{fig:6x6}. Once again, none of these elements can be constructed as combinations of the others. Other configurations that could appear as basis elements, $S_{1}^{5}(x)$ and $S_{1}^{6}(x)$, are excluded from Eq.~\eqref{eq:A5} because they are linear combinations,
\begin{equation}
	S_{1}^{5}(x) = x^{5} + x = S_{1}^{1}(x) + S_{1}^{3}(x)
	\hspace{0.05\linewidth}  \text{and} \hspace{0.05\linewidth}
	S_{1}^{6}(x) = x^{6} + x^{2} = S_{1}^{2}(x) + S_{1}^{4}(x).
\end{equation}
Any other possible first line configuration that can be constructed as combinations of the elements of \eqref{eq:A5}, such as
\begin{equation} \label{eq:A6}
	x+x^{5} = S_{1}^{1}(x)+S_{1}^{3}
	\hspace{0.07\linewidth}  \text{and} \hspace{0.07\linewidth}
	x^{2}+x^{3}+x^{4}+x^{5} = S_{1}^{2}(x)+S_{1}^{3}(x),
\end{equation}
are periodic, as depicted in Fig.~\ref{fig:6x6others}.

Configurations that cannot be expressed as combination of basis elements are excluded from the set of first lines because they do not form closed structures under PBC. For example, consider the configuration $\tilde{S}_{1}(x) = x+x^{2}$, which cannot  be construct combining elements of Eq.~\eqref{eq:A5}. This configuration does not satisfy $\tilde{S}_{7}(x) = \tilde{S}_{1}(x)$, 
\begin{eqnarray}\label{eq:A7}
	\tilde{S}_{1+6}(x) &=& \left(1+x\right)^{6} \tilde{S}_{1}(x) \nonumber \\
	&=& \left(1+x^{2}+x^{4}+x^{6}\right) \left(x+x^{2}\right) \nonumber \\
	&=& x^{3}+x^{4}+^{5}+x^{6} \neq \tilde{S}_{1}(x),
\end{eqnarray}
as depicted in Fig.~\ref{fig:6x6others}.

\begin{figure}
	\begin{centering}
		\includegraphics[width=0.2\linewidth]{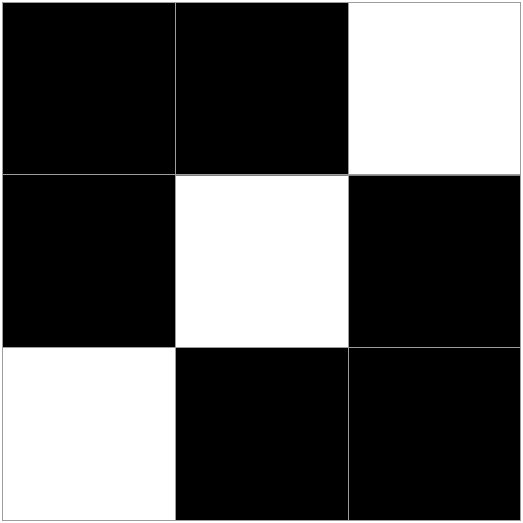} \hspace{0.02\linewidth} \includegraphics[width=0.2\linewidth]{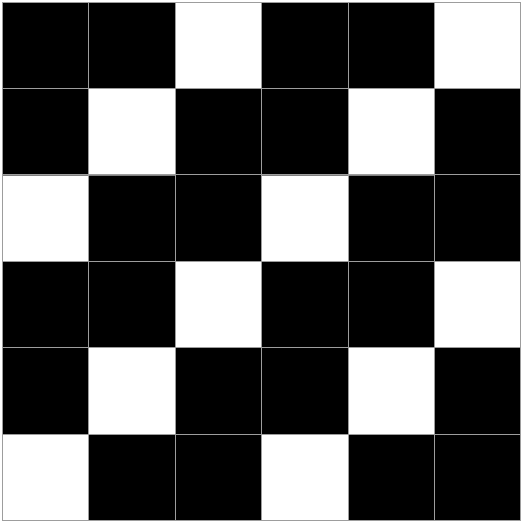} \hspace{0.02\linewidth} \includegraphics[width=0.2\linewidth]{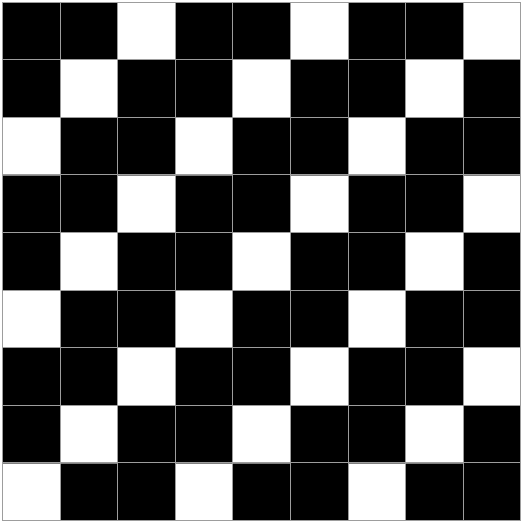}
		\caption{Periodic configurations for $3 \times 3$, $6 \times 6$, and $9 \times 9$, respectively.} \label{fig:9x9}
	\end{centering}
\end{figure}

\textit{Example 3:}  For system sizes where $k\neq 1$, the first line is constructed with $k$ copies of the $2^{n}-2^{m}$ configuration. Let us discuss an example to clarify this point. Consider $L=9$, corresponding to $k=3, n=2, m=0$. For this system size, the first line configuration, given by Eq. \eqref{eq:basis}, is equivalent to three copies of the $S_{1}^{i}(x)$ of $L=3$, as depicted in Fig.~\ref{fig:9x9}.

In a size $6 \times 6$ it is also possible to construct as copies of $3 \times 3$ configurations. However, the basis elements are given by the size $6$ configuration, instead of two cpoies of size $3$. Indeed, the $6 \times 6$ configuration depicted in Fig.~\ref{fig:9x9} is generated by $S_{1}^{2}(x) + S_{1}^{2} (x) + S_{1}^{3}(x)$ from Eq.~\eqref{eq:A5}.

\section{Boundary Anomalies  \label{sec:MS2}}

\begin{figure}
	\centering
	\includegraphics[width=0.9\linewidth]{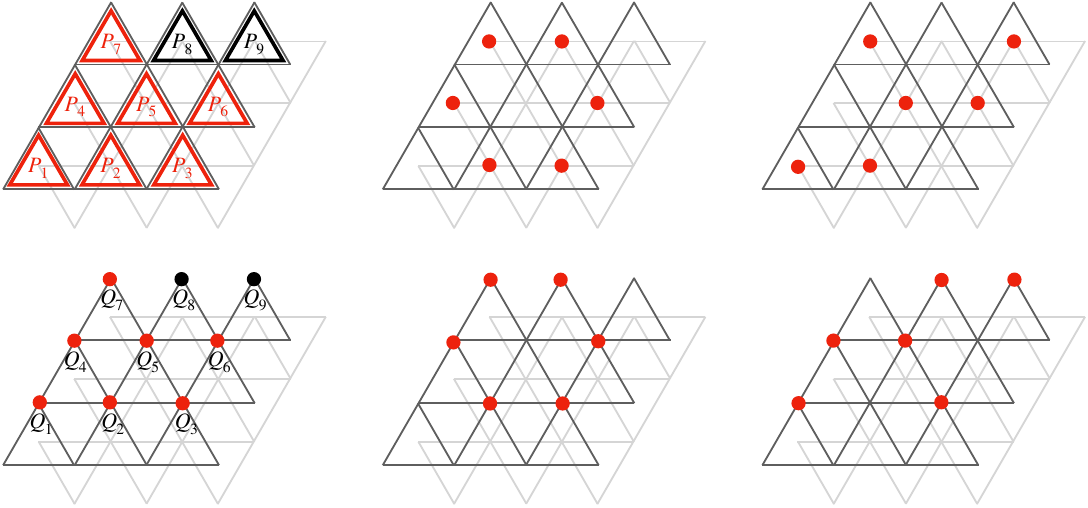}
	\caption{Representation of all boundary operators $P_{i}$ and $Q_{i}$, and the membranes operators on a $3 \times 3$ lattice.}
	\label{fig:tildeq1}
\end{figure}

In this section, we describe the construction of the operators $\tilde{Q}_{\bar{I}}$ that exhibit a one-to-one non-trivial commutation rule with $P_{\bar{I}}$ for degenerate system sizes. We will begin with an example before providing a general demonstration. Let us consider a boundary of a $3 \times 3$ lattice with operators $P_{i}$ and $Q_{i}$, in which $i = 1,\,2,\,\dots \, ,\,9$, as shown in Fig.~\ref{fig:tildeq1}.

\begin{figure}
	\centering
	\includegraphics[width=0.3\linewidth]{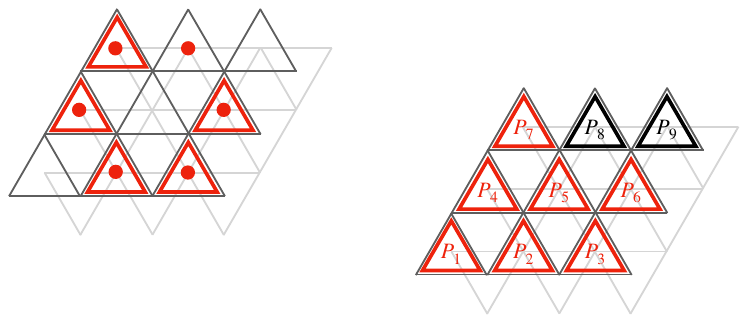}
	\caption{This figure demonstrates that a membrane combined with $P_{\bar{I}}$ operators results in $P_{8}$.}
	\label{fig:tildeq2}
\end{figure}

First, as discussed in Sec.~\ref{sec:boundanom}, notice that there are some constraints among the operators. Specifically, there are $N_{c} = 2$ constraints among $P_{i}$ operators and also $N_{c} = 2$ constraints among $Q_{i}$ operators. This means that two $P_{i}$ operators, namely $P_{8}$ and $P_{9}$, can be constructed as combinations of $P_{\bar{I}}$, $\bar{I} = 1$, $2$, $\dots$, $7$, and membranes. Similarly, two of $Q_{i}$ operators, $Q_{8}$ and $Q_{9}$, can be expressed as combinations of $Q_{\bar{I}}$ and membranes. In Fig.~\ref{fig:tildeq2} there is an example of these combinations. Therefore, 
\begin{equation}\label{eq:compset}
	\left\lbrace M^{1}_{I}, M^{2}_{I}, P_{\bar{I}}, Q_{\bar{I}} \right\rbrace
\end{equation}
forms a complete basis.

\begin{figure}
	\centering
	\includegraphics[width=0.9\linewidth]{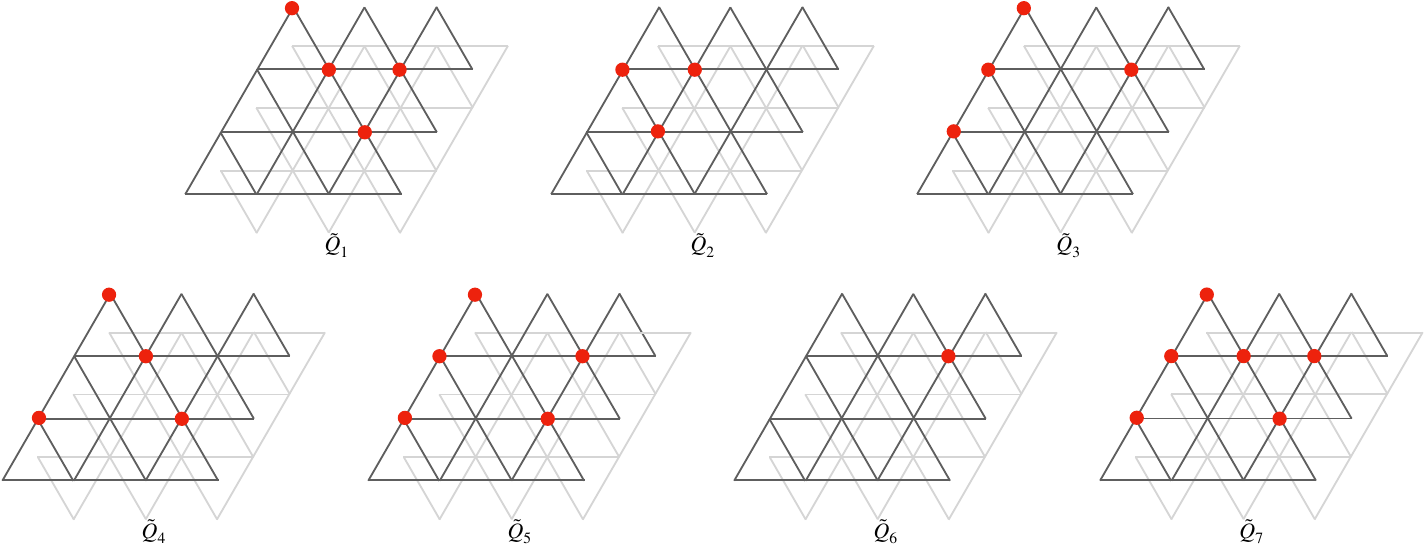}
	\caption{Representation of the $\tilde{Q}_{\bar{I}}$ operators on a $3 \times 3$ lattice.}
	\label{fig:tildeq3}
\end{figure}

Using only the $Q_{\bar{I}}$ operators, it is possible to define $\tilde{Q}_{\bar{I}}$ to have a one-to-one commutation rule with $P_{\bar{I}}$. The explicit definition of these operators are 
\begin{equation}
	\begin{aligned}
		&\tilde{Q}_{1} \equiv Q_{3} \, Q_{5} \, Q_{6} \, Q_{7}  
		& \qquad &\tilde{Q}_{2} \equiv Q_{2} \, Q_{4} \, Q_{5} & \qquad &\tilde{Q}_{3} \equiv Q_{1} \, Q_{4} \, Q_{6} \, Q_{7} & \\
		&\tilde{Q}_{4} \equiv Q_{1} \, Q_{3} \, Q_{5} \, Q_{7}  & \qquad & \tilde{Q}_{5} \equiv Q_{1} \, Q_{3} \, Q_{4} \, Q_{6} \, Q_{7}  & \qquad & \tilde{Q}_{6} \equiv  Q_{6} \\
		&\tilde{Q}_{7} \equiv Q_{1} \, Q_{3} \, Q_{4} \, Q_{5} \, Q_{6} \, Q_{7},
	\end{aligned}
\end{equation}
and they are graphically represented in Fig.~\ref{fig:tildeq3}. One can verify that these operators satisfy the commutation rules
\begin{equation}
	P_{\bar{I}} \tilde{Q}_{\bar{J}} = \left(-1\right)^{\delta_{\bar{I},\bar{J}}} 	\tilde{Q}_{\bar{J}} P_{\bar{I}}.
\end{equation}

\begin{figure}
	\centering
	\includegraphics[width=0.9\linewidth]{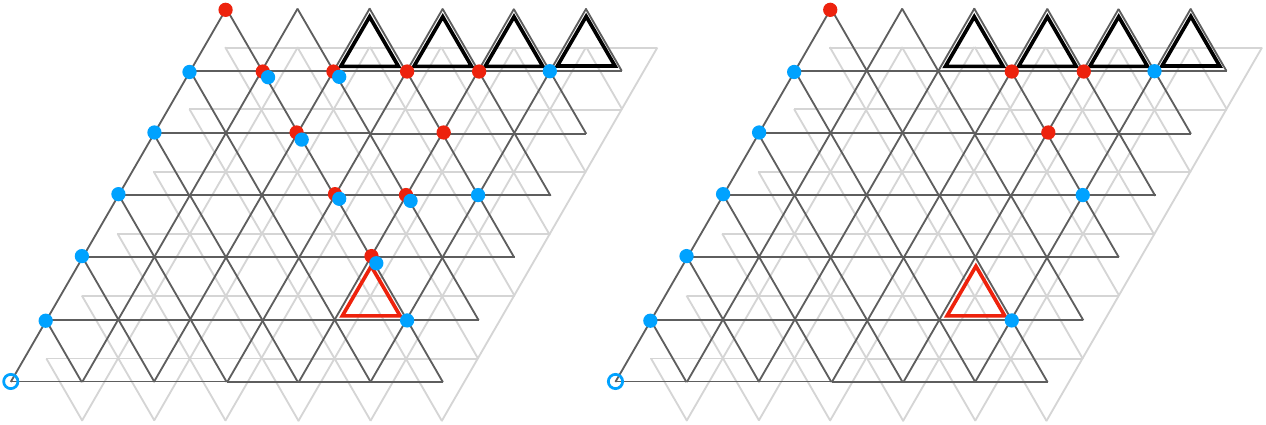}
	\vspace{20 pt}
	\includegraphics[width=0.9\linewidth]{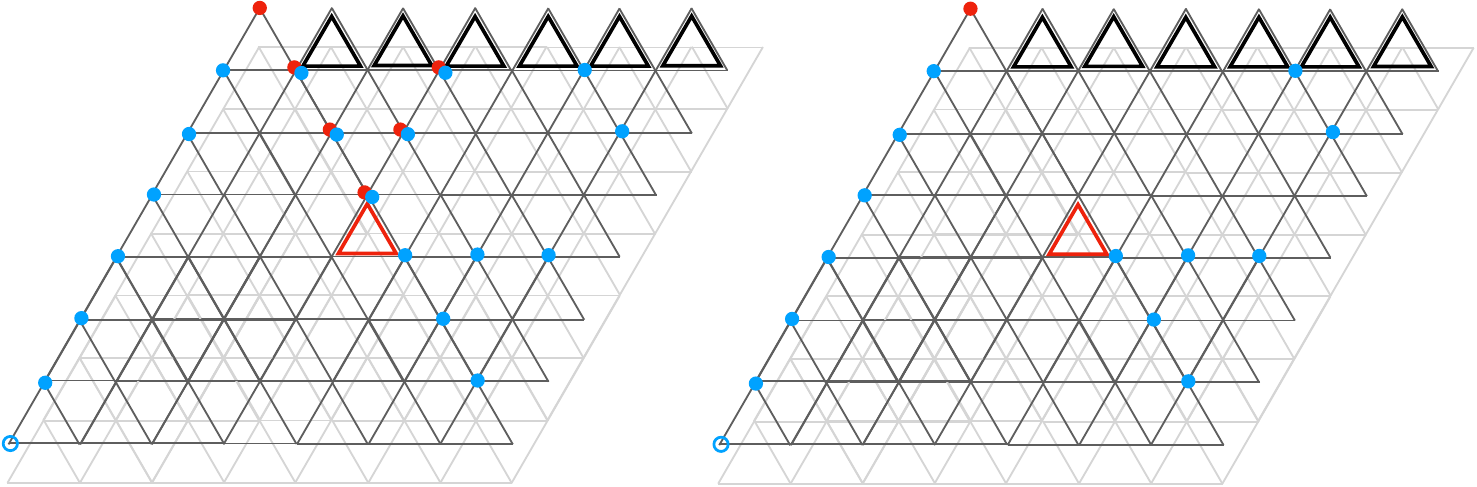}
	\caption{Construction of a $\tilde{Q}_{\bar{I}}$ operator in lattice sizes $6 \times 6$ and $7 \times 7$. Black triangles represent the $P_{i}$ operators that can be constructed as combination of $P_{\bar{I}}$ and the fractal membranes. On the left, the fractal structure begins with a red dot at the point $\bar{I}$, and  after winding, the blue dots indicate the subsequent points. On the right, it is shown the resulting structure.}
	\label{fig:tildeq5}
\end{figure}

There is a way to construct all $\tilde{Q}_{\bar{I}}$'s for a general system sizes, which we describe succinctly. It consists in combining the $Q_{\bar{I}}$'s into fractal structures, as illustrated in Fig.~\ref{fig:tildeq5}. The fractal structure begins with $Q_{\bar{I}}$ positioned on top of the $P_{\bar{I}}$ and winds around the plane. The points with $i \neq \bar{I}$ do not contribute to this product, and any superpositions that may come from the winding results in identities. The resulting extend operators have a non-trivial commutation rule only with the $P_{\bar{I}}$ where the fractal begins.

\bibliography{CC_refs-2}

\end{document}